\documentclass[journal=jctcce,manuscript=article]{achemso}
\usepackage{graphicx}
\usepackage{amsmath}
\usepackage{wrapfig}
\usepackage{physics}
\usepackage{caption}
\usepackage{hyperref}
\usepackage{subfiles}
\usepackage{longtable}

\usepackage[table]{xcolor}

\usepackage{xr}
\usepackage{multicol}

\graphicspath{ {figures/} }

\title{Statistical mechanical approximations to more efficiently determine polymorph free energy differences for small organic molecules}

\author{Nathan S. Abraham}
\affiliation{Department of Chemical and Biological Engineering, University of Colorado Boulder, Boulder, CO 80309, USA}
\author{Michael R. Shirts}
\affiliation{Department of Chemical and Biological Engineering, University of Colorado Boulder, Boulder, CO 80309, USA}
\email{michael.shirts@colorado.edu}

\begin{document}
\begin{singlespace}
\begin{multicols}{2}

\maketitle
\bibliographystyle{plain}

\section{abstract}
   Methods to efficiently determine the relative stability of polymorphs of organic crystals are highly desired in crystal structure predictions (CSPs). Current methodologies include use of static lattice phonons, quasi-harmonic approximation (QHA), and computing the full thermodynamic cycle using replica exchange molecular dynamics (REMD). We found that 13 out of the 29 systems minimized from experiment restructured to a lower energy minima when heated using REMD, a phenomena that QHA cannot capture. Here, we present a series of methods that are intermediate in accuracy and expense between QHA and computing the full thermodynamic cycle which can save 42--80\% of the computational cost and introduces, on this benchmark, a relatively small ($0.16\pm0.04$ kcal/mol) error relative to the full pseudosupercritical path approach. In particular, a method that Boltzmann weights the harmonic free energy of the trajectory of an REMD replica appears to be an appropriate intermediate between QHA and full thermodynamic cycle using MD when screening crystal polymorph stability.

\section{Introduction}
    Predicting the solid-solid stability of organic crystals \textit{a priori} is of great interest in the materials design of pharmaceuticals, energetic materials, and small molecule semi-conductors. Polymorphism, the presence of alternate packings in the solid phase, can lead to significant difference in the physiochemical and mechanical properties. It has been shown that solubility,~\cite{censiPolymorphImpactBioavailability2015,chemburkarDealingImpactRitonavir2000,romeroSolubilityBehaviorPolymorphs1999,yuScientificConsiderationsPharmaceutical2003,milosovichDeterminationSolubilityMetastable1964,singhalDrugPolymorphismDosage2004,sniderPolymorphismGenericDrug2004,llinasPolymorphControlPresent2008,brittainPolymorphismPharmaceuticalSolids2016,higuchiPolymorphismDrugAvailability1963,haleblianPharmaceuticalApplicationsPolymorphism1969,millerIdentifyingStablePolymorph2005} reactivity,~\cite{millarCrystalEngineeringEnergetic2012,vanderheijdenCrystallizationCharacterizationRDX2004,fabbianiHighpressureStudiesPharmaceutical2006} and charge transport~\cite{haasHighChargecarrierMobility2007,giriTuningChargeTransport2011,stevensTemperatureMediatedPolymorphismMolecular2015,chithambararajHydrothermallySynthesizedHMoO32016,valleOrganicSemiconductorsPolymorphism2004} can all be affected by the solid form. The appearance of new polymorphs in the late stages of development can be particularly problematic and potentially require reformulation.

    A class of prediction algorithms generally classified as ``crystal structure predictions'' (CSP) provide one avenue to determine solid stability, with recent emphasis on using free energy to rank the generated crystals. Given a molecule of interest, a CSP exhaustively generates possible solid arrangements and estimates their stability. The most common way to generate structures is through random or quasi-random generation techniques,~\cite{reillyReportSixthBlind2016,eddlestonDeterminationCrystalStructure2013, caseConvergencePropertiesCrystal2016,a.fosterPharmaceuticalPolymorphControl2017,shtukenbergPowderDiffractionCrystal2017} but annealing and genetic algorithms have also been used.~\cite{reillyReportSixthBlind2016,shtukenbergPowderDiffractionCrystal2017} Once crystals are generated, they are ranked on their thermodynamic stability as an approximation as to how likely they are to crystallize. Early development of CSPs relied on the static lattice energy to rank crystals, which would underestimate the importance of entropic affects. But more recently free energy techniques that approximate the entropy using harmonic phonons have been used for CSPs~\cite{shtukenbergPowderDiffractionCrystal2017,nymanAccuracyReproducibilityCrystal2019} and a significant focused has been placed on developing and probing what free energy techniques are necessary.~\cite{nymanStaticLatticeVibrational2015,nymanModellingTemperaturedependentProperties2016,dybeckEffectsMoreAccurate2016,dybeckCapturingEntropicContributions2017,abrahamThermalGradientApproach2018,abrahamAddingAnisotropyStandard2019}
  
    The harmonic approximation (HA) is the easiest way to include entropy by computing the static lattice harmonic phonons, which can be improved by considering thermal expansion through the quasi-harmonic approximation (QHA). Given a lattice minimum crystal, the harmonic phonons can be computed to determine the free energy of the crystal at all temperatures of interest.~\cite{maherSolubilityMetastablePolymorph2012,hasegawaReevaluationSolubilityTolbutamide2009,stolarSolidStateChemistryPolymorphism2016,grzesiakComparisonFourAnhydrous2003,sacchettiThermodynamicAnalysisDSC2000,cherukuvadaPyrazinamidePolymorphsRelative2010,vemavarapuCrystalDopingAided2002,yoshinoContributionHydrogenBonds1999,cesaroThermodynamicPropertiesCaffeine1980,torrisiSolidPhasesCyclopentane2008,badeaFusionSolidtosolidTransitions2007,alcobeTemperatureDependentStructuralProperties1994,cansellPhaseTransitionsChemical1993,boldyrevaEffectHighPressure2002,boldyrevaHighpressureDiffractionStudies2008,seryotkinHighpressurePolymorphChlorpropamide2013,nymanStaticLatticeVibrational2015} QHA more accurately represents the temperature-dependent change in the crystal than HA by including thermal expansion, which is done by determining the crystal volume that minimizes the free energy at the specified temperature. Typically QHA has been limited to isotropic expansion or quasi-anisotropic expansion,~\cite{nymanModellingTemperaturedependentProperties2016,ramirezPhaseDiagramIce2013,erbaThermalPropertiesMolecular2016,heitHowImportantThermal2016,dybeckCapturingEntropicContributions2017} but recently we have developed methods to perform anisotropic expansion.~\cite{abrahamThermalGradientApproach2018} For a number of systems we have shown approximations that can be made to anisotropic QHA so that QHA only costs the user 7$\times$ the cost of performing isotropic QHA.~\cite{abrahamAddingAnisotropyStandard2019} The simplicity and low computational cost of harmonic approaches is desirable in a CSP where 50--200 structures could exist within 2.5 kcal/mol of the global minimum and therefore screening at some level for entropy contributions to stability is useful. 

    The exact free energy difference between two crystals, given a molecular model, can be determined by relating them along a thermodynamic cycle using molecular dynamics (MD), but is prohibitively expensive. In molecular dynamics two polymorphs are simulated with NPT simulations over a specified temperature range, and the free energy as a function of temperature is estimated.  These functions of $G(T)$ for each polymorph can then be related by determining their free energy difference at a reference temperature.~\cite{dybeckEffectsMoreAccurate2016,dybeckCapturingEntropicContributions2017} One way that a reference free energy can be computed is by simulating the crystal along the pseudosupercritical path (PSCP), which restrains the atoms and removes all molecular interactions.~\cite{dybeckEffectsMoreAccurate2016,jayaramanComputingMeltingPoint2007,eikeRobustGeneralMolecular2005,eikeAtomisticSimulationSolidliquid2006,paluchMethodComputingSolubility2010} At this final state the free energy difference between ``crystals'' will be zero. The full PSCP with MD approach is being used in a CSP framework in an industrial settings (currently, only described in conference proceedings), but the use of this technique is limited due to the computational expense. A full free energy determination using the PSCP approach with MD can cost up to three orders of magnitude more than QHA with even classical fixed-charge potentials, let alone polarizable potentials.

\begin{figure*}
  \begin{center}
  \includegraphics[width=14cm]{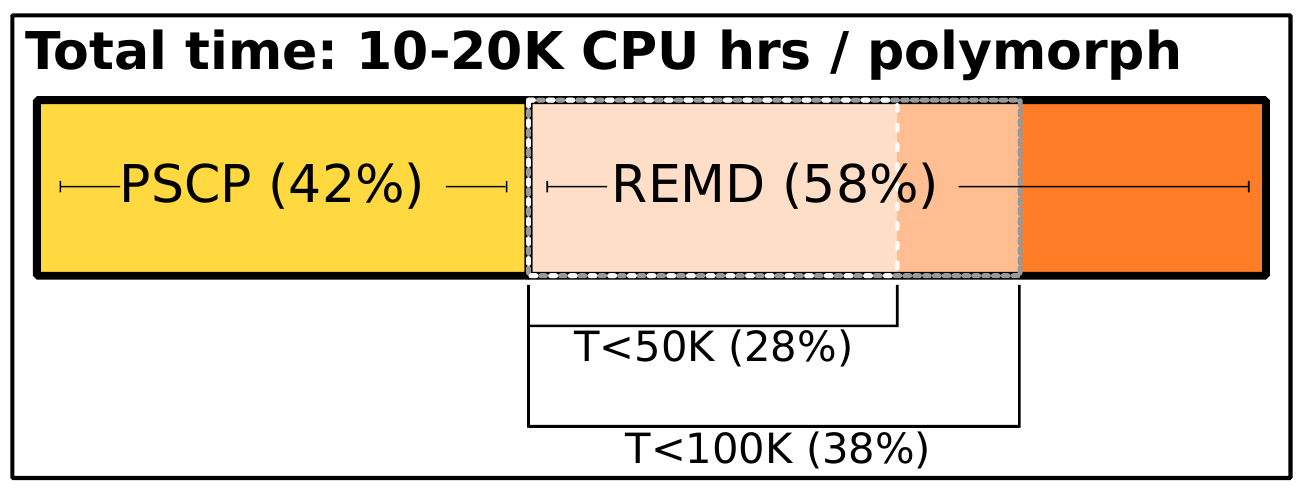}
  \end{center}
  \caption{Computing the free energy differences of polymorphs using OPLS with PSCP and REMD from 10 K up to 250--400 K costs 10,000--20,000 CPU hrs. On average, 42\% of the cost in $\Delta G$ is from PSCP and 38\% is from running REMD from 10--100 K. In later sections we discuss and show approximations to help reduce the overall computational cost by 80\%. These values are based on the actual output of the systems ran here and are dependent on simulation length, force field used, scaling, and other factors. Percentages are based off of total computational time.
  \label{figure:MD_time_split}}
\end{figure*}

    The ability of MD, especially with replica exchange or other enhanced sampling techniques, to overcome crystal restructuring and converge to a free energy minimum configurational ensemble is the largest benefit obtained from spending the comparatively large computational cost (10--20K CPU hrs / polymorph). For a given potential, we have shown that irreversible crystal restructuring away from direct minimization of the experimental crystal is a common phenomena when polymorphs are heated in MD.~\cite{dybeckCapturingEntropicContributions2017,dybeckExploringMultiminimaBehavior2019,abrahamAddingAnisotropyStandard2019} Further investigation by Dybeck \textit{et al.}~found that configurational ensembles obtained from MD include many static lattice minimum, being found in 8 of 12 polymorphic systems studied.~\cite{dybeckExploringMultiminimaBehavior2019} Approaches like QHA cannot overcome this barrier of restructuring, which can lead to inaccurate stability ranking or large errors ($>$1.0 kcal/mol) in the thermodynamics computed with QHA relative to MD.~\cite{abrahamAddingAnisotropyStandard2019} Determining the structures characteristic of the correct free energy minima is critical for a CSP, motivating the use of MD as a part of the search. In particular, replica exchange molecular dynamics (REMD) provides the most robust way to overcome crystal restructuring because it allows the crystal to be constantly heated up and annealed throughout the simulation. We have found that QHA run from a minimum consistent with the restructured crystal in REMD introduces error less than 0.12 kcal/mol,~\cite{abrahamAddingAnisotropyStandard2019} leaving room for an intermediate approach between QHA and computing the full thermodynamic cycle. The breakdown of the computational cost of REMD+PSCP is reported in Figure~\ref{figure:MD_time_split} and shows clear places where time can potentially be saved.

    Due to the frequency at which we see crystal restructuring occur for pairs of polymorphs, we should logically conclude that this behavior will be more common in a CSP, which generates far more crystal structures than are found in experimental databases. Therefore, using REMD seems to be advisable when screening CSPs, but we can reduce the temperature range over which REMD is performed to save computational cost. In previous work we ran REMD from 10 K to 250--400 K depending on the system of interest. The benefit of running temperature REMD below 100 K is to find an appropriate lattice minimum for QHA and to compare how QHA performs relative to REMD at low temperatures, as discussed in our previous work.~\cite{abrahamAddingAnisotropyStandard2019} 
Running REMD down to 10 K may not be necessary for all polymorphs, but previous work shows that for some systems energy minimizing trajectories at 10 K produces energetically indistinguishable lattice minima while minimizing from 50 K can produce structures that differ by up to 0.24 kcal/mol.~\cite{dybeckExploringMultiminimaBehavior2019} Figure~\ref{figure:MD_time_split} shows that the time spent running replicas below 50 K (28\%) and between 50--100 K (10\%) has a large contribution to the overall computational cost, as lower temperature replicas must be more closely spaced together. By excluding temperatures below 100 K, which are generally not of practical interest, the computational cost can be reduced by roughly 38\% if one only calculates down to the temperature at which the the PSCP is performed.

    Performing the PSCP contributes to 42\% of the computational cost of the determination of $\Delta G(T)$ on average. This percent will change if different PSCP paths are approaches are used, but the approximate percentage will in most cases be similar.
If a harmonic-based approach could be used instead to connect the polymorph $G(T)$ curves, there could be significant speed up with potentially little error introduced if the systems behaved harmonically below a given $T$. Even if an approach did not perfectly capture the exact MD-derived free energies, it could help provide a screen to reduced the need for the more computationally expensive full PSCP approach. Ideally, we would be able to remove both the low temperature REMD simulations and the PSCP to reduce the computational cost by 80\% with little cost to the accuracy in the polymorph free energy differences. In the remainder of this paper, we explore 
various approximations to lower the cost to compute $\Delta G(T)$ and evaluate any sacrifices to the accuracy of PSCP.

\section{Methods}
\subsection{Quasi-Harmonic Approximation}
    The quasi-harmonic approximation (QHA) assumes that there is a single static lattice that can be used to determine the free energy at any given temperature and pressure of interest. In QHA we determine the lattice geometry $\boldsymbol{C}$ that minimizes the free energy $F$ in eq~\ref{eq:QHA_raw} to determine the Gibbs free energy at the temperature and pressure of interest.~\cite{abrahamThermalGradientApproach2018,abrahamAddingAnisotropyStandard2019}
  \begin{eqnarray}
      G(T,P) &=& \min_{\boldsymbol{C}} F(\boldsymbol{C}, T, P) \label{eq:QHA} \\
      F(\boldsymbol{C}, T, P) &=& U(\boldsymbol{C}) + A_{v}(\boldsymbol{C}, T) \nonumber \\
			      &+& PV \label{eq:QHA_raw}
  \end{eqnarray}
Where $U$ is the crystal's potential energy and $A_{v}$ is the Helmholtz free energy of a harmonic oscillator. Here we use the classical equation for the vibrational free energy ($A_{v}$) in eq~\ref{eq:classical} because we are using a classical potential. 
  \begin{eqnarray}
      A_{v} &=& \sum_{k} \beta^{-1} \ln{\left(\beta \hbar \omega_{k}(V)\right)} \label{eq:classical}
  \end{eqnarray}
The vibrational free energy is a summation of the 3$\times$N vibrational phonons ($\omega_{k}$) of the crystal lattice, where N is the number of atoms. At higher temperatures the free energy difference between of QHA using classical and quantum statistics should be negligible. To solve eq~\ref{eq:QHA} we determine the crystal thermal expansion, $\frac{\partial \boldsymbol{C}}{\partial T} = (\frac{\partial^{2} G}{\partial \boldsymbol{C}^{2}})^{-1} (\frac{\partial S}{\partial T})$,~\cite{abrahamThermalGradientApproach2018,abrahamAddingAnisotropyStandard2019} and iteratively integrate and expand the lattice minimum up to a temperature of interest.

    The expense of QHA can be simplified by assuming a one-dimensional anisotropic thermal expansion in conjunction with the constant Gr\"{u}neisen parameter approximation. For each set of lattice parameters ($\boldsymbol{C}$) the lattice phonons ($\boldsymbol{\omega}$) must be computed, which can be expensive if done exactly through diagonalization of the mass-weighted Hessian. It has been shown that assuming a constant Gr\"{u}neisen parameter ($\gamma_{k} = -\frac{1}{\omega_{k}} \frac{\partial \omega_{k}}{\partial \eta}$) can be used to determine how a particular phonon changes due to a strain placed on the crystal. This approach is shown to introduce error $<$ 0.01 kcal/mol for the computed solid--solid free energy differences.~\cite{abrahamThermalGradientApproach2018} When we first presented the method to compute the crystal thermal expansion, we found that a 1D thermal expansion based on a scaling of the gradients at 0 K was a sufficient substitute for the full anisotropic expansion. In the 1D-expansion approach, the anisotropic expansion at 0 K and the ratio of those rates 
($\kappa_{i}$) are used to determine the crystal expansion ($C_{i}(\lambda) = C_{i}(\lambda=0) + \kappa_{i} \lambda(T)$), which are scaled by $\lambda$. To save on computational resources, QHA is performed using both the Gr\"{u}neisen parameter and the anisotropic 1D-expansion method.

\subsection{Molecular Dynamics with the Pseudo-Supercritical Path}
    An absolute free energy difference between polymorphs can be determined by computing the free energy for restraining the atoms and turning off the inter- and intramolecular interactions at a reference temperature to complete the thermodynamic cycle. In the current work we use eq~\ref{eq:PSCP} to describe changes to the potential energy function due to changes in the harmonic restraints or interactions.
\end{multicols}
  \begin{eqnarray}
  \begin{aligned}
      U(\lambda_{1}, \lambda_{2}, \gamma_{1}, \gamma_{2}) &= \gamma_{1}^{2} (U_{inter} + U_{tors.}) \\ 
      &+  \gamma_{2} (U_{bond} + U_{angle})  \\ 
      &+ \left((1 - \lambda_{1})^{2} \frac{1}{2} k_{x,1} + (1 - \lambda_{2})^{4} \frac{1}{2} k_{x,2}\right)(x - x_{0})^{2} \label{eq:PSCP}
  \end{aligned}
  \end{eqnarray}
\begin{multicols}{2}
Where the path the crystal follows to the restrained non-interacting ideal gas state is obtained by:
\begin{enumerate}
  \item restraining the atoms ($\lambda_{1}=1\rightarrow0$) with a force constant of $k_{x,1}$;
  \item turning off ($\gamma_{1}=1\rightarrow0$) the intermolecular interactions ($U_{inter}$) and torsions ($U_{tors}.$);
  \item applying a second larger restraint ($\lambda_{2}=1\rightarrow0$) to the atoms with a force constant of $k_{x,1} + k_{x,2}$; and then
  \item turning off ($\gamma_{2}=1\rightarrow0$) the bonds and angles.
\end{enumerate}

    Temperature replica exchange molecular dynamics (REMD) allows us to quickly converge to the thermodynamic minimum at all temperatures of interest. REMD will periodically exchange the states of the replica based using the probability of observing the state in the other replica. By exchanging neighboring temperatures, replicas, the crystal is constantly annealing from high to low temperature, allowing the system to reach a stable thermodynamic state faster.~\cite{zhangConvergenceReplicaExchange2005} The reduced free energy ($f$) of each polymorph can be computed using the Multistate Bennett Acceptance Ratio (MBAR), which estimates the difference of $\beta$ times the free energy most consistent with the energies of the samples generated at each temperature.~\cite{shirtsStatisticallyOptimalAnalysis2008} In eq~\ref{eq:MD_dG} the Gibbs free energy difference between polymorphs as a function of temperature ($\Delta G(T)$) is determined by relating the reduced free energies to the Gibbs free energy difference compute with PSCP.
\end{multicols}
  \begin{eqnarray}
    \Delta G_{ij} (T) = k_{B} T (\Delta f_{ij} (T) - \Delta f_{ij} (T_{ref})) + \frac{T}{T_{ref}} \Delta G_{ij} (T_{ref}) \label{eq:MD_dG}
  \end{eqnarray}
\begin{multicols}{2}

\subsubsection{Replacing PSCP with QHA}
    Once an appropriate lattice minimum is found from REMD, one could in theory replace the reference free energy difference between polymorphs computed with PSCP by performing QHA and using this value in eq~\ref{eq:MD_dG}, assuming that the crystal was sufficiently harmonic below a given temperature. Since an accelerated sampling method such as temperature REMD is required to overcome the free energy barriers to even local restructuring, it introduces significant error to use QHA started from CSP- or experimentally-minimized structures to determine the polymorph free energy differences without additional REMD.~\cite{abrahamAddingAnisotropyStandard2019} However, we have shown that the error in QHA is relatively small (0.03--0.22 kcal/mol at the failure temperature of QHA) if QHA starts from a lattice minimum that is energy minimized from a 10 K (or potentially somewhat higher $T$) replica of REMD.~\cite{abrahamAddingAnisotropyStandard2019} In this approximation, the speed of QHA is combined with the ability of REMD to converge on a thermodynamic minimum and the statistical ensemble that a static lattice representation like QHA fails to capture.

\subsubsection{Replacing PSCP with conformationally weighted HA}
    Using QHA as an approximation for PSCP does save computational time, but requires us to perform REMD down to low temperatures to find a lattice minimum and may be limited to systems that vibrate around a single molecular conformation. In our experience, finding a restructured lattice minimum requires us to run REMD down to 10 K, which can be expensive as outlined above. Additionally, at temperatures as low as 10 K, the conformationally flexible molecules studied here are able to access multiple low-energy configurations which have slightly different vibrational energies.~\cite{dybeckExploringMultiminimaBehavior2019,abrahamAddingAnisotropyStandard2019} The propensity of the molecules to visit a number of low-energy conformations will only increase with larger and more complex molecules than those tested here. Since QHA only considers a single conformation, it is likely to diverge significantly from the exact result obtained with high-quality MD for systems with larger molecular flexibility.

    The reference free energy difference can be directly estimated by averaging the probabilities of the harmonic free energy of the configurations minimized from samples of a molecular dynamics simulation. The full configurational ensemble in a crystal can be reasonably represented by an ensemble of molecular and geometric configurations, each of which belongs to a particular harmonic well. By Boltzmann weighting those harmonic free energies, we can estimate the absolute harmonic energy of the system. Since we can collect these configurations from a moderately high temperature REMD simulation, we can overcome issues of disorder that limit other harmonic methods.~\cite{purohitImplementationHarmonicallyMapped2020} In eq~\ref{eq:boltz_HA} we show how a reference free energy can be computed for a single polymorph, which can be combined for two polymorphs to approximate a $\Delta G(T_{ref})$ for eq~\ref{eq:MD_dG}.
\end{multicols}
  \begin{eqnarray}
    G(T_{ref}) = - \beta^{-1} \ln({\sum_{i=1}^{N} \exp\left({-\beta F(\boldsymbol{C}, T_{ref}, P)}\right)}) \label{eq:boltz_HA}
  \end{eqnarray}
\begin{multicols}{2}
Where $F$ is the harmonic free energy computed from eq~\ref{eq:QHA_raw} for some specified $N$ frames of the trajectory at $T_{ref}$. Each frame is geometry minimized by allowing the molecular coordinates to relax but keeping the box fixed. When approximating $\Delta G(T_{ref})$ with this method $N$ must be large that $\Delta G(T_{ref})$ has converged within a desired precision, and we explore the choice of $N$ later in this paper.

This method also does not require a lattice minimum to be found, but should be run at a temperature such the error in the approximate $\Delta G(T_{ref})$ with respect to full molecular dynamics approaches is minimized, or at least is smaller than the level of accuracy the user requires. We point out that this method likely over-counts states sampled within the simulation, as we may incorporate multiple configurations that may minimize to the same minimum structure, but leave that approximation uncorrected at the present time.

\subsection{Simulation Details}
\begin{figure*}
  \begin{center}
  \includegraphics[width=16cm]{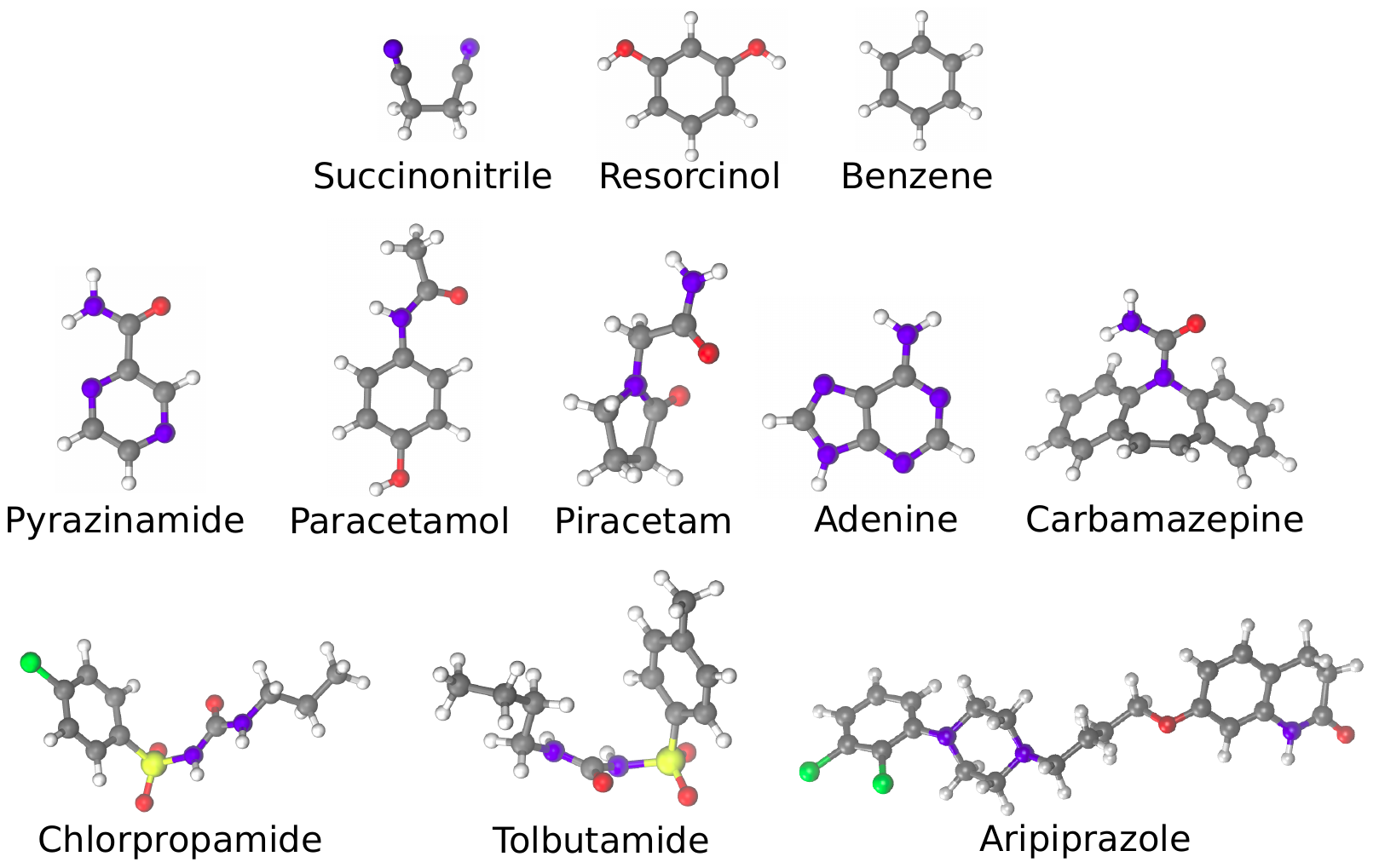}
  \end{center}
  \caption{Molecules evaluated in this study.
  \label{figure:molecules}}
\end{figure*}

    We evaluate our methods with the polymorphs of the molecules in Figure~\ref{figure:molecules}. The experimental polymorphs are reported in table S\ref{table:systems} with their CCDC refcodes. Each molecule was parameterized using the Maestro \textit{ffld\_server} for the OPLS2 fixed charge potential. Topologies for all molecules are available in supporting information of our previous publication.~\cite{abrahamAddingAnisotropyStandard2019} All molecular systems used supercells with between 6 to 48 unit cells, as we have shown that there is significant (add number range) size dependence in the QHA approximation, requiring at least a moderate number of unit cells to capture~\cite{abrahamAddingAnisotropyStandard2019}

    For temperature REMD we chose temperatures between 10 K up to 250--400 K with replica spacing such that the exchange probability between replicas was approximately 0.3.  All crystals were run for 20 ns with a time step of 0.5 fs as flexible harmonic bonds were used. PSCP was run at a reference temperature of 200 K for all polymorphs. A reference structure was selected from PSCP and NVT simulations were run at the following spacing for each section of the path: $\Delta \lambda_{1} =$ 0.05, $\Delta \gamma_{1} = 0.05$, $\Delta \lambda_{2} = 0.01$, and $\Delta \gamma_{1} = 0.02$. The values of the harmonic restraint force constants are $k_{x,1} = 10^3$ kJ / (mol $\cdot$ nm$^{2}$) and $k_{x,2}= 10^6$  kJ / (mol $\cdot$ nm$^{2}$). We have developed a code package to setup, run, and analyze all results for this method, which is available on GitHub at \url{http://github.com/shirtsgroup/finite-temperature-crystal-scripts}.~\cite{dybeckComparisonMethodsReweight2016,dybeckCapturingEntropicContributions2017} All simulations were performed with GROMACS 2019.3.

    The quasi-harmonic approximation is performed on the lattice minimum found by energy minimizing the experimental crystal and  found by energy minimizing the 10 K replica of temperature REMD. We selected 5 random configurations from the equilibrated 10 K NPT simulation from REMD and energy minimized each one. For all polymorphs, all 5 lattice minimum were within 0.015 kcal/mol of one another and the lowest energy structure was used to run QHA for the restructured crystal. All harmonic approximation calculations were run using the Tinker molecular modeling package 8.7. 
Here the lattice minimum were found using \textit{xtalmin} to a tolerance of 10$^{-5}$ and the lowest energy minimum was selected for QHA. QHA was performed using our Python wrapper package that is available on GitHub at \url{http://github.com/shirtsgroup/Lattice\_dynamics}.~\cite{abrahamThermalGradientApproach2018,abrahamAddingAnisotropyStandard2019} An example input file is provided with the Supporting Information.

Boltzmann-weighted HA was completed by performing multiple instances of HA on the fixed geometry configurations found with REMD of the NPT molecular simulations. For the reference temperature of interest, $N$ configurations were selected from the equilibrated replica and the coordinates were optimized using Tinker's \textit{minimize} command, which holds the lattice geometry fixed. The harmonic energy was then computed using our GitHub package \url{Lattice\_dynamics} and then \url{finite-temperature-crystal-scripts} was used to compute eq~\ref{eq:boltz_HA} and eq~\ref{eq:MD_dG}.  Convergence plots for each polymorph pair can be found in the Supporting Information (Figures S~\ref{figure:converge_0}--S~\ref{figure:converge_10}).

\section{Results and Discussion}
\subsection{The prevalence of crystal restructuring impairs the use of QHA independent of MD} \label{section:restructure}
    Of the crystals studied, almost half of them restructured during the REMD simulation. In previous work we have discussed the prevalence of crystal restructuring due to REMD simulations and shown its affect on the error between QHA and REMD+PSCP.~\cite{abrahamAddingAnisotropyStandard2019} With a larger sub of crystals, we have found that 13 out of the 29 (45\%) restructure to a more stable thermodynamic crystal structure when heated up and cooled down with REMD. Crystal restructuring is generally due to symmetry breaking between unit cells in the supercell.~\cite{dybeckCapturingEntropicContributions2017,abrahamAddingAnisotropyStandard2019,dybeckExploringMultiminimaBehavior2019} Since restructuring occurred for polymorphs of piracetam, adenine, pyrazinamide, succinonitrile, chlorpropamide, aripiprazole, and tolbutamide, REMD is required to find stable thermodynamic minima regardless of molecular flexibility. The specific polymorphs to undergo restructuring due to REMD are listed in the Supporting Information (Table S~\ref{table:systems}).

    The error in $\Delta G_{QHA}(T=300K)$ for QHA relative to REMD+PSCP is large (RMSE$=$0.81$\pm$0.21 kcal/mol), but can be significantly reduced (RMSE$=$0.15$\pm$0.03 kcal/mol) if the restructured lattice minimum is used. We ran QHA on the lattice minimum found from energy minimizing both the experimental crystal and the 10 K REMD trajectory. In previous work we compared the error in QHA to REMD+PSCP at $T_{max}$, which is the maximum temperature QHA can reach while still staying at a free energy minimum (satisfying eq~\ref{eq:QHA_raw}).~\cite{abrahamAddingAnisotropyStandard2019} Since the polymorph free energy differences for QHA are fairly linear and we want to compare all our results at 300 K, we linearly extrapolated $\Delta G$ using the last 50 K of QHA before $T_{max}$. Using eq~\ref{eq:RMSE} we can determine the error in $\Delta G$ computed with QHA to the full thermodynamic cycle at 300 K.
\end{multicols}

  \begin{eqnarray}
    RMSE_{approx.} = \sqrt{\frac{1}{n} \sum_{i}^{n} \left(\Delta G_{approx.}^{T_{ref}}(T=300 K) - \Delta G_{PSCP}(T=300 K) \right)^{2}} \label{eq:RMSE}
  \end{eqnarray}
\begin{multicols}{2}
Here, $\Delta G_{approx.}^{T_{ref}}(T=300 K)$ is the polymorph free energy difference at 300 K computed with one of the approximations and $\Delta G_{PSCP}(T=300 K)$ is the absolute free energy difference at 300 K using REMD+PSCP.

\subsection{Characterizing reference temperatures that minimize error in the approximation}
\begin{figure*}
  \begin{center}
  \includegraphics[width=16cm]{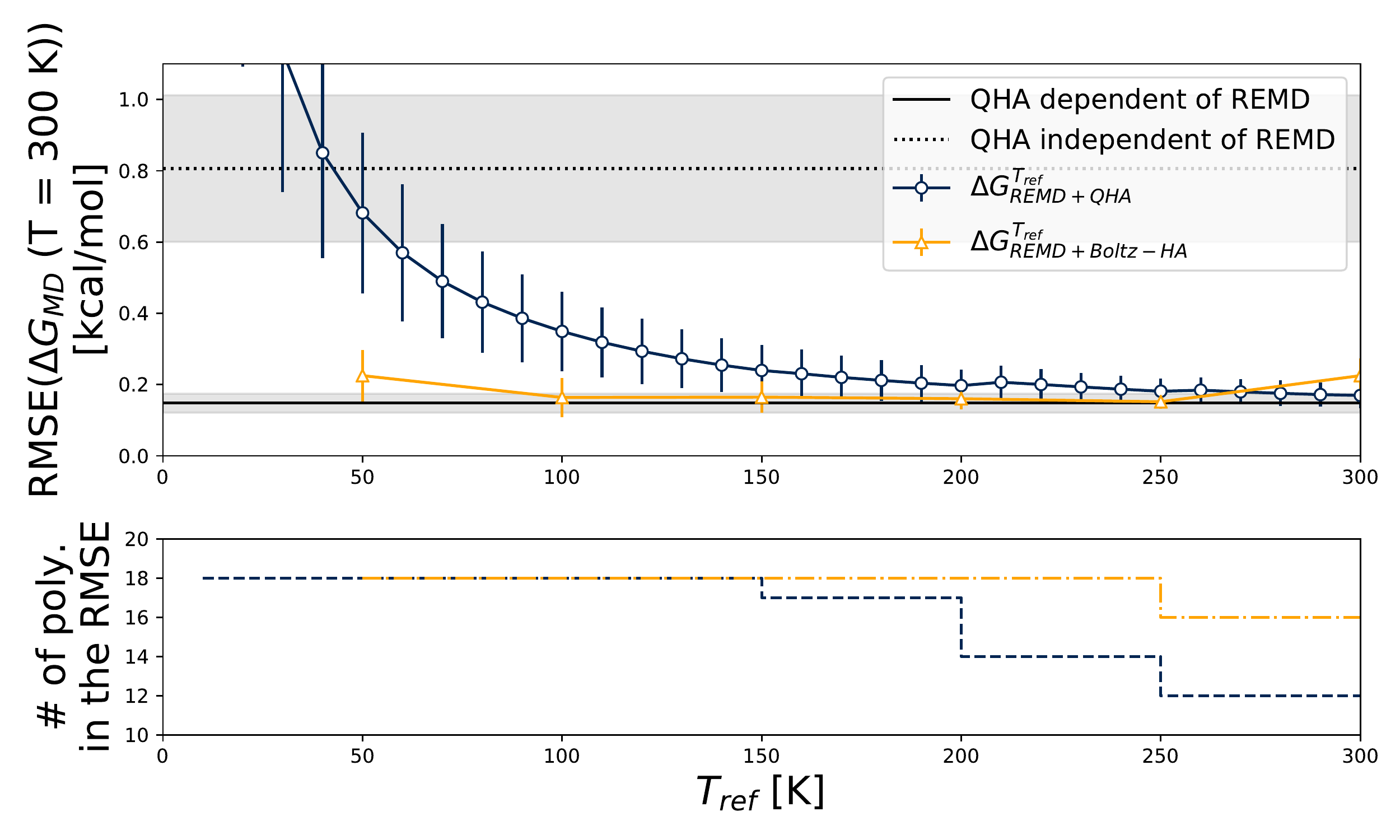}
  \end{center}
  \caption{The RMSE of the polymorph free energy differences at 300 K for all approximations relative to REMD+PSCP is shown (top) and the number of polymorph pairs in that RMSE ($n$) is reported in the corresponding step plot (bottom) for different reference temperatures ($T_{ref}$). We can determine a reference free energy difference at all temperatures using REMD+Boltzmann-HA, whereas QHA becomes unusable due to numeric instability, as shown in the step plot. The drop in number of free energy differences $n$ for Boltzmann-HA is solely because benzene is liquid above 250 K. REMD+Boltzmann-HA and (extrapolated) QHA dependent of REMD produce the lowest error (RMSE$=$0.15 kcal/mol) at reference temperatures that include all polymorph pairs ($n=$18). The solid line is solely QHA with REMD restructuring and the dashed line is for QHA run from structures minimized from experiment. Errors are the standard deviation of the bootstrapped error of the RMSE over all $n$ systems.
  \label{figure:RMSE_Tref}}
\end{figure*}

\begin{figure*}
  \begin{center}
    \includegraphics[width=8cm]{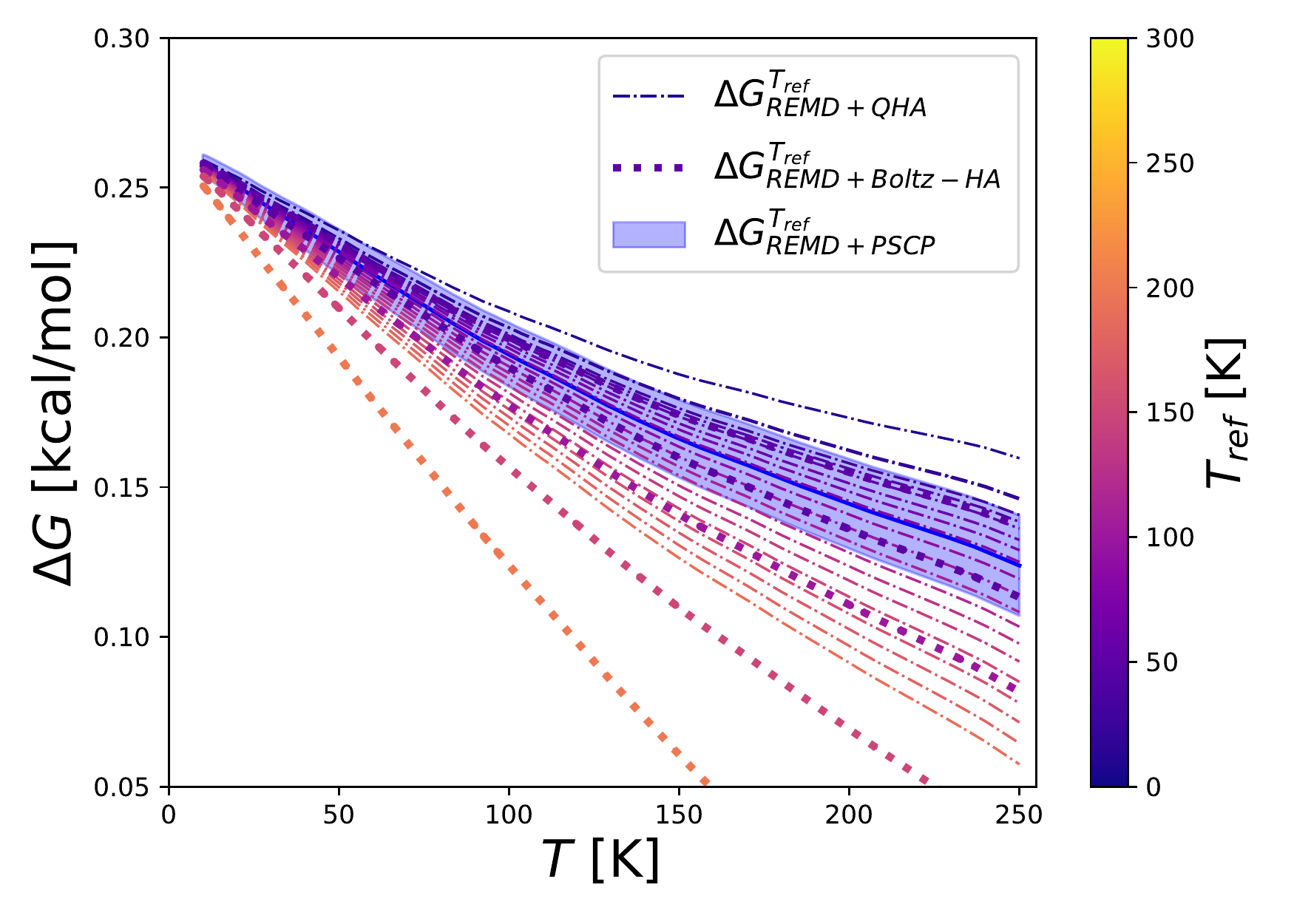}
    \includegraphics[width=8cm]{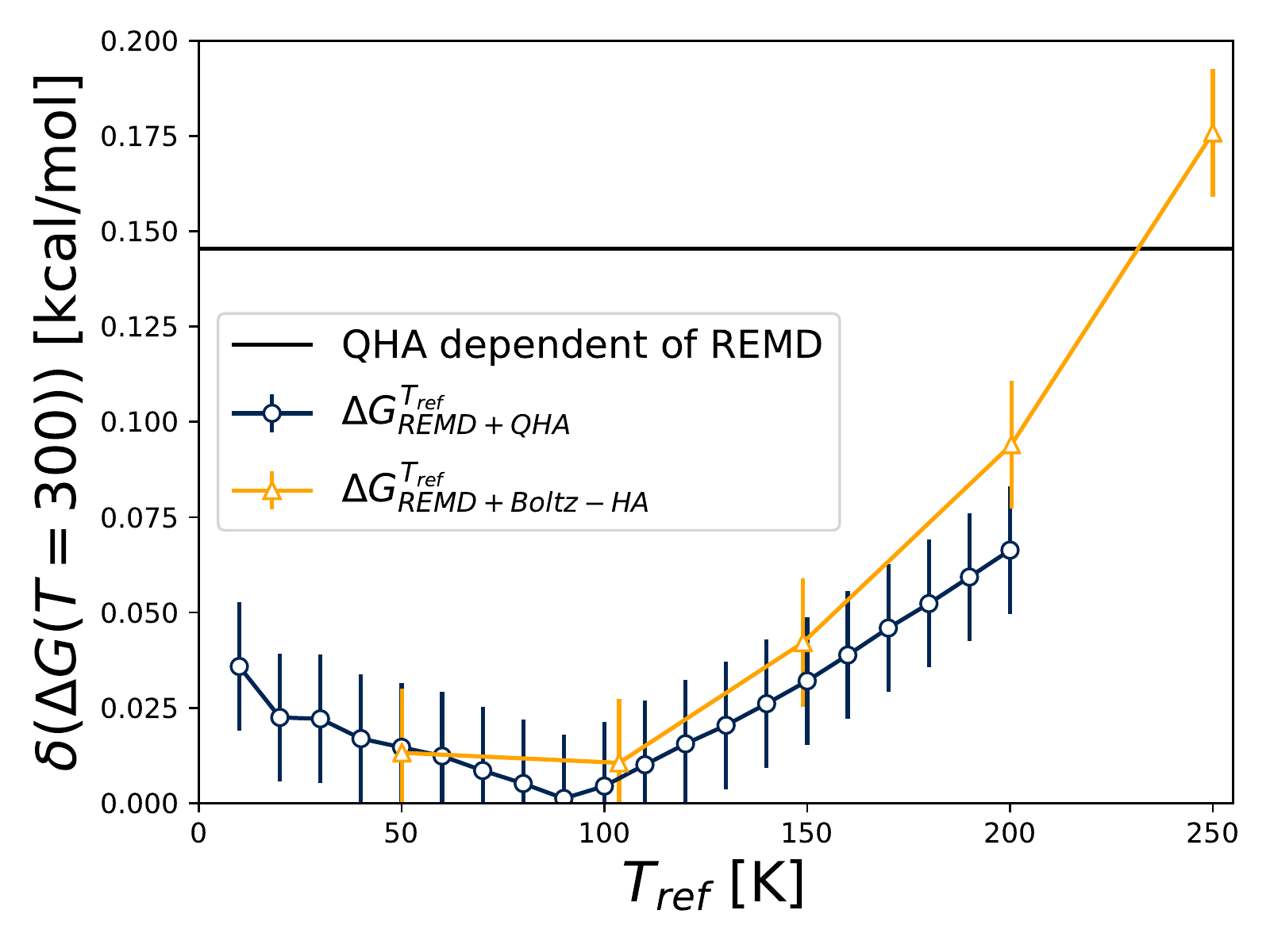}
  \end{center}
  \caption{Left) The polymorph free energy difference of benzene (form II relative to form I) using MD, QHA, and replacing $\Delta G (T_{ref})$ from PSCP in equation ~\ref{eq:MD_dG} with the QHA values for $T_{ref}$ every 10 K is shown. Right) The error in the $\Delta G(300K)$ is reported for replacing $\Delta G (T_{ref})$ from QHA with PSCP and using Boltzmann weighted HA. Each line of the approximation in the left figure is color coded based off the reference temperature used.
  \label{figure:dG_error_start}}
\end{figure*}

\begin{figure*}
  \begin{center}
    \includegraphics[width=8cm]{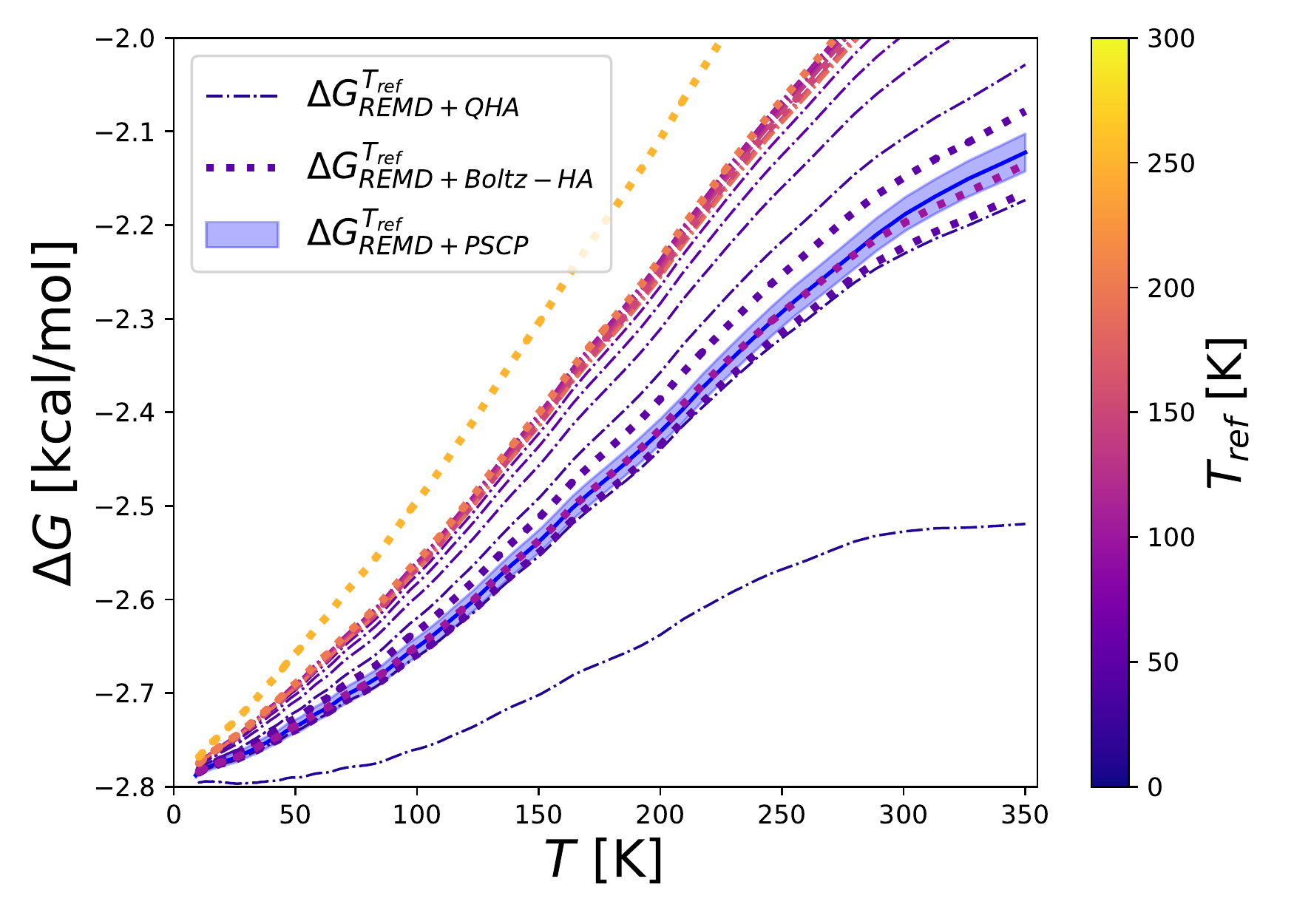}
    \includegraphics[width=8cm]{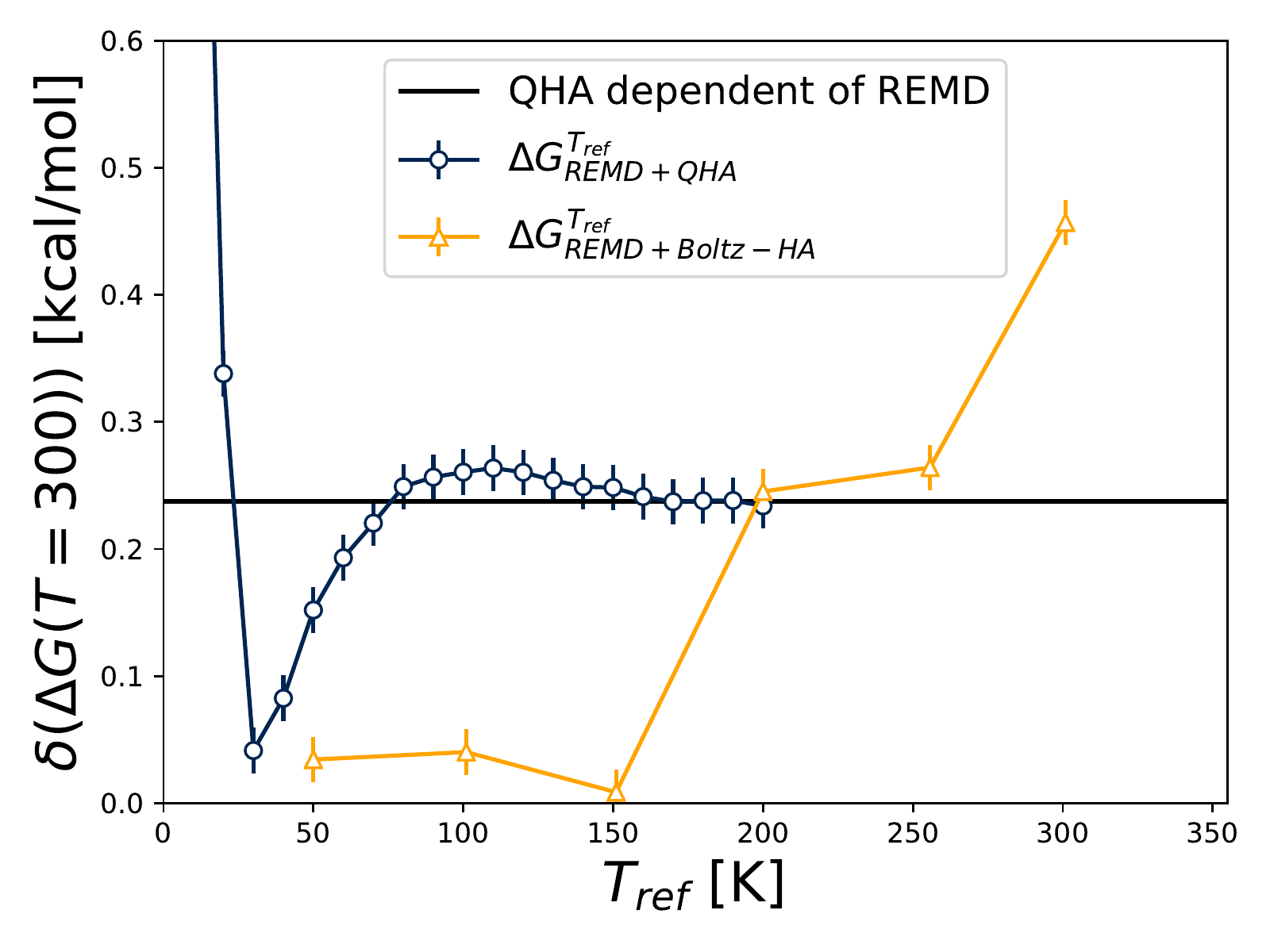}
  \end{center}
  \caption{Left) The polymorph free energy difference of tolbutamide (form V relative to form I) using MD, QHA, and replacing $\Delta G (T_{ref})$ from PSCP in equation ~\ref{eq:MD_dG} with the QHA values for $T_{ref}$ every 10 K is shown. Right) The error in the $\Delta G(300K)$ is reported for replacing $\Delta G (T_{ref})$ from QHA with PSCP and using Boltzmann weighted HA. Each line of the approximation in the left figure is color coded based off the reference temperature used.
  \label{figure:dG_error_finish}}
\end{figure*}

    The optimal temperature to compute a reference free energy difference between polymorphs is dependent on the approximation used. In REMD+PSCP the free energy difference is independent of the reference temperature ($T_{ref}$) that PSCP is computed at. However, this is not the case for the approximations. In Figure~\ref{figure:RMSE_Tref} (top), the RMSEs (eq~\ref{eq:RMSE}) for the approximations are plotted as a function of the reference temperature used. For REMD+QHA the RMSE decreases as $T_{ref}$ increases and converges to the RMSE of QHA performed on the restructured crystals. On contrast, the RMSE of REMD+Boltzmann-HA method remains at a roughly constant minimum for reference temperatures of 100--250 K, though there is variation for different polymorphs shown in the Supporting Information (Figures S~\ref{figure:dG_error_start}--S~\ref{figure:dG_error_finish}).  

    Boltzmann-weighted HA is a more robust method since the free energy difference can be computed at all temperatures in which REMD can be performed. In Figure~\ref{figure:RMSE_Tref}(bottom) the number of polymorph pairs($n$) included in the RMSE (eq~\ref{eq:RMSE}) is shown as a step plot. To highlight methodological errors in the approximations, we have chosen not to use reference free energy difference of polymorphs greater than $T_{max}$ in REMD+QHA. At temperatures greater than 150 K the number of systems with usable QHA, as shown in the step plot of REMD+QHA, starts to decrease, whereas REMD+Boltzmann-HA can be performed at all temperatures. There is a drop in the REMD+Boltzmann-HA step plot at 250 K, but this is because benzene cannot be simulated at higher temperatures as it starts to approach its simulated melting point. 
  
\begin{figure*}
  \begin{center}
    \includegraphics[width=8cm]{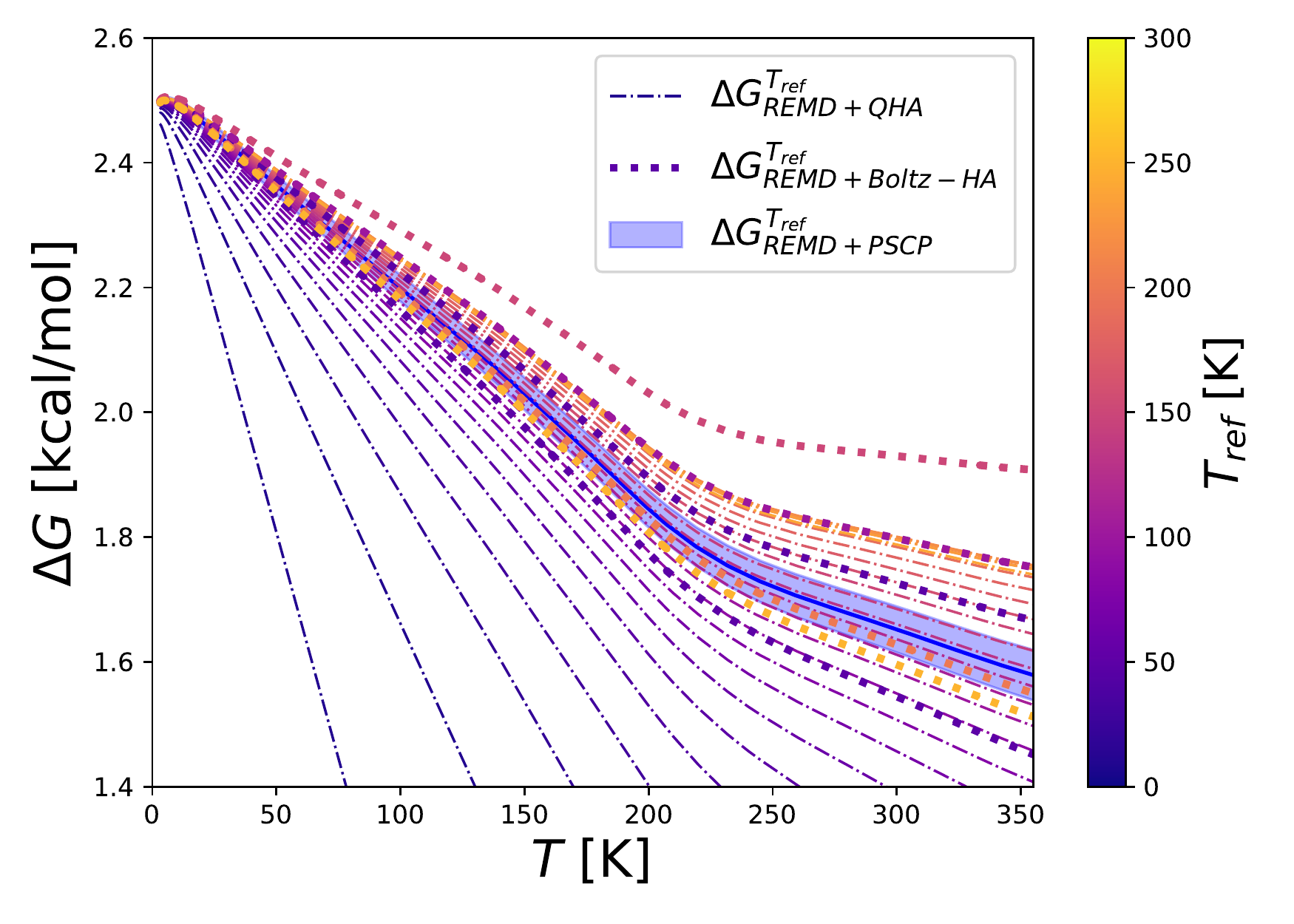}
    \includegraphics[width=8cm]{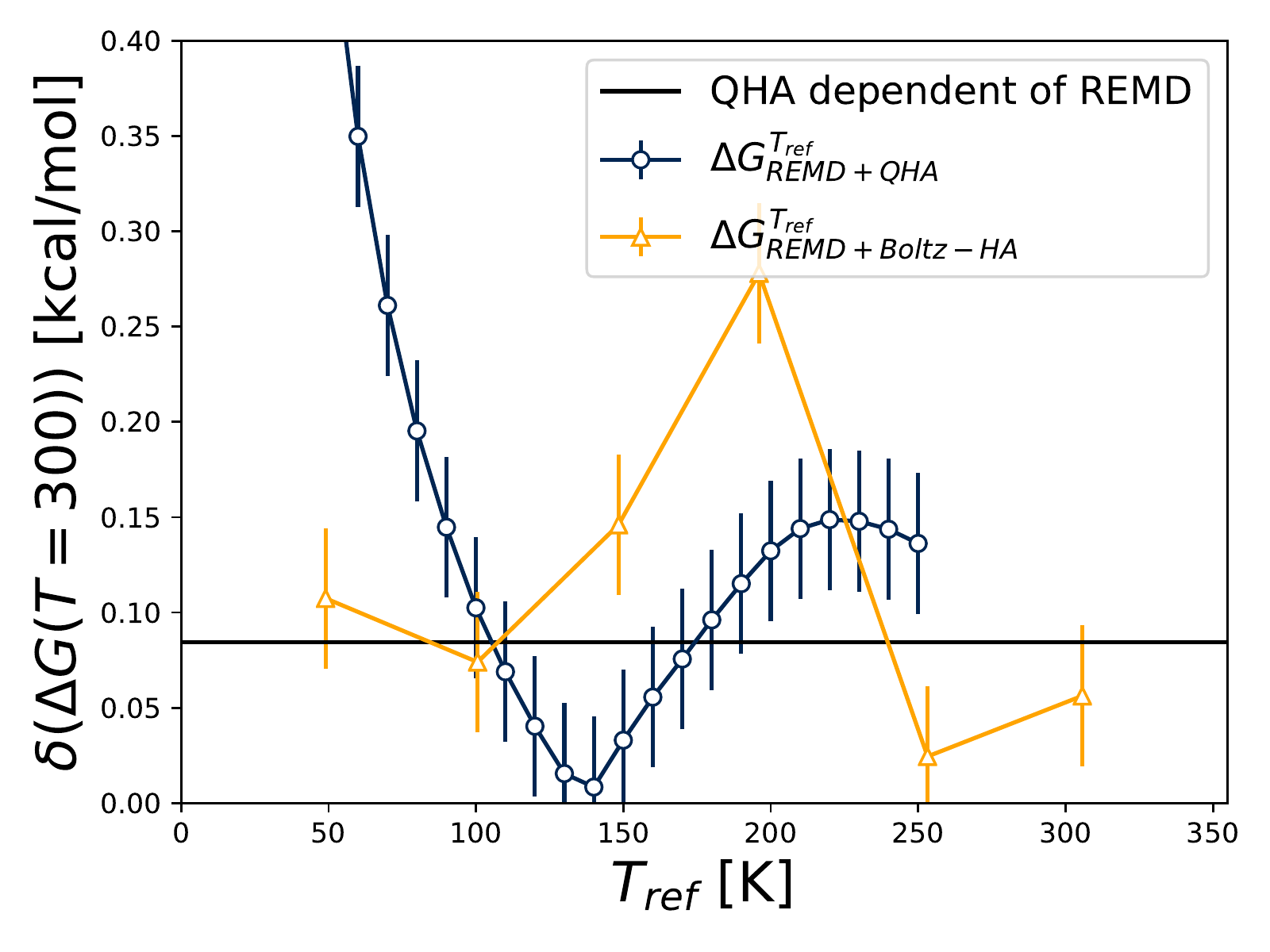}
  \end{center}
  \caption{Left) The polymorph free energy difference of chlorpropamide (form V relative to form I) using MD, QHA, and replacing $\Delta G (T_{ref})$ from PSCP in equation ~\ref{eq:MD_dG} with the QHA values for $T_{ref}$ every 10 K is shown. Right) The error in the $\Delta G(300K)$ is reported for replacing $\Delta G (T_{ref})$ from QHA with PSCP and using Boltzmann weighted HA. Each line of the approximation in the left figure is color coded based off the reference temperature used.
  \label{figure:dG_error_13}}
\end{figure*}

    Using REMD+Boltzmann-HA or extrapolated QHA produces the smallest error (RMSE$=$0.15 kcal/mol) relative to REMD+PSCP, with REMD+Boltzmann-HA being the most stable. For all approximations we want to look at the minimum value of the RMSE that is inclusive of all polymorph pairs ($n=$18). In the case of REMD+QHA this is at 150 K where the RMSE is 0.24$\pm$0.07 kcal/mol, which is 0.09 kcal/mol greater than the RMSE of REMD+Boltzmann-HA and extrapolated QHA, 0.15$\pm$0.02 kcal/mol. REMD+Boltzmann-HA is the most robust method because it 1) doesn't require extrapolation, and 2) errors remain with 0.02 kcal/mol of the minimum RMSE using reference temperatures between 100--250 K. The optimal reference temperature and generated error is both molecule and polymorph dependent as seen in the Supporting Information (Figures S~\ref{figure:dG_error_start} -- S~\ref{figure:dG_error_finish}). Chlorpropamide (Figure S~\ref{figure:dG_error_13}) is a good example of this phenomenon where the largest error (0.28 kcal/mol) for REMD+Boltzmann-HA is for $T_{ref}=$ 200 K, but the smallest error (0.02 kcal/mol) is at $T_{ref}=$ 300 K, falling outside of the general range for all systems. If it is feasible, determining a reference free energy difference at a couple of temperatures with REMD+Boltzmann-HA may be beneficial to determine the robustness the result to the temperature reference.

\begin{figure*}
  \begin{center}
  \includegraphics[width=10cm]{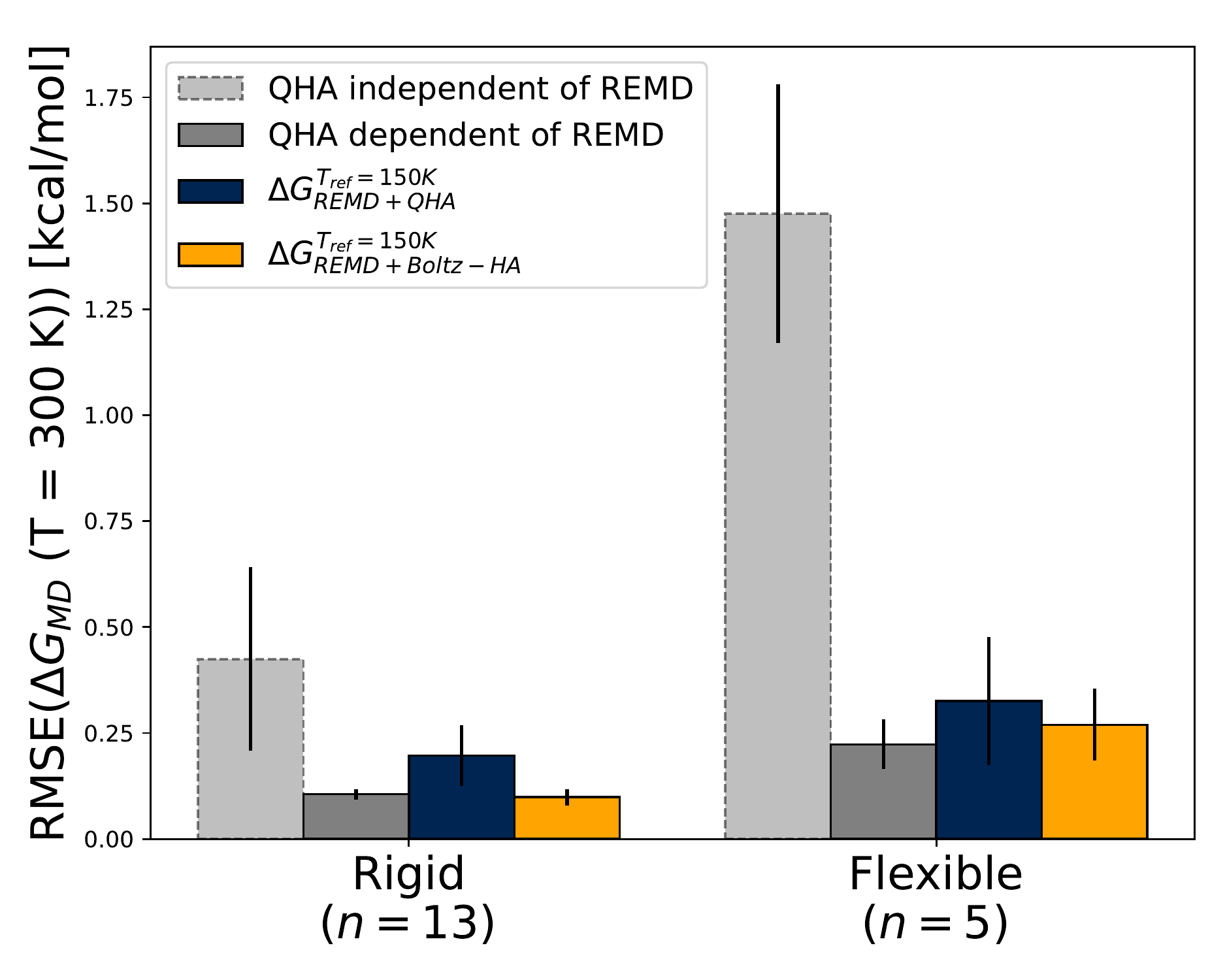}
  \end{center}
  \caption{The RMSE of the polymorph free energy differences at 300 K for all approximations relative to REMD+PSCP is shown for the rigid and flexible molecules. The flexible molecules are aripiprazole, chlorpropamide, and tolbutamide. The error in restructuring has the greatest difference between rigid and flexible molecules because 7 of the 8 flexible polymorphs restructure, whereas only 6 of the 21 rigid polymorphs restructure. The value $n$ reported is the number of polymorph pairs between which free energies are calculated that are included in the RMSE.
  \label{figure:RMSE_150K}}
\end{figure*}

    The RMSE of the various approximations that use REMD are between 0.10--0.33 kcal/mol at 300 K, and are functions of the size and flexibility. In Figure~\ref{figure:RMSE_150K} the RMSE for each approach is reported using $T_{ref}=$ 150 K in the bar plots and the systems have been split by molecular flexibility, where the flexible molecules are aripiprazole, chlorpropamide, and tolbutamide. The largest trend of error due to molecular flexibility is for the error in the polymorph free energy differences of QHA run independently of REMD. For the rigid systems 6 of the 21 polymorphs restructure, whereas 7 of the 8 more flexible polymorphs restructure leading to the larger RMSE for the flexible systems. The error in the approximations that use REMD are much smaller, but also trend with molecular flexibility. We know that the flexible molecules are able to access a greater number of molecular conformations, which are most evidently anharmonic since even the error in  REMD+Boltzmann-HA is increasing.
    
\subsection{There are several approaches to choose from approximate the free energy before computing the full thermodynamic cycle}
    Boltzmann-weighted HA provides the most efficient way to compute the polymorph energy differences to still include the full statistical ensemble from REMD. In table~\ref{table:method_summary} we report the error for each approximation with their relative computational times. Both approximations (REMD+QHA \& REMD+Boltzmann-HA) introduce similar amounts of error ($RMSE=$0.15--0.17 kcal/mol) relative to REMD+PSCP, but REMD+Boltzmann-HA outperforms REMD+QHA because it is able to eliminate low temperature replica from REMD in addition to the PSCP.  If the speed/accuracy trade-off favors speed, QHA independent of REMD is the fastest, but introduces the most error relative to REMD+PSCP. REMD+Boltzmann-HA may be ideal for an initial screening of a CSP followed by using PSCP on select low energy systems.
\end{multicols}
\begin{center}
\begin{table}
\begin{tabular}{l|c|c}
  \hline \hline
  Method & $\delta (\Delta G(T=300 K))$ [kcal/mol] & CPU hrs/poly. \\ \hline
  REMD+PSCP (10--300 K) & 0.0  & 10--20K \\
  REMD+PSCP (100--300 K) & 0.0 & 6.2--12.4K \\
  REMD+QHA (10--300 K) & 0.14$\pm$0.07 & 5.8--11.6K \\
  REMD+Boltzmann-HA (100--300 K) & 0.16$\pm$0.04 & 2--4K \\
  QHA with REMD minima & 0.15$\pm$0.03 & 5.8--11.6K \\  
  QHA without REMD minima & 0.81$\pm$0.20 & $<$ 0.1K \\  
  \hline \hline
\end{tabular} 
\caption{Comparison of error in the polymorph $\Delta G$ at 300 K from REMD+PSCP and the computational time for methods examined in this study. The RMSE is reported with the bootstrapped error using a reference temperature of 150 K for REMD+Boltzmann-HA and REMD+QHA. Boltzmann-weighted HA provides the most efficient method to determine the free energy difference between polymorphs.
\label{table:method_summary}}
\end{table}
\end{center}
\begin{multicols}{2}


\section{Conclusions}
    We present a range of approximations for predicting the free energy differences of polymorphs which can be easily be implemented as a screening method for CSPs. We have previously used the PSCP to determine an absolute reference free energy difference to complete a thermodynamic cycle between two polymorphs.~\cite{dybeckEffectsMoreAccurate2016,dybeckCapturingEntropicContributions2017,abrahamAddingAnisotropyStandard2019} Here we approximate the reference free energy difference using QHA and a method that Boltzmann weights the harmonic energy of frames from an NPT simulation. Two advantages of the Boltzmann-weighted HA approach over QHA are: 1) that low temperature replicas do not need to be used and 2) that it can sample multiple minima that may exist in the ensemble.
   
    Boltzmann-weighted HA is a cheaper method than PSCP that can be used to screen crystal structures in  a CSP. Free energy approximations have increasingly been used in CSPs to rank generated structures. The ``gold standard'' for free energy rankings given a potential energy model is using REMD with the PSCP to compute a full thermodynamic cycle between crystals, but can be prohibitively expensive. By combining the ability of REMD to overcome barriers of crystal restructuring and the speed of harmonic approaches we have developed a method to more efficiently screen crystal structures. Our Boltzmann-weighted HA method is 42--80\% cheaper than using REMD+PSCP and introduces a relatively small error of 0.15$\pm$0.04 kcal/mol. 
We do note that this error is dependent on the molecular flexibility, with 0.10$\pm$0.02 and 0.19$\pm$0.05 kcal/mol error in $\Delta G$ for rigid and flexible molecules in our small benchmark, respectively. This Boltzmann-weighted harmonic method performs roughly comparably to using free energies from QHA from REMD-discovered minima to provide the reference free energy, but it is both more robust and faster. In contrast, QHA free energies minimized directly from experiment can introduce error as large as 0.81$\pm$0.20 kcal/mol relative to REMD+PSCP. The Boltzmann-weighted HA method appears to be a potentially useful choice for initial screening of crystals in a CSP to eliminate entropically disfavorable crystals when PSCP is performed.

\section*{Acknowledgments}
    All results were performed on the Extreme Science and Engineering
    Discovery Environment (XSEDE), which is supported by National
    Science Foundation grant number ACI-1548562. Specifically, it used
    the Bridges system, which is supported by NSF award number
    ACI-1445606, at the Pittsburgh Supercomputing Center (PSC). This
    work also utilized the RMACC Summit supercomputer, which is
    supported by the National Science Foundation (awards ACI-1532235
    and ACI-1532236), the University of Colorado Boulder, and Colorado
    State University. The Summit supercomputer is a joint effort of
    the University of Colorado Boulder and Colorado State University.
    This work was also supported financially by NSF through the grant
    CBET-1351635 and a Graduate Assistance in Areas of National Need
    (GAANN) fellowship which is funded by the U.S. Department of
    Education.

\newpage
\newpage

\newpage
\newpage

\end{multicols}
\end{singlespace}

\begin{singlespace}
\begin{multicols}{2}
\bibliography{citations}
\end{multicols}
\end{singlespace}
\newpage

\maketitle

\begin{singlespace}
\section{Systems Studied}
    In Table \ref{table:systems} we provide the systems studied, their polymorphs name/numbering, corresponding Cambridge Structure Database reference code (CSD Refcode), and the dimensions of the supercells relative to the symmetric cell pulled from the database. Those systems that are highlighted blue have different lattice minimum when energy minimized from experiment and from the 10 K REMD simulation.
\singlespacing
      \begin{longtable}{l|l|l|l|l}
      \hline \hline
      \textbf{Piracetam} & Form I & \cellcolor{blue!25}Form II & Form III & \\
      CSD Refcode & BISMEV03 & \cellcolor{blue!25}BISMEV & BISMEV02 &  \\
      Space Group & P2$_{1}$/n & \cellcolor{blue!25}P$\bar{1}$& P2$_{1}$/n &  \\
      Supercell Dimensions & $ 4\times 2\times 3$ & \cellcolor{blue!25}$2\times 4 \times 3$ & $ 2\times 3\times 4$ & \\
      \hline \hline
      
      \textbf{Benzene} & Form I & Form II & Form III & \\
      CSD Refcode & BENZEN11 & BENZEN16 & BENZEN04 & \\
      Space Group & Pbca & P2$_{1}$/c & P2$_{1}$/c &  \\
      Supercell Dimensions & $3\times 2\times 3$ & $3\times 3\times 4$ & $3\times 4\times 3$ & \\
      \hline \hline
      
      \textbf{Carbamazepine} & Form I & Form III & Form IV  & \\
      CSD Refcode & CBMZPN11 & CBMZPN02 & CBMZPN12  & \\
      Space Group & P$\bar{1}$ & P2$_{1}$/n & C2/c  & \\
      Supercell Dimensions & $ 4\times 2\times 1$ & $ 4\times 2\times 2$ & $ 1\times 4\times 2$  & \\
      \hline \hline

      \textbf{Adenine} & Form I & \cellcolor{blue!25}Form II &  & \\
      CSD Refcode & KOBFUD  & \cellcolor{blue!25}KOBFUD01 &  & \\
      Space Group & P2$_{1}$/c & \cellcolor{blue!25}Fdd2  & & \\
      Supercell Dimensions & $ 3\times 1\times 4$ & \cellcolor{blue!25}$ 3\times 1\times 2$ &  & \\
      \hline \hline

      \textbf{Pyrazinamide} & Form $\alpha$ & \cellcolor{blue!25}$\beta$ & \cellcolor{blue!25}Form $\gamma$ & \cellcolor{blue!25} Form $\delta$ \\
      CSD Refcode & PYRZIN14 & \cellcolor{blue!25}PYRZIN01 &\cellcolor{blue!25} PYRZIN20  & \cellcolor{blue!25}PYRZIN16 \\
      Space Group & P2$_{1}$/a & \cellcolor{blue!25}P2$_{1}$/c & \cellcolor{blue!25}Pc  & \cellcolor{blue!25}P$\bar{1}$\\
      Supercell Dimensions & $ 1\times 4\times 6$ &\cellcolor{blue!25} $ 2\times 6\times 2$ & \cellcolor{blue!25}$ 4\times 6\times 2$  &\cellcolor{blue!25} $4\times 4\times 3$\\
      \hline \hline

      \textbf{Resorcinol} & Form $\alpha$ & Form $\beta$ & &  \\
      CSD Refcode & RESORA03 & RESORA08 &  & \\
      Space Group & Pna2$_{1}$ & Pna2$_{1}$ &  & \\
      Supercell Dimensions & $ 2\times 2\times 4$ & $ 2\times 2\times 4$ &  & \\
      \hline \hline

      \textbf{Succinonitrile} & Form I &\cellcolor{blue!25} Form III &  & \\
      CSD Refcode & QOPBED  & \cellcolor{blue!25}N/A &  & \\
      Space Group & P2$_{1}$/a & \cellcolor{blue!25}N/A & &  \\
      Supercell Dimensions & $ 2\times 3\times 4$ & \cellcolor{blue!25}N/A &  & \\
      \hline \hline

      \textbf{Paracetamol} & Form I & Form II &  & \\
      CSD Refcode & HXACAN01 & HXACAN &  & \\
      Space Group & P2$_{1}$/a & Pcab &  & \\
      Supercell Dimensions & $ 2\times 2\times 3$ & $ 2\times 1\times 3$ &  & \\
      \hline \hline

      \textbf{Aripiprazole} & \cellcolor{blue!25}Form I & \cellcolor{blue!25}Form X &  & \\
      CSD Refcode &\cellcolor{blue!25} MELFIT01 &\cellcolor{blue!25} MELFIT05 &  & \\
      Space Group & \cellcolor{blue!25}P2$_{1}$ & \cellcolor{blue!25}P2$_{1}$ &  & \\
      Supercell Dimensions &\cellcolor{blue!25} $ 3\times 4\times 2$ &\cellcolor{blue!25} $ 3\times 4\times 2$ &  & \\
      \hline \hline

      \textbf{Tolbutamide} & \cellcolor{blue!25}Form I & \cellcolor{blue!25}Form II & \cellcolor{blue!25}Form III &\cellcolor{blue!25} Form V \\
      CSD Refcode & \cellcolor{blue!25}ZZZPUS04 & \cellcolor{blue!25}ZZZPUS05 & \cellcolor{blue!25}ZZZPUS09  & \cellcolor{blue!25}ZZZPUS10\\
      Space Group & \cellcolor{blue!25}Pna2$_{1}$ & \cellcolor{blue!25}Pc & \cellcolor{blue!25}P2$_{1}$/n  & \cellcolor{blue!25}Pbcn\\
      Supercell Dimensions & \cellcolor{blue!25}$ 2\times 3\times 2$ &\cellcolor{blue!25} $ 3\times 2\times 1$ & \cellcolor{blue!25}$ 2\times 3\times 2$ &\cellcolor{blue!25} $2\times 3\times 1$\\
      \hline \hline

      \textbf{Chlorpropamide} & Form I & \cellcolor{blue!25}Form V &  & \\
      CSD Refcode & BEDMIG &\cellcolor{blue!25} BEDMIG04 &  & \\
      Space Group & P2$_{1}$2$_{1}$2$_{1}$ & \cellcolor{blue!25}Pna2$_{1}$ &  & \\
      Supercell Dimensions & $ 1\times 4\times 2$ & \cellcolor{blue!25}$ 1\times 4\times 2$ &  & \\
      \hline \hline
      \caption{Table of systems studied in the main paper. We have included the polymorph names, CSD reference codes, and the cell dimensions relative to the symmetric unit cell pulled down from the database. If the cell is blue it indicates that the crystal energy minimized from experiment is different from the crystal minimized from REMD.
      \label{table:systems}}
      \end{longtable}

\section{Convergence of Boltzmann weighted HA frames} \label{section:boltz-convergence}
\begin{figure*}
  \begin{center}
    \includegraphics[width=8cm]{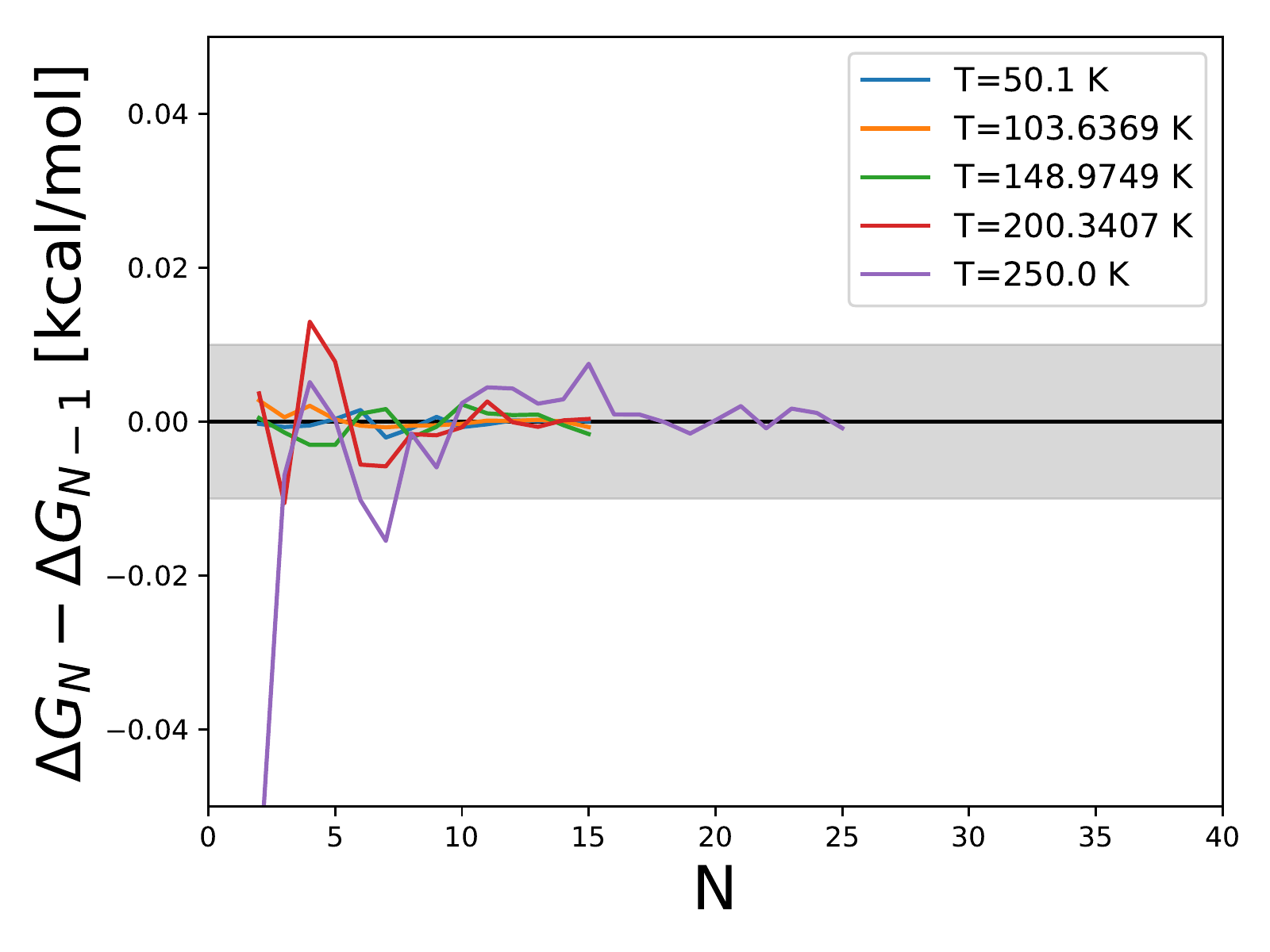} 
    \includegraphics[width=8cm]{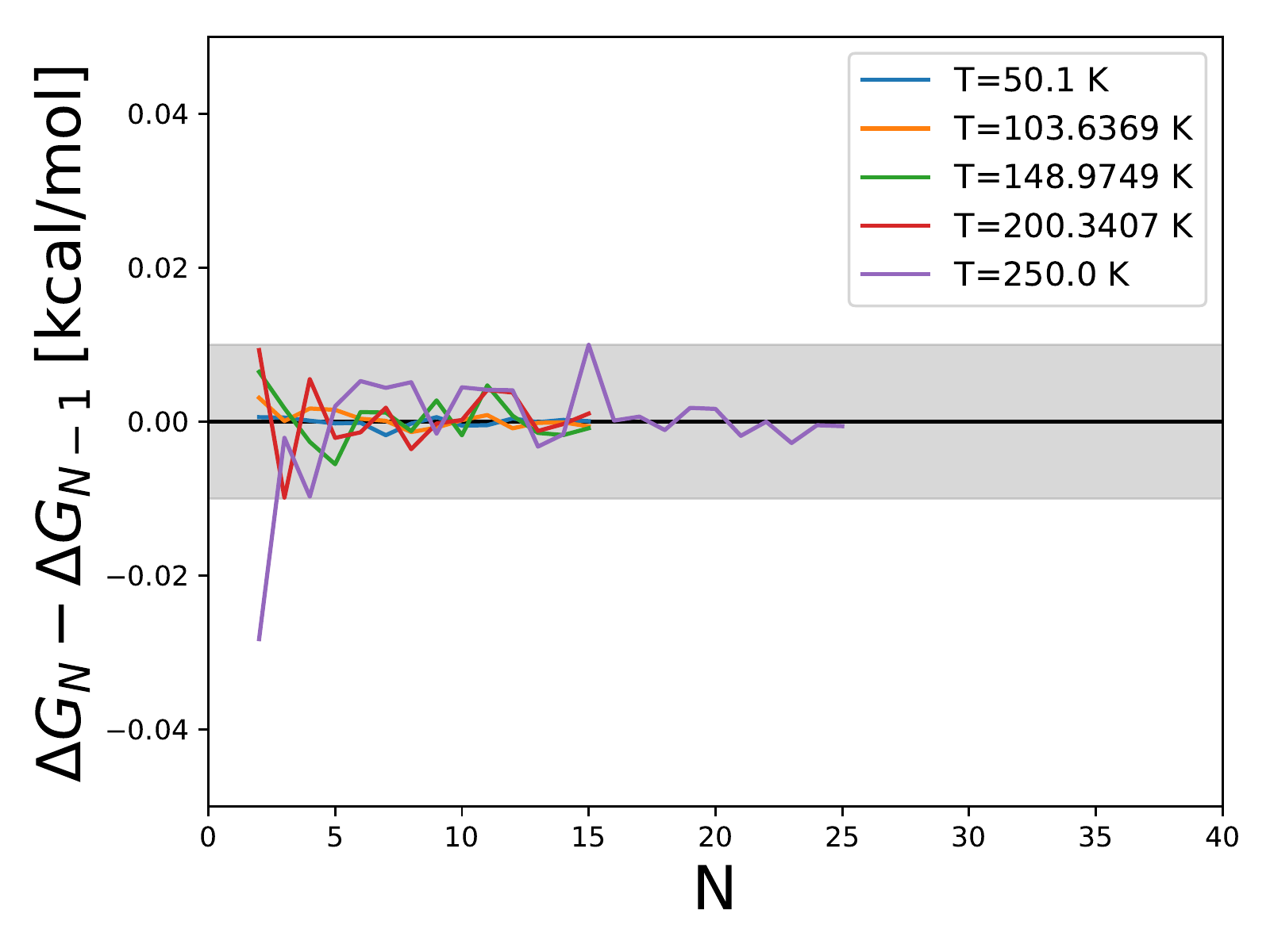} 
  \end{center}
  \caption{The change in the polymorph free energy difference when using $N$ and $N-1$ frames using Boltzmann-weighted HA to complete the thermodynamic cycle for polymorph II (left) and III (right) of benzene relative to form I. 
  \label{figure:converge_0}}
\end{figure*}

\begin{figure*}
  \begin{center}
    \includegraphics[width=8cm]{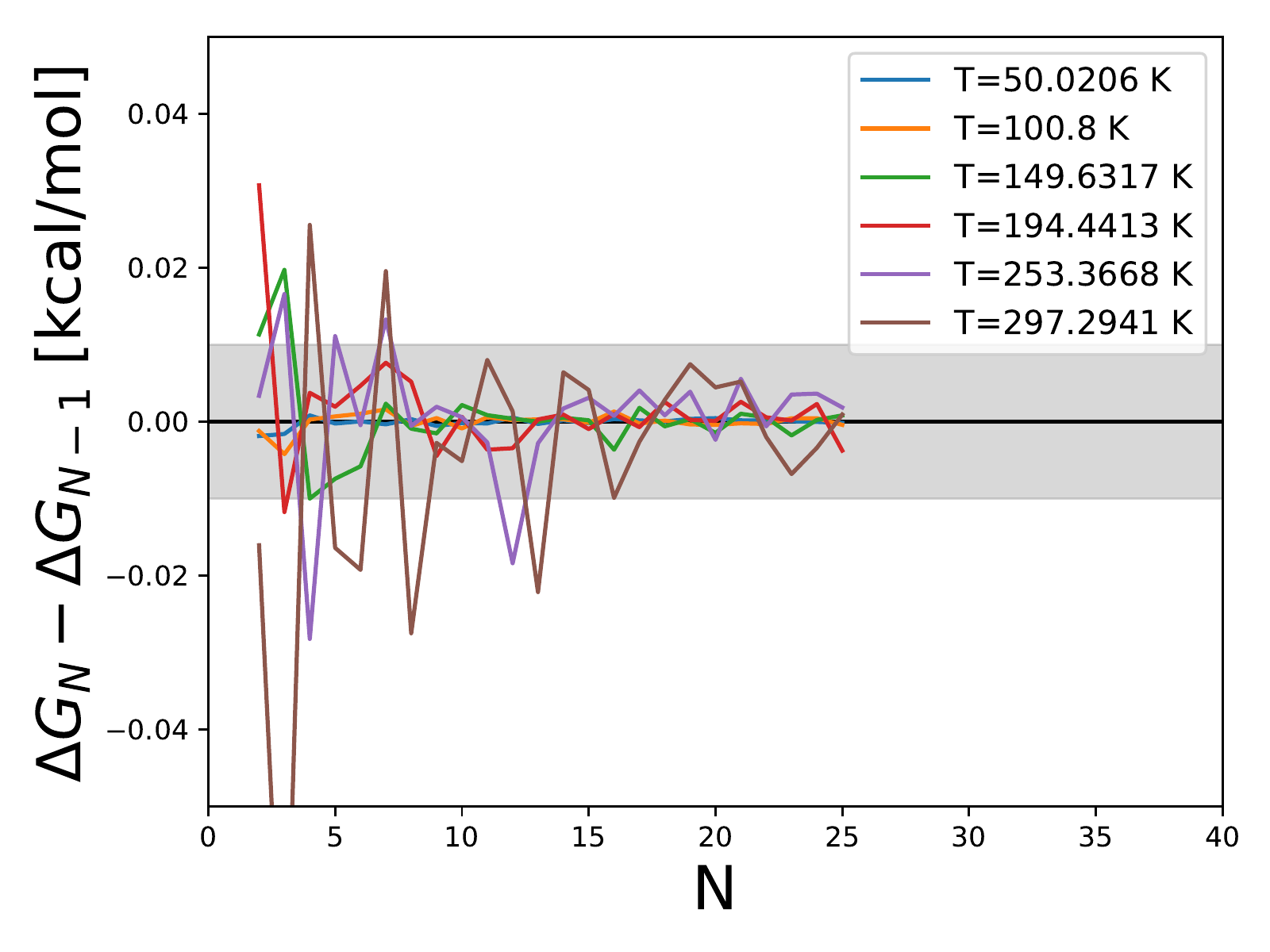}   
    \includegraphics[width=8cm]{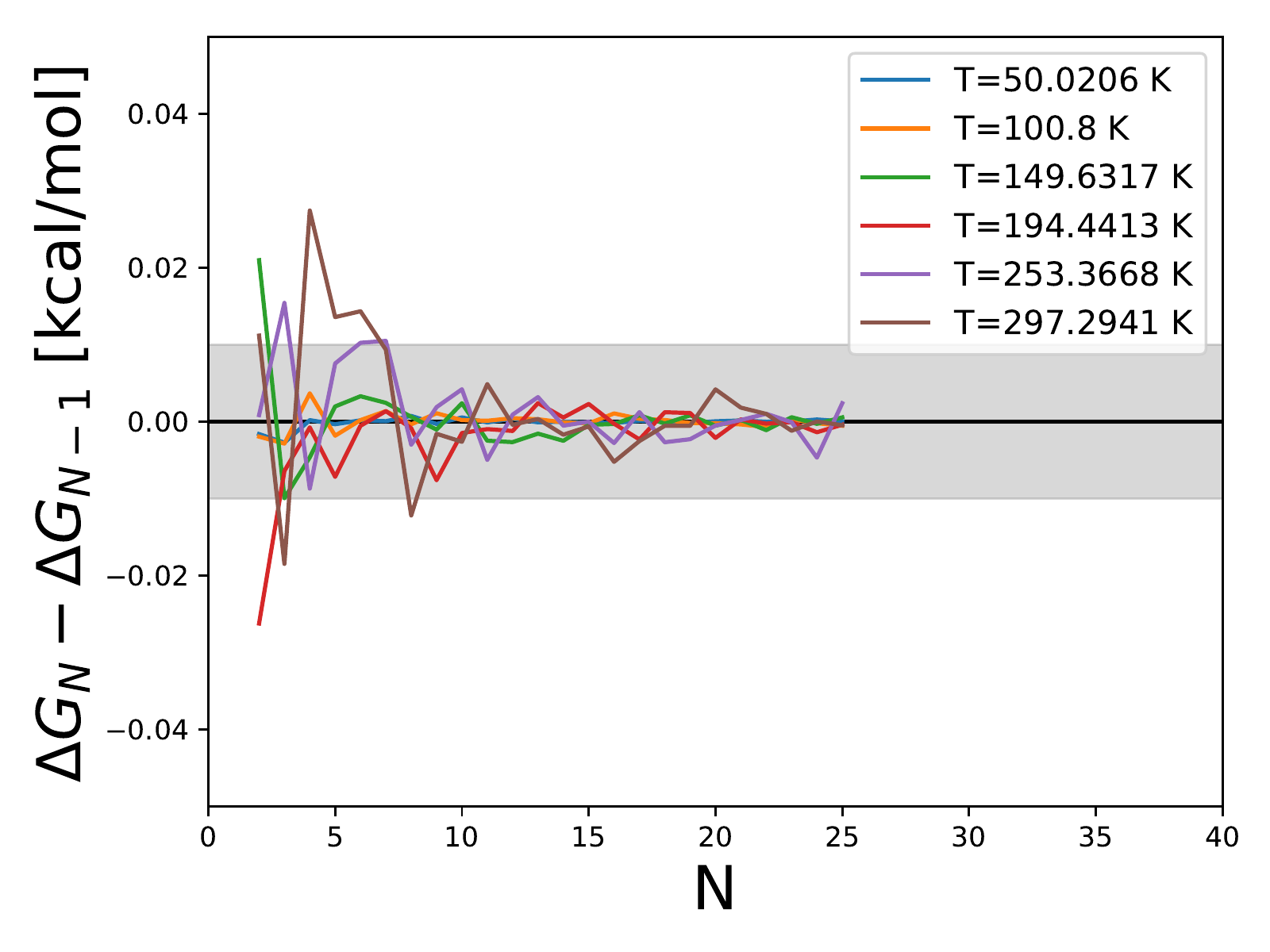}   
  \end{center}
  \caption{The change in the polymorph free energy difference when using $N$ and $N-1$ frames using Boltzmann-weighted HA to complete the thermodynamic cycle for polymorph II (left) and III (right) of piracetam relative to form I. 
  \label{figure:converge_1}}
\end{figure*}

\begin{figure*}
  \begin{center}
    \includegraphics[width=8cm]{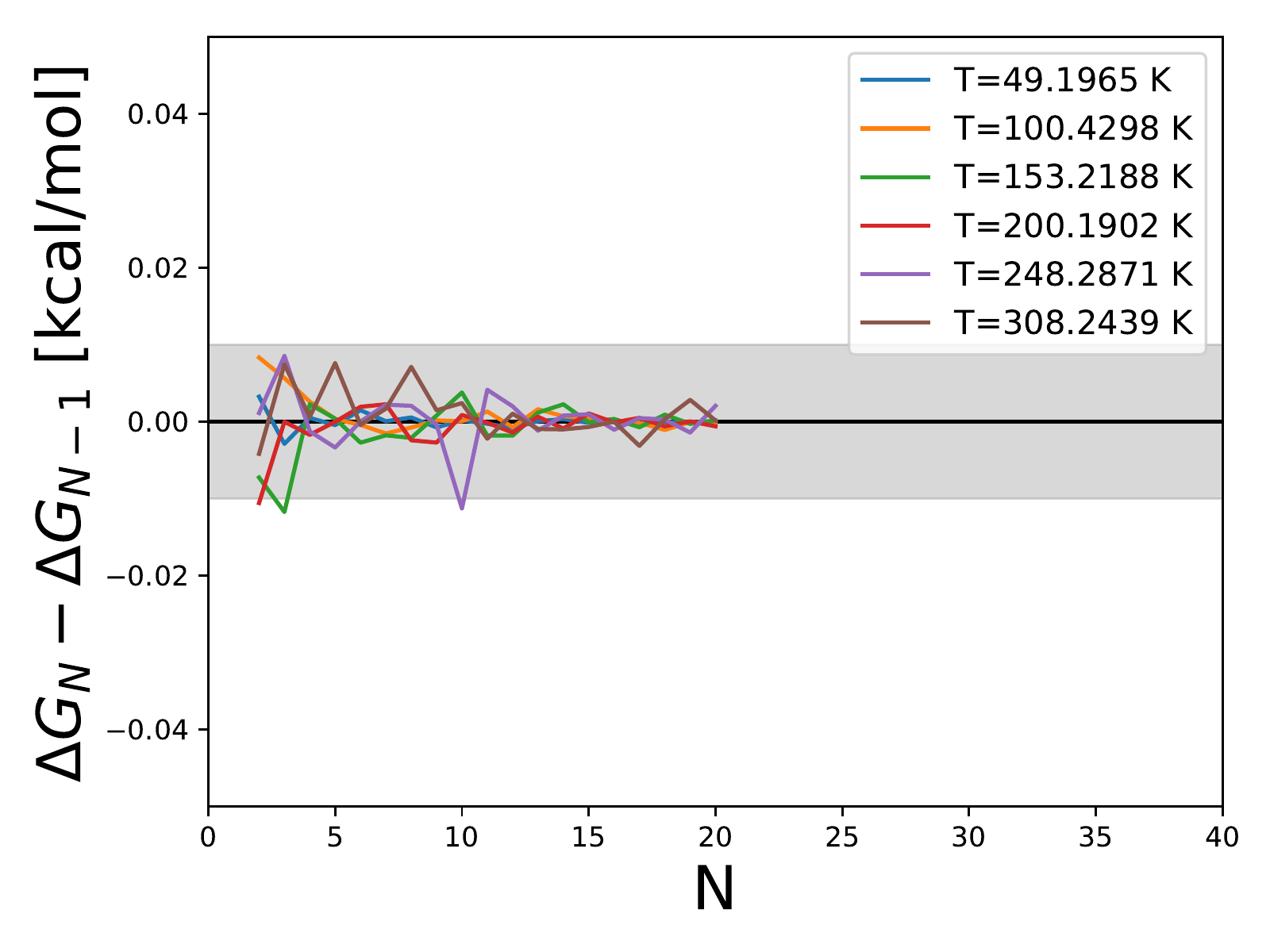}  
    \includegraphics[width=8cm]{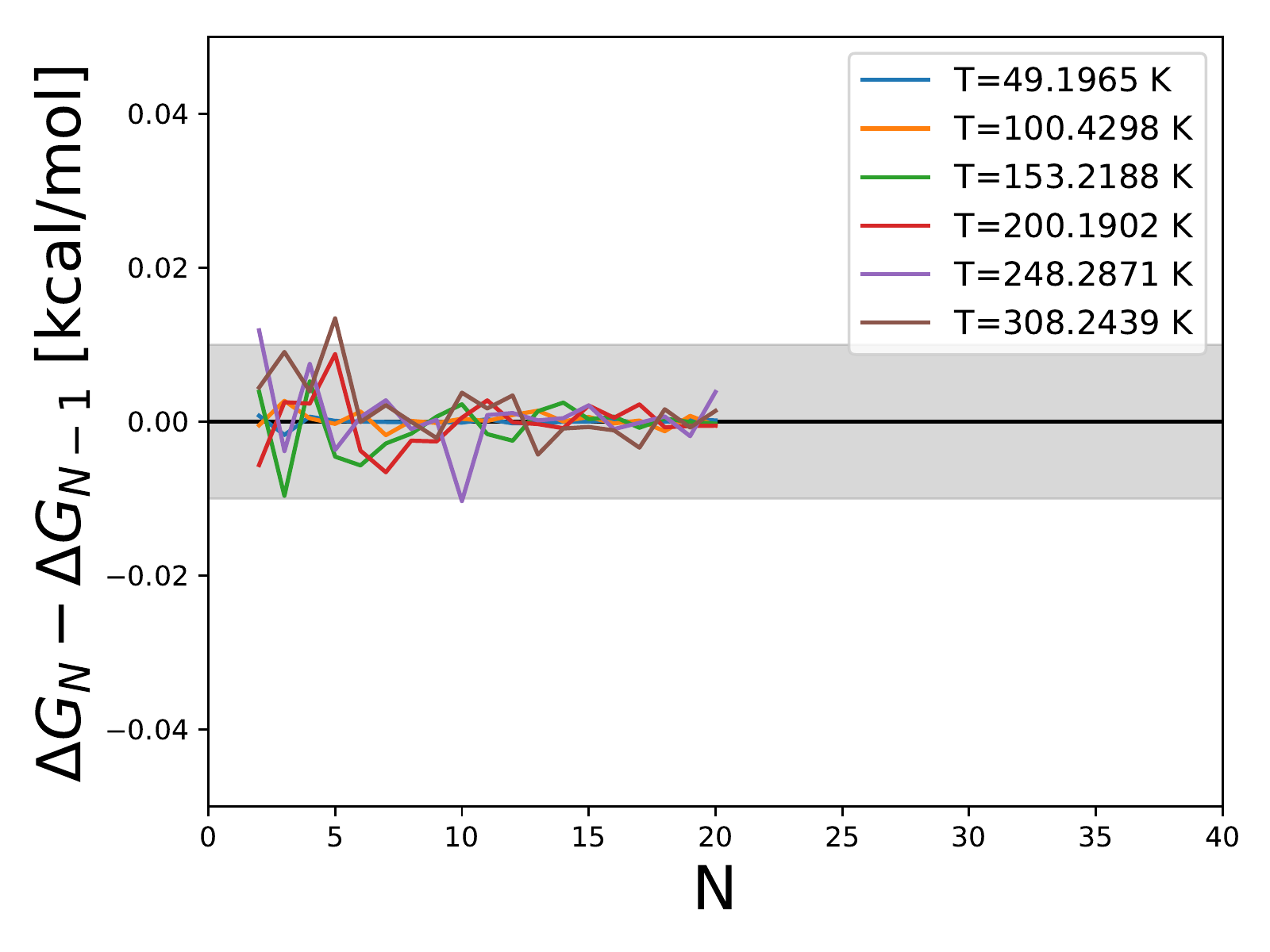}  
  \end{center}
  \caption{The change in the polymorph free energy difference when using $N$ and $N-1$ frames using Boltzmann-weighted HA to complete the thermodynamic cycle for polymorph III (left) and IV (right) of carbamazepine relative to form I. 
  \label{figure:converge_2}}
\end{figure*}

\begin{figure*}
  \begin{center}
    \includegraphics[width=8cm]{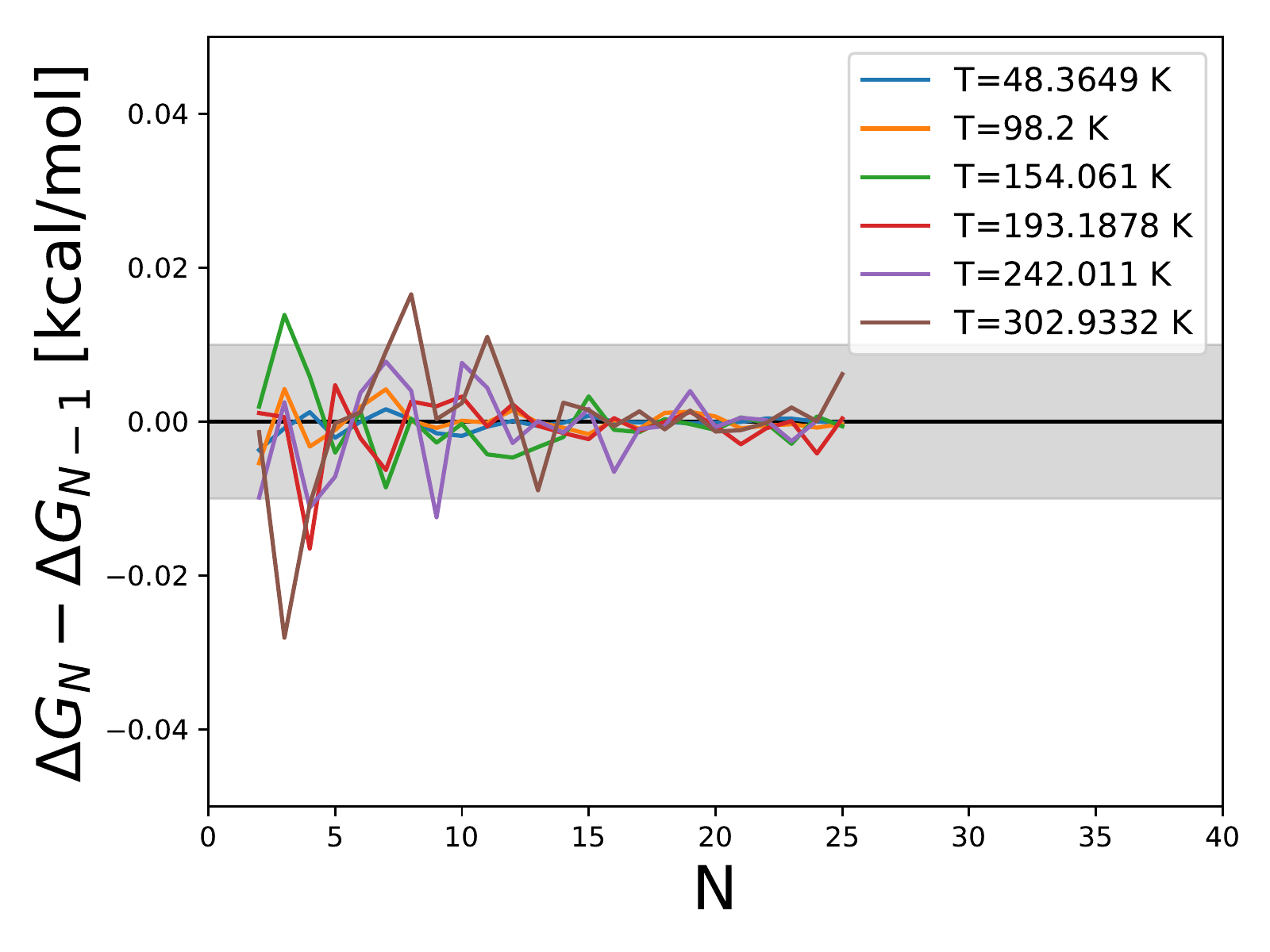}  
  \end{center}
  \caption{The change in the polymorph free energy difference when using $N$ and $N-1$ frames using Boltzmann-weighted HA to complete the thermodynamic cycle for polymorph II of paracetamol relative to form I. 
  \label{figure:converge_3}}
\end{figure*}

\begin{figure*}
  \begin{center}
    \includegraphics[width=8cm]{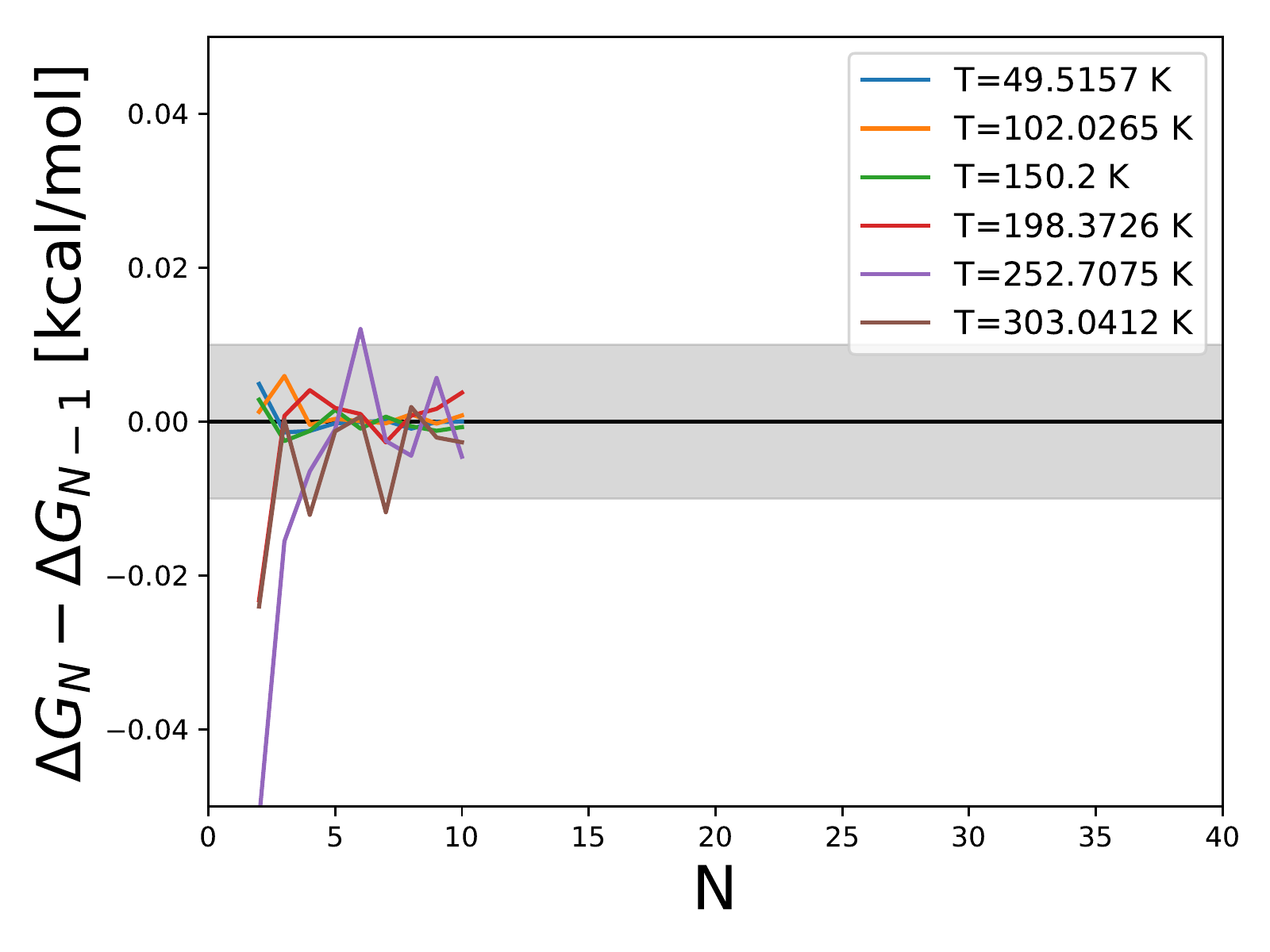}  
  \end{center}
  \caption{The change in the polymorph free energy difference when using $N$ and $N-1$ frames using Boltzmann-weighted HA to complete the thermodynamic cycle for polymorph II of adenine relative to form I. 
  \label{figure:converge_4}}
\end{figure*}

\begin{figure*}
  \begin{center}
    \includegraphics[width=8cm]{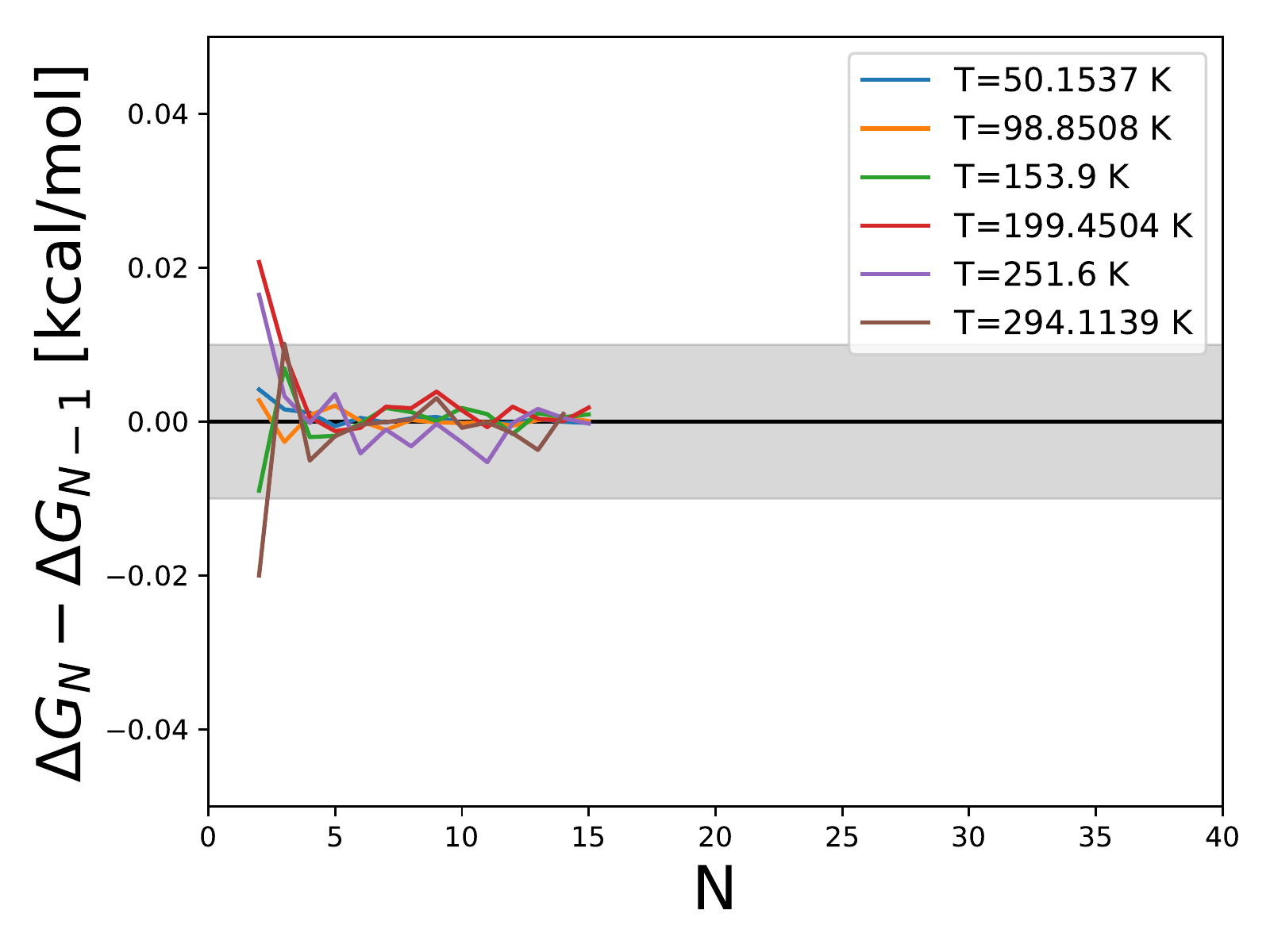}
    \includegraphics[width=8cm]{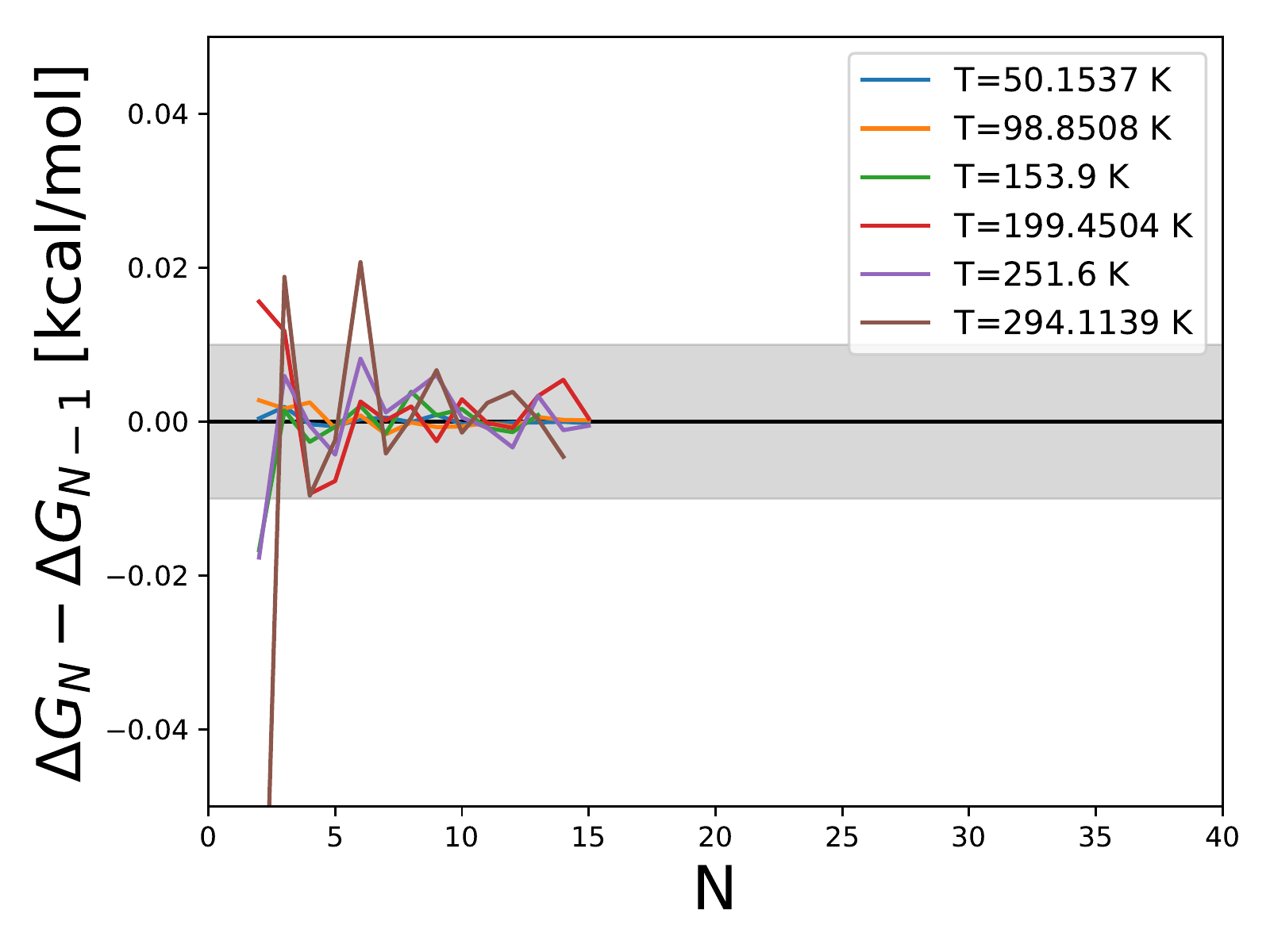} \\
    \includegraphics[width=8cm]{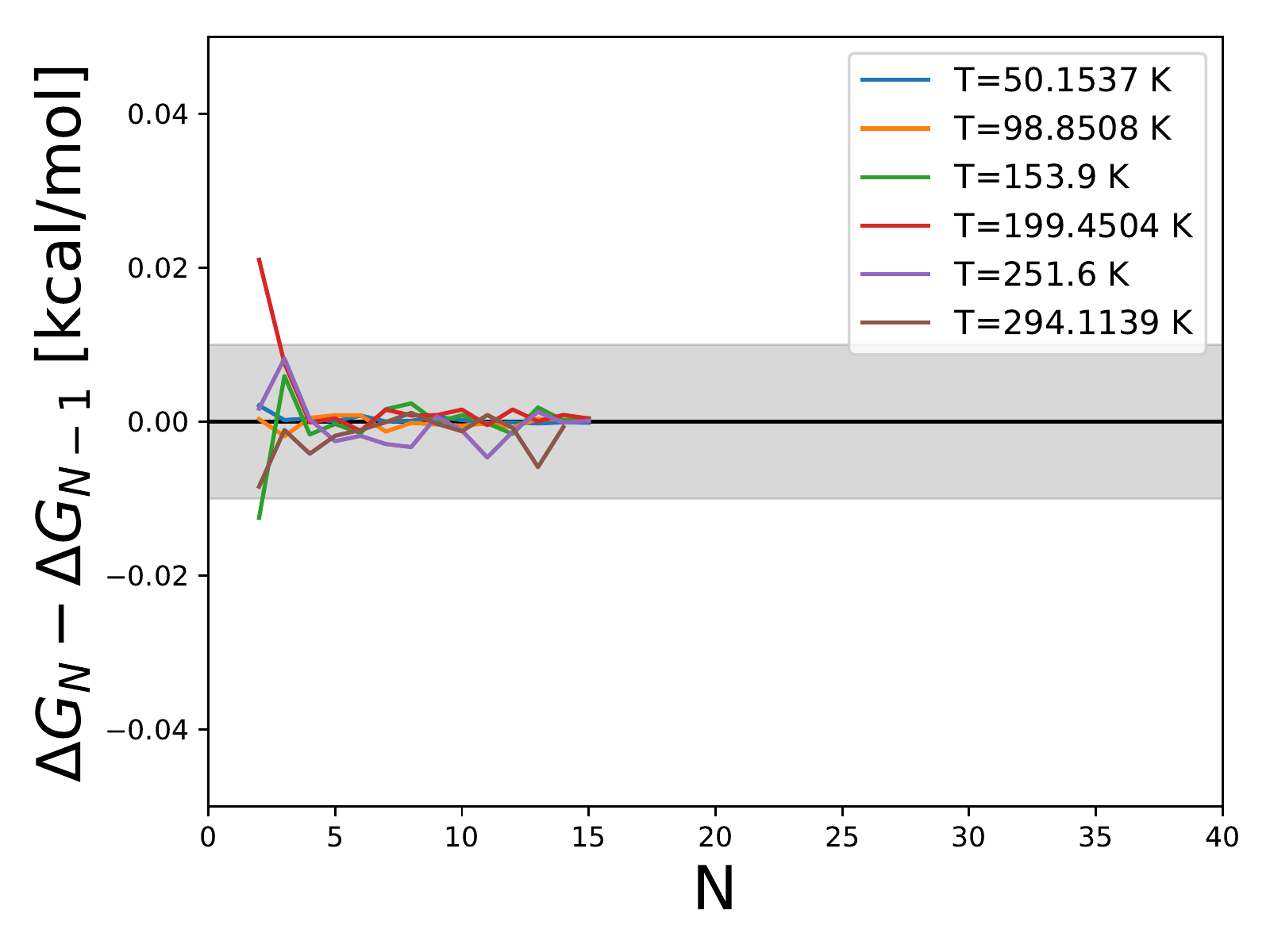}
  \end{center}
  \caption{The change in the polymorph free energy difference when using $N$ and $N-1$ frames using Boltzmann-weighted HA to complete the thermodynamic cycle for polymorph $\beta$ (top-left), $\gamma$ (top-right), and $\delta$ (bottom) of pyrazinamide relative to form $\alpha$. 
  \label{figure:converge_5}}
\end{figure*}

\begin{figure*}
  \begin{center}
    \includegraphics[width=8cm]{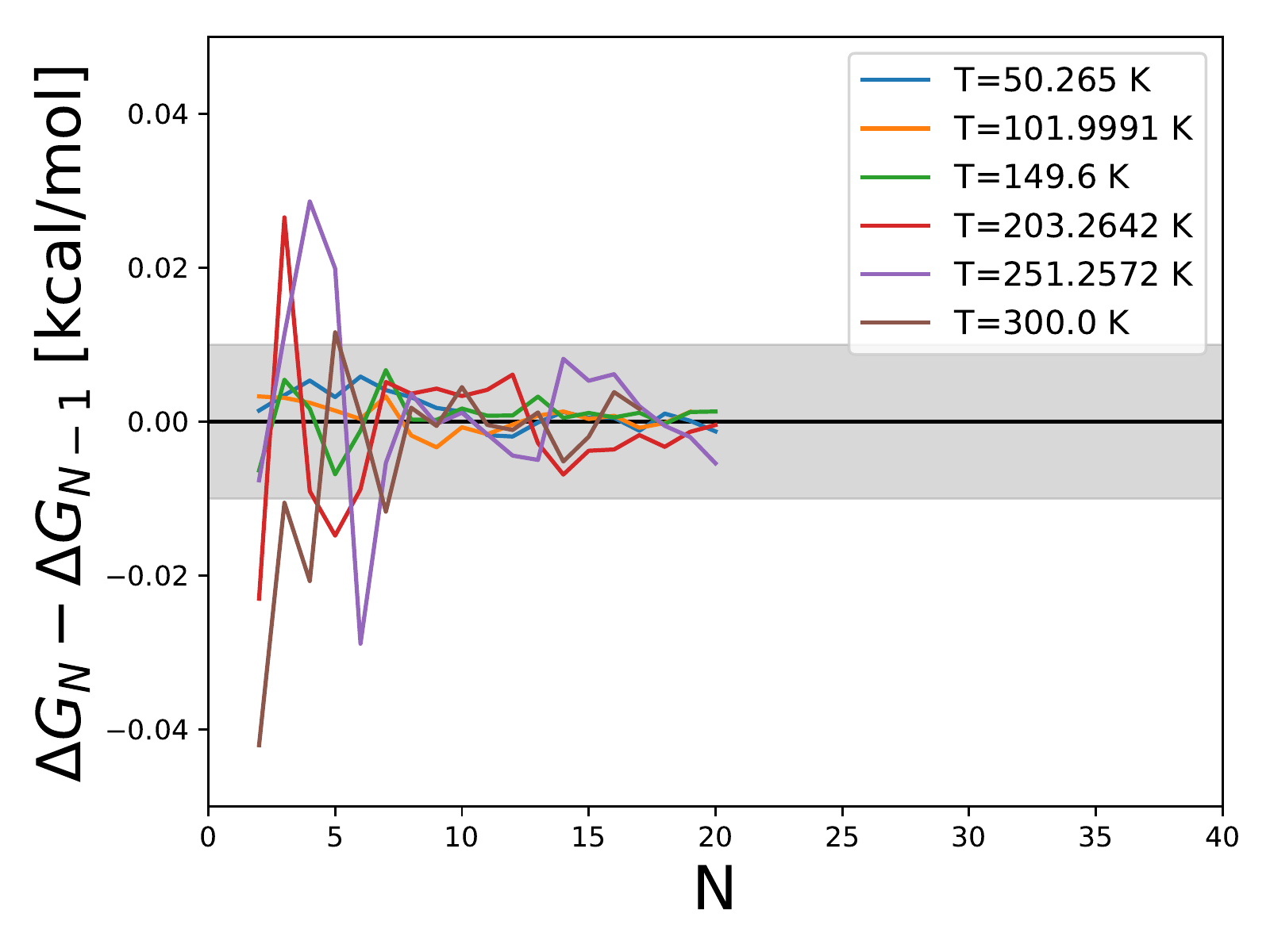}
  \end{center} 
  \caption{The change in the polymorph free energy difference when using $N$ and $N-1$ frames using Boltzmann-weighted HA to complete the thermodynamic cycle for polymorph III of succinonitrile relative to form I. 
  \label{figure:converge_6}}
\end{figure*}

\begin{figure*}
  \begin{center}
    \includegraphics[width=8cm]{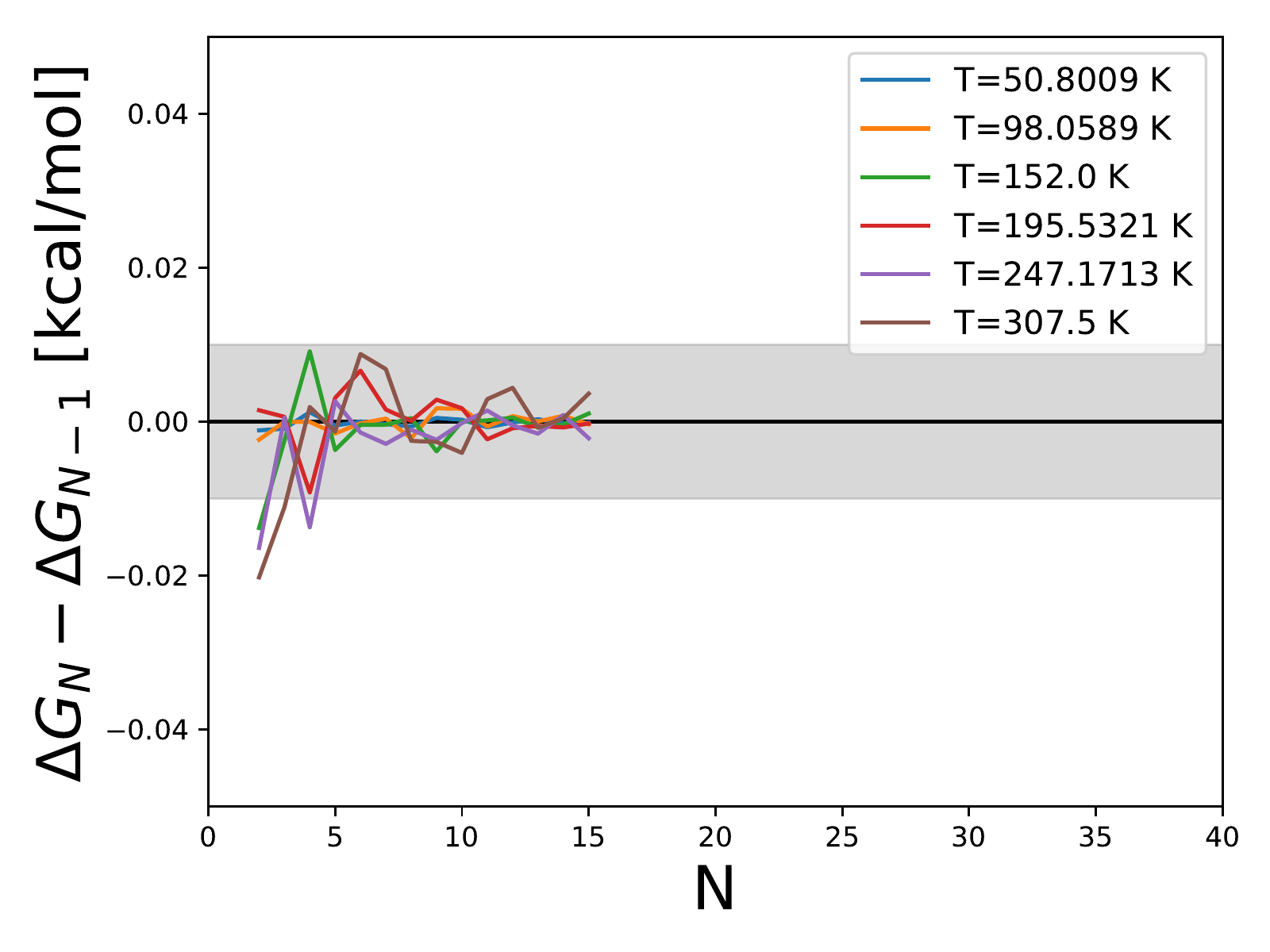}  
  \end{center}
  \caption{The change in the polymorph free energy difference when using $N$ and $N-1$ frames using Boltzmann-weighted HA to complete the thermodynamic cycle for polymorph $\beta$ of resorcinol relative to form $\alpha$. 
  \label{figure:converge_7}}
\end{figure*}

\begin{figure*}
  \begin{center}
    \includegraphics[width=8cm]{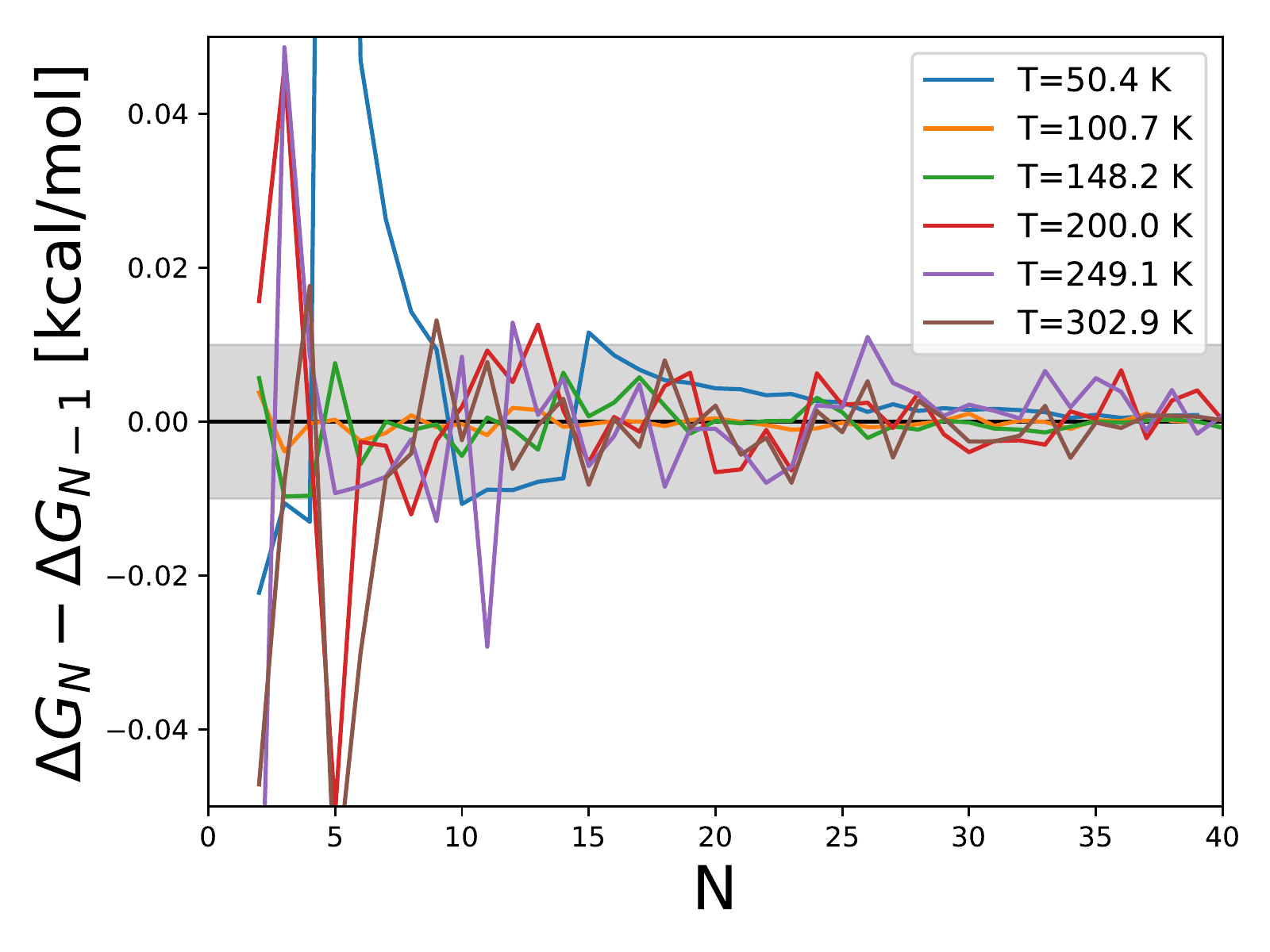}  
  \end{center}
  \caption{The change in the polymorph free energy difference when using $N$ and $N-1$ frames using Boltzmann-weighted HA to complete the thermodynamic cycle for polymorph X of aripiprazole relative to form I. 
  \label{figure:converge_8}}
\end{figure*}

\begin{figure*}
  \begin{center}
    \includegraphics[width=8cm]{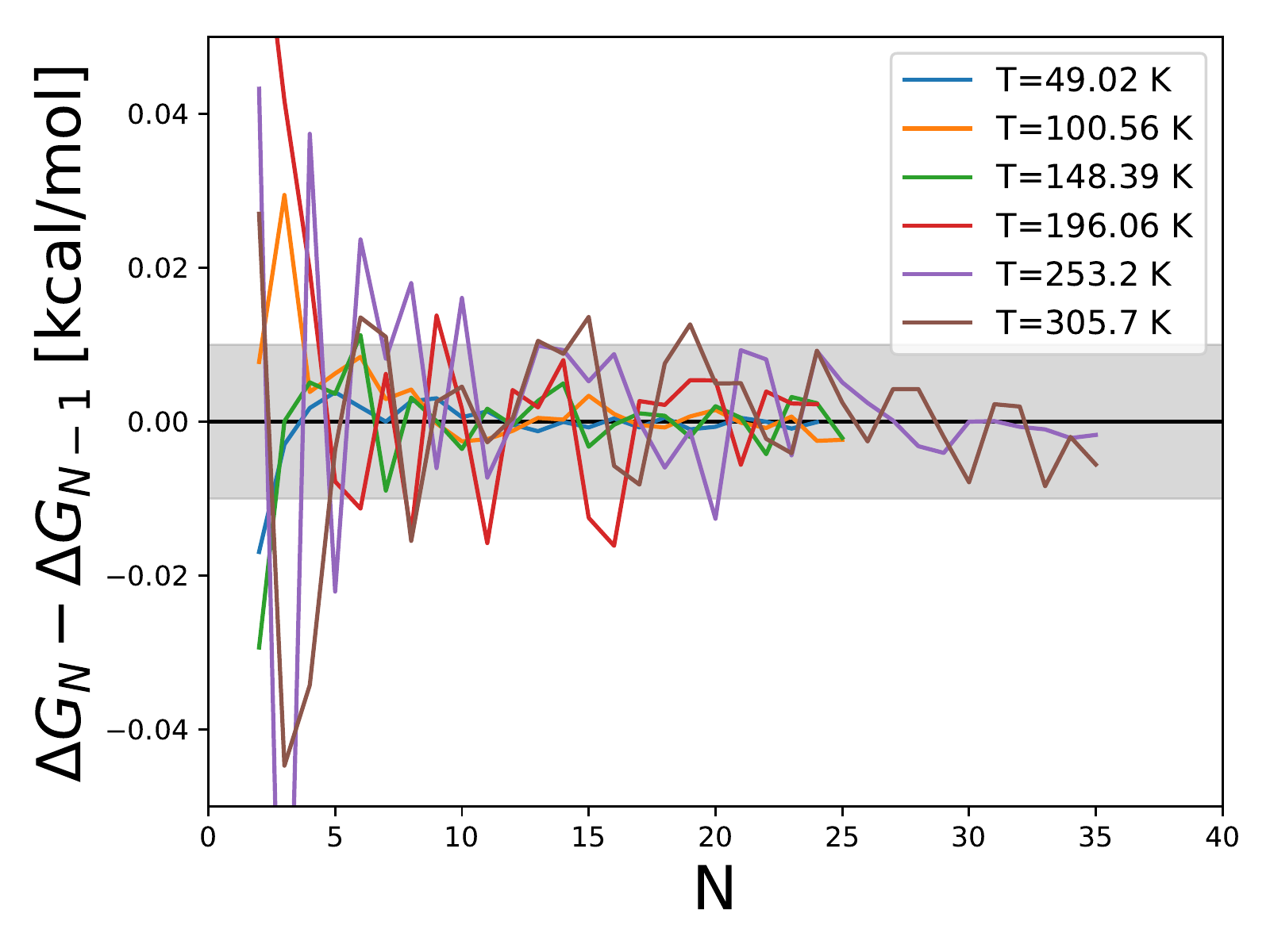}
  \end{center}
  \caption{The change in the polymorph free energy difference when using $N$ and $N-1$ frames using Boltzmann-weighted HA to complete the thermodynamic cycle for polymorph V of chlorpropamide relative to form I. 
  \label{figure:converge_9}}
\end{figure*}

\begin{figure*}
  \begin{center}
    \includegraphics[width=8cm]{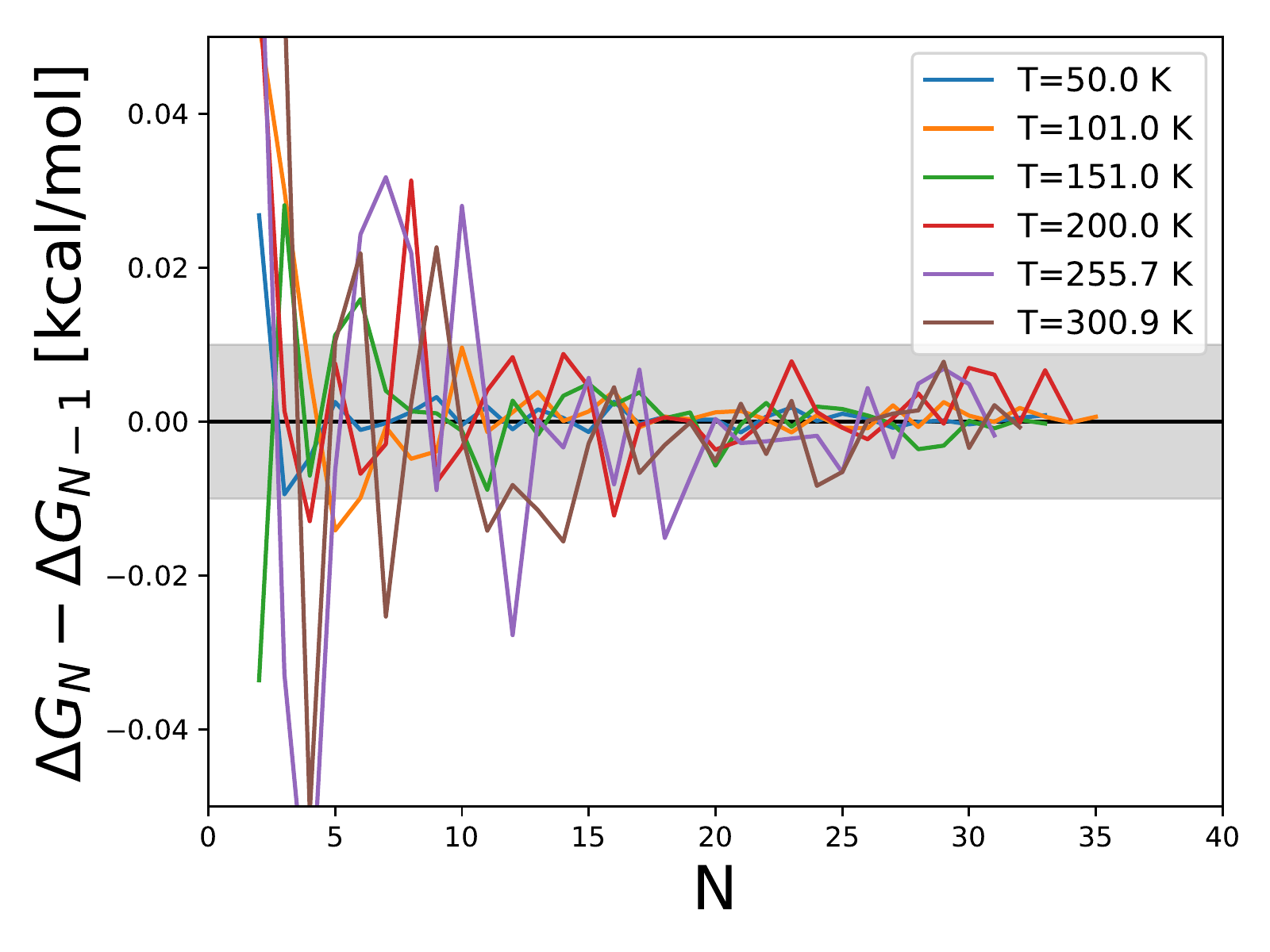}  
    \includegraphics[width=8cm]{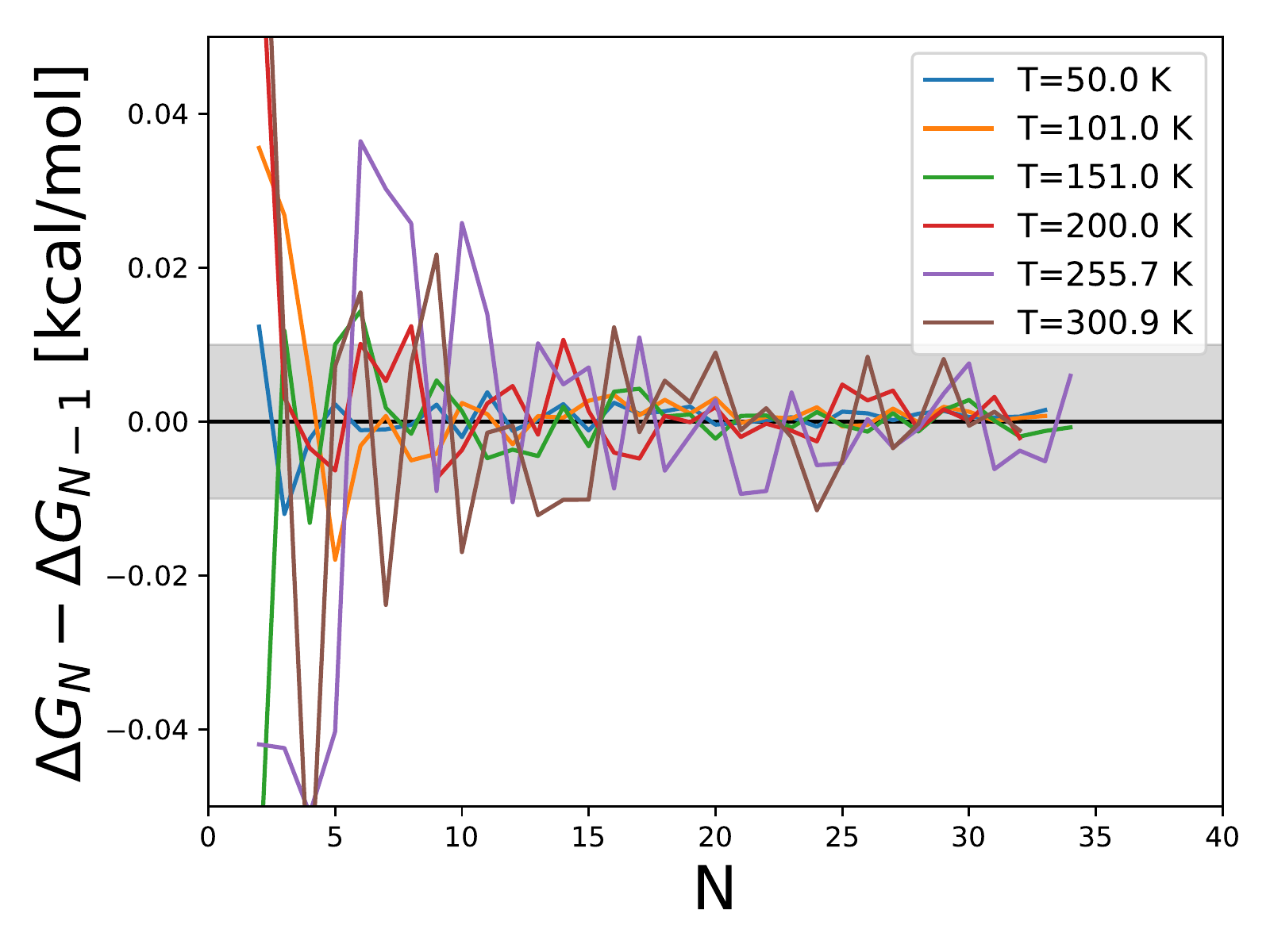} \\
    \includegraphics[width=8cm]{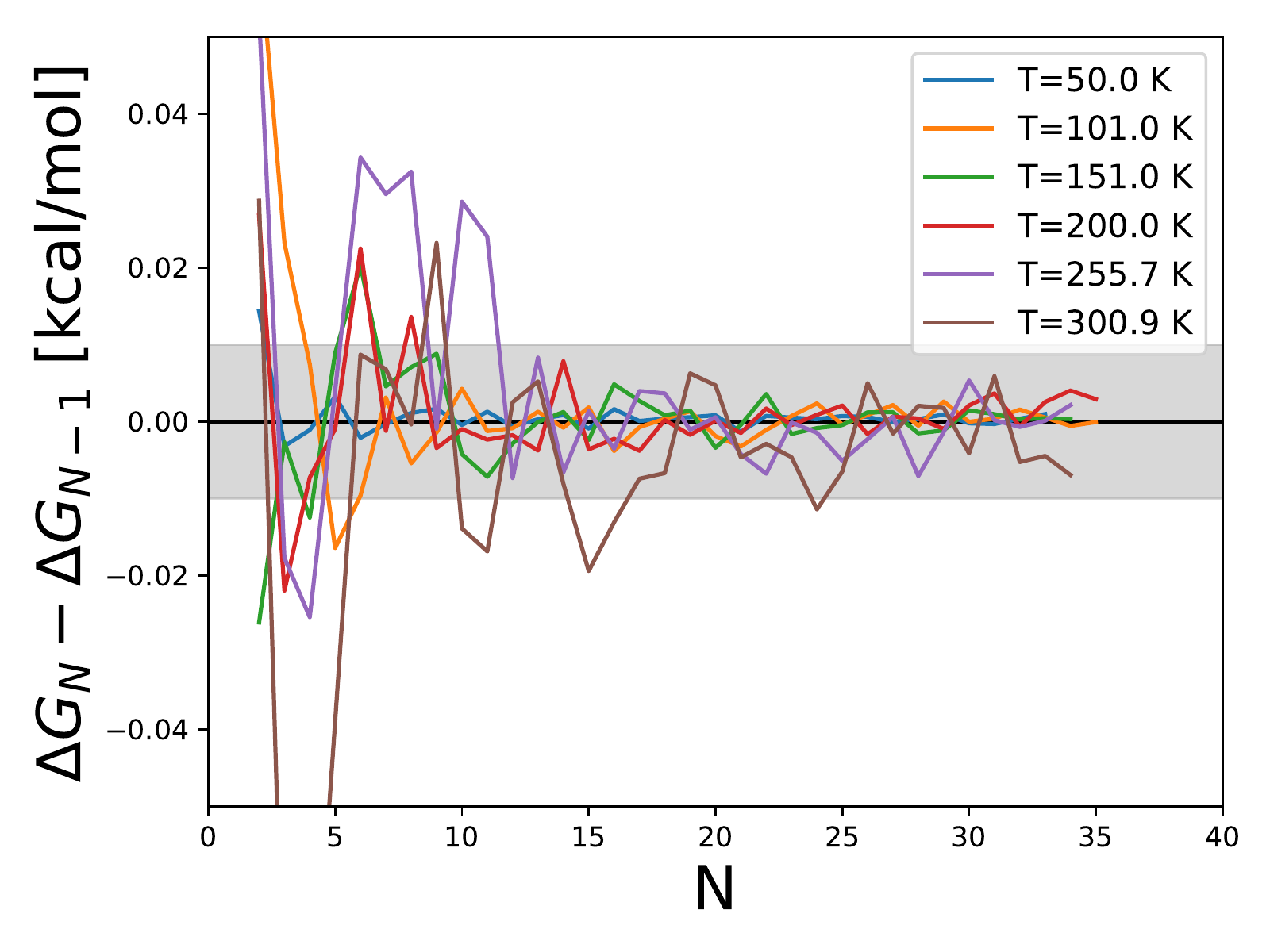}
  \end{center}
  \caption{The change in the polymorph free energy difference when using $N$ and $N-1$ frames using Boltzmann-weighted HA to complete the thermodynamic cycle for polymorph II (top-left), III (top-right), and V (bottom) of tolbutamide relative to form I. 
  \label{figure:converge_10}}
\end{figure*}

    Determining a polymorph free energy difference with Boltzmann weighted HA for these systems requires 10 -- 40 frames from an NPT trajectory when using a reference temperature between 50 -- 300 K. Figures ~\ref{figure:converge_0} -- ~\ref{figure:converge_10} show the change in the polymorph free energy differences of each pair of polymorphs using our Boltzmann weighted HA method. The error in potential energy differences of polymorphs when computed with GROMACS (MD simulations) and Tinker (HA/QHA) is between 0.01--0.005 kcal/mol. If the change in the free energy difference computed with Boltzmann weighted HA is fluctuating between -0.01--0.01 kcal/mol (spanned by the grey bar) as $N$ is increased, we can assume that the method is converged within the error in the simulation packages.

\newpage
\newpage

\section{Error evaluation of the approximations} \label{section:dG_approx_error}
    In equation M\ref{eq:MD_dG} the error for using an approximate $\Delta G(T_{ref})$ can be computed as:
  \begin{eqnarray}
    \delta(\Delta(G_{ij}(T)) &= |\Delta G_{ij}^{aprox.}(T) - \Delta G_{ij}(T)| \\
    &= Tk_{B}\left(\Delta f_{ij}(T_{ref}) - \Delta f_{ij}(T_{ref}^{aprox.})\right) \nonumber \\
    &+ T \left( \frac{\Delta G_{ij}^{aprox.}(T_{ref}^{aprox.})}{T_{ref}^{aprox.}} - \frac{\Delta G_{ij}(T_{ref})}{T_{ref}} \right)
  \end{eqnarray}
Here we can move all temperature dependence to the left side of the equation to get a constant value of the error for the approximation:
  \begin{eqnarray}
    \frac{\delta(\Delta(G_{ij}(T))}{T} =& k_{B}\left(\Delta f_{ij}(T_{ref}) - \Delta f_{ij}(T_{ref}^{aprox.})\right) \nonumber \\ 
    &+  \left( \frac{\Delta G_{ij}^{aprox.}(T_{ref}^{aprox.})}{T_{ref}^{aprox.}} - \frac{\Delta G_{ij}(T_{ref})}{T_{ref}} \right)
  \end{eqnarray}
Once we solve this, we can evaluate the error in the free energy at any temperature of interest.

\section{System Dependent Results} \label{section:system_results}

\begin{figure*}
  \begin{center}
    \includegraphics[width=8cm]{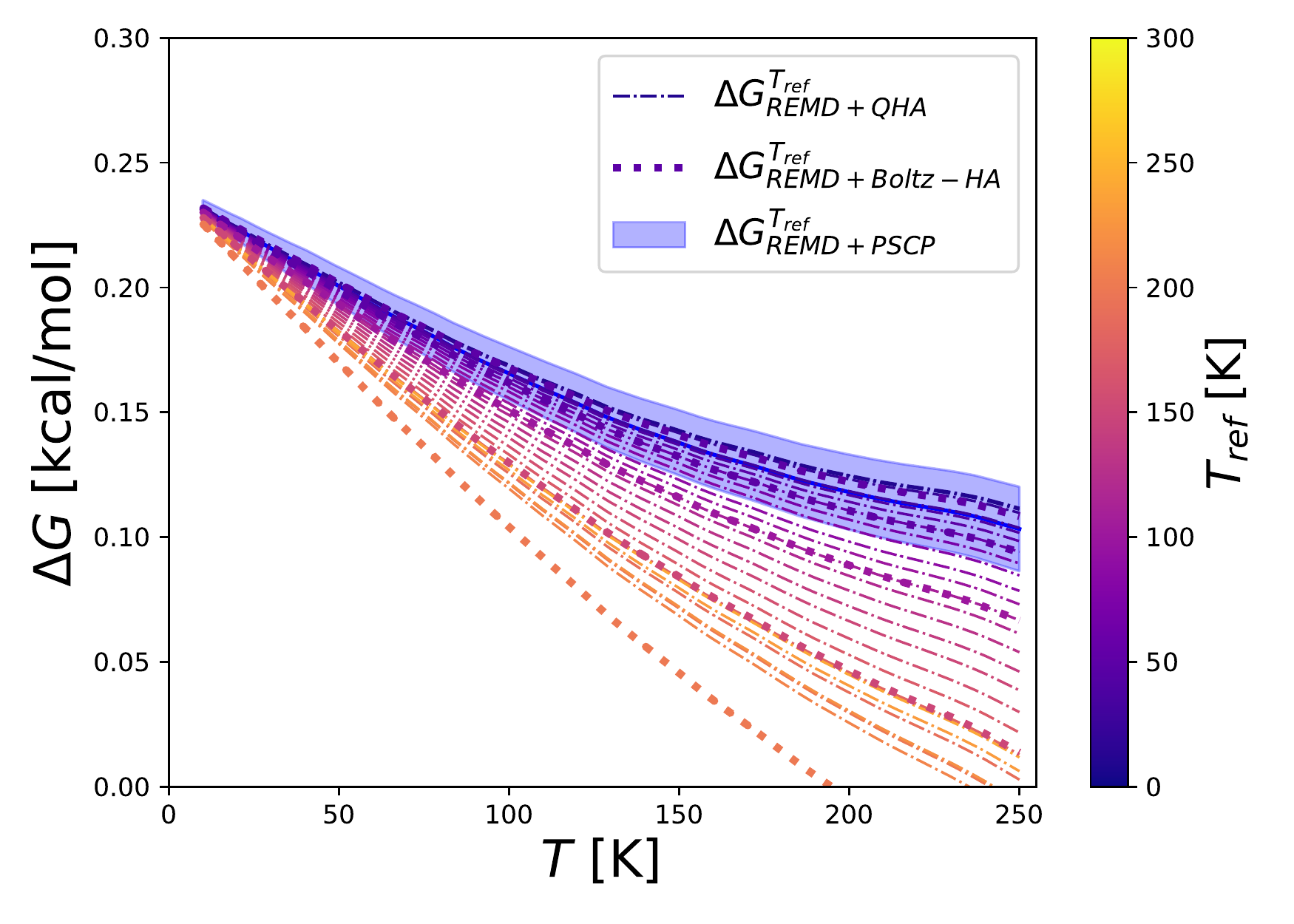}
    \includegraphics[width=8cm]{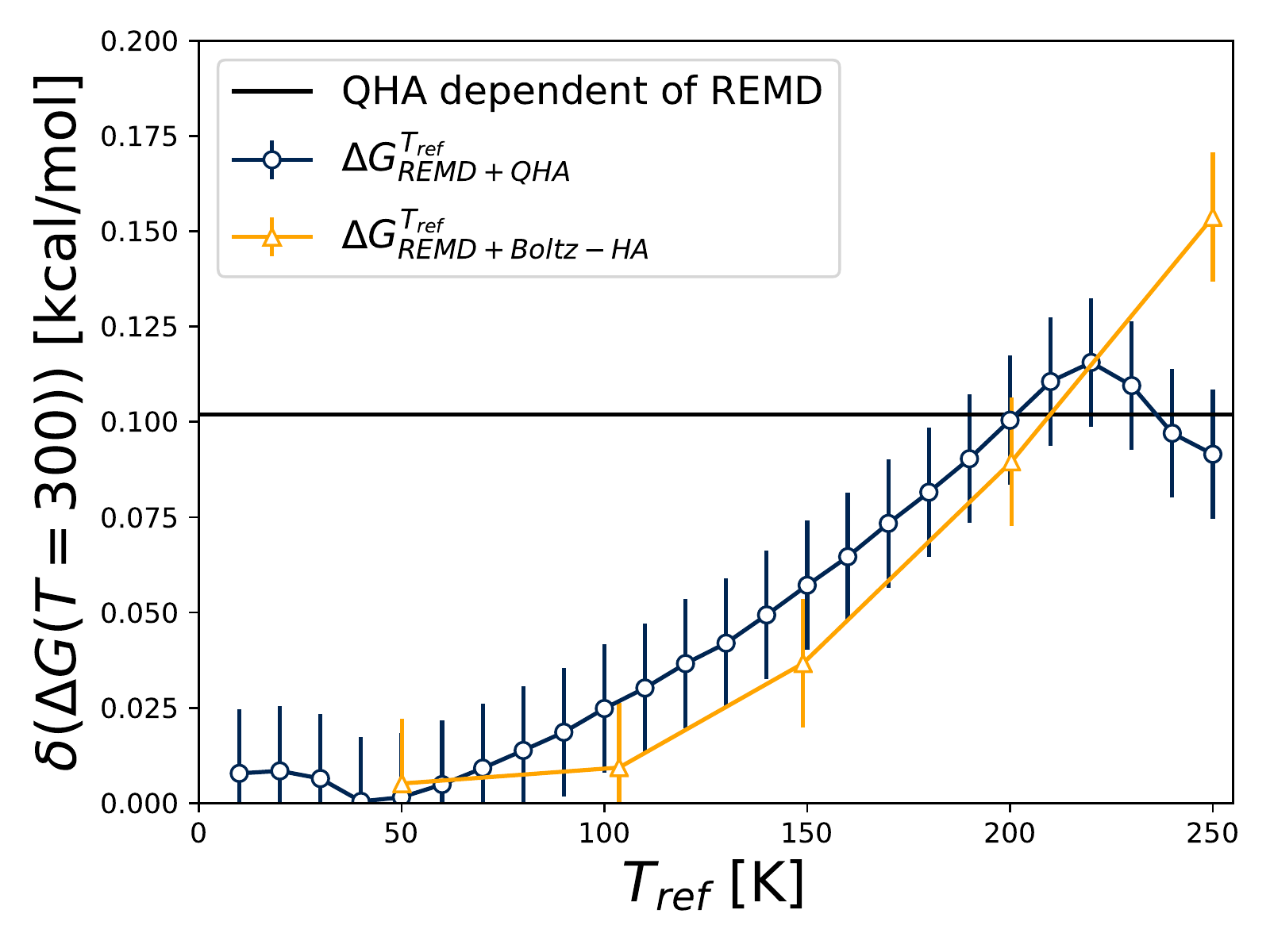}
  \end{center}
  \caption{Left) The polymorph free energy difference of benzene (form III relative to form I) using MD, QHA, and replacing $\Delta G (T_{ref})$ from PSCP in equation ~\ref{eq:MD_dG} with the QHA values for $T_{ref}$ every 10 K is shown. Right) The error in the $\Delta G(300K)$ is reported for replacing $\Delta G (T_{ref})$ from QHA with PSCP and using Boltzmann weighted HA. Each line of the approximation in the left figure is color coded based off the reference temperature used.
  \label{figure:dG_error_1}}
\end{figure*}

\begin{figure*}
  \begin{center}
    \includegraphics[width=8cm]{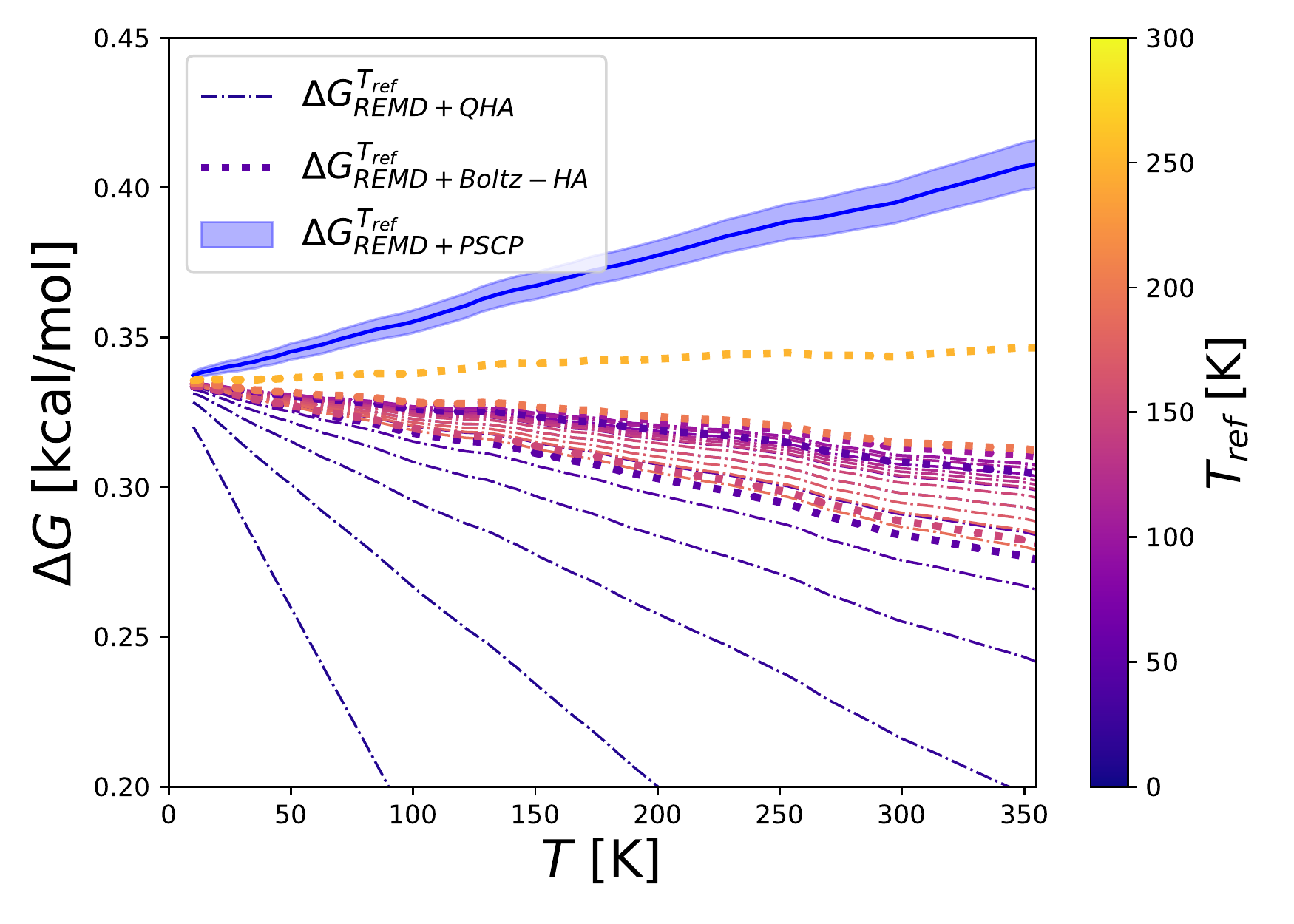}
    \includegraphics[width=8cm]{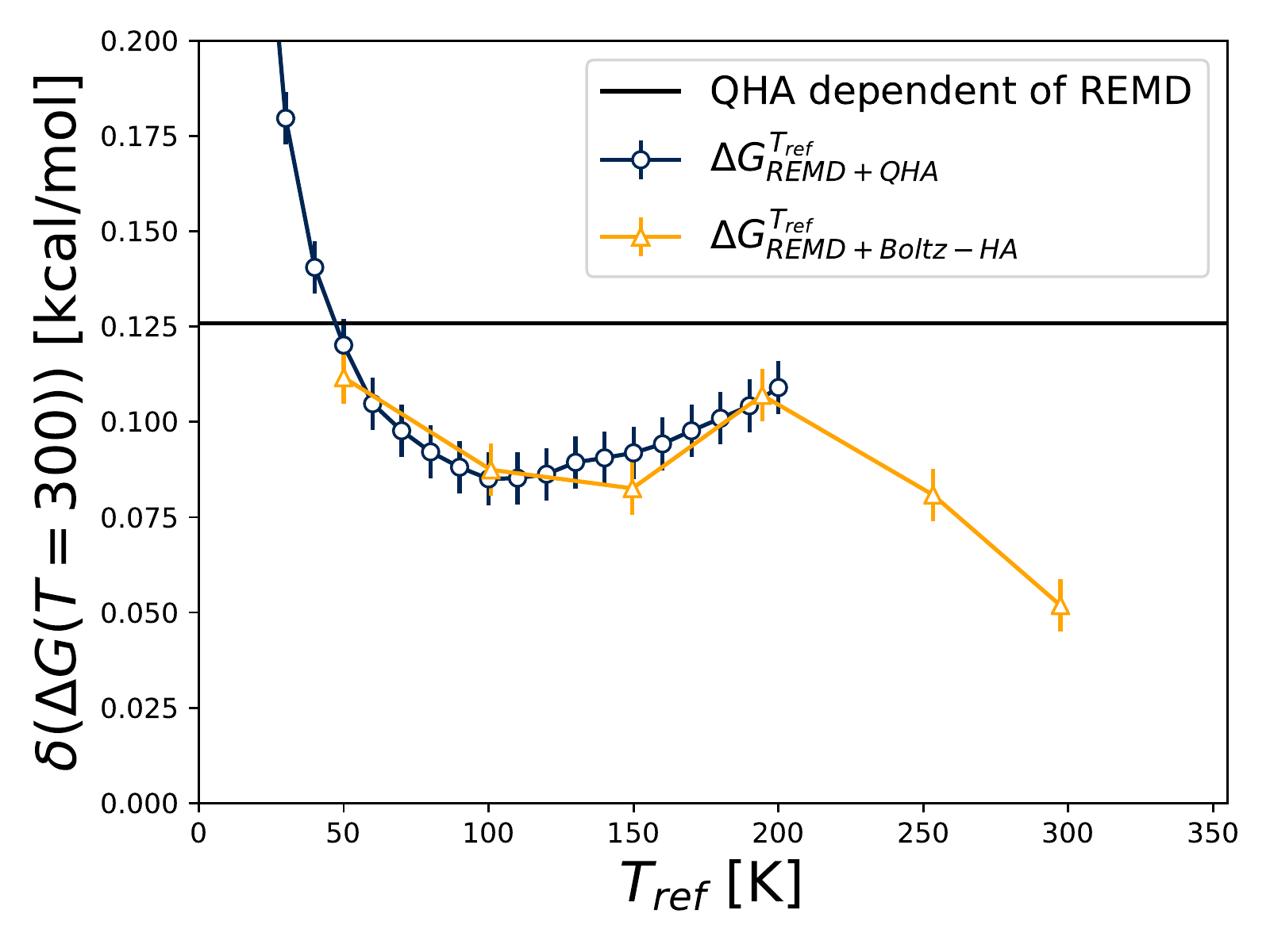}
  \end{center}
  \caption{Left) The polymorph free energy difference of piracetam (form II relative to form I) using MD, QHA, and replacing $\Delta G (T_{ref})$ from PSCP in equation ~\ref{eq:MD_dG} with the QHA values for $T_{ref}$ every 10 K is shown. Right) The error in the $\Delta G(300K)$ is reported for replacing $\Delta G (T_{ref})$ from QHA with PSCP and using Boltzmann weighted HA. Each line of the approximation in the left figure is color coded based off the reference temperature used.
  \label{figure:dG_error_2}}
\end{figure*}

\begin{figure*}
  \begin{center}
    \includegraphics[width=8cm]{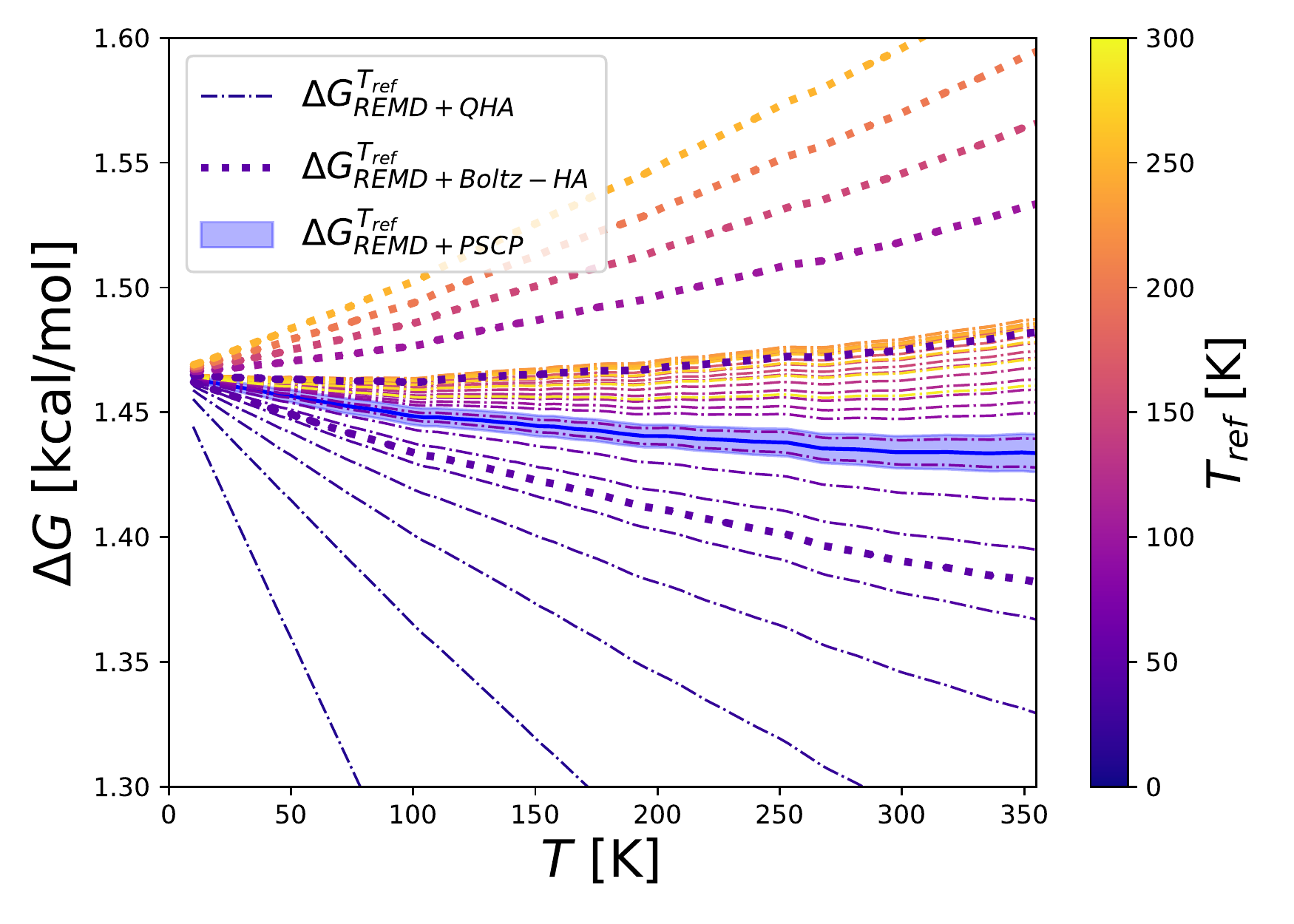}
    \includegraphics[width=8cm]{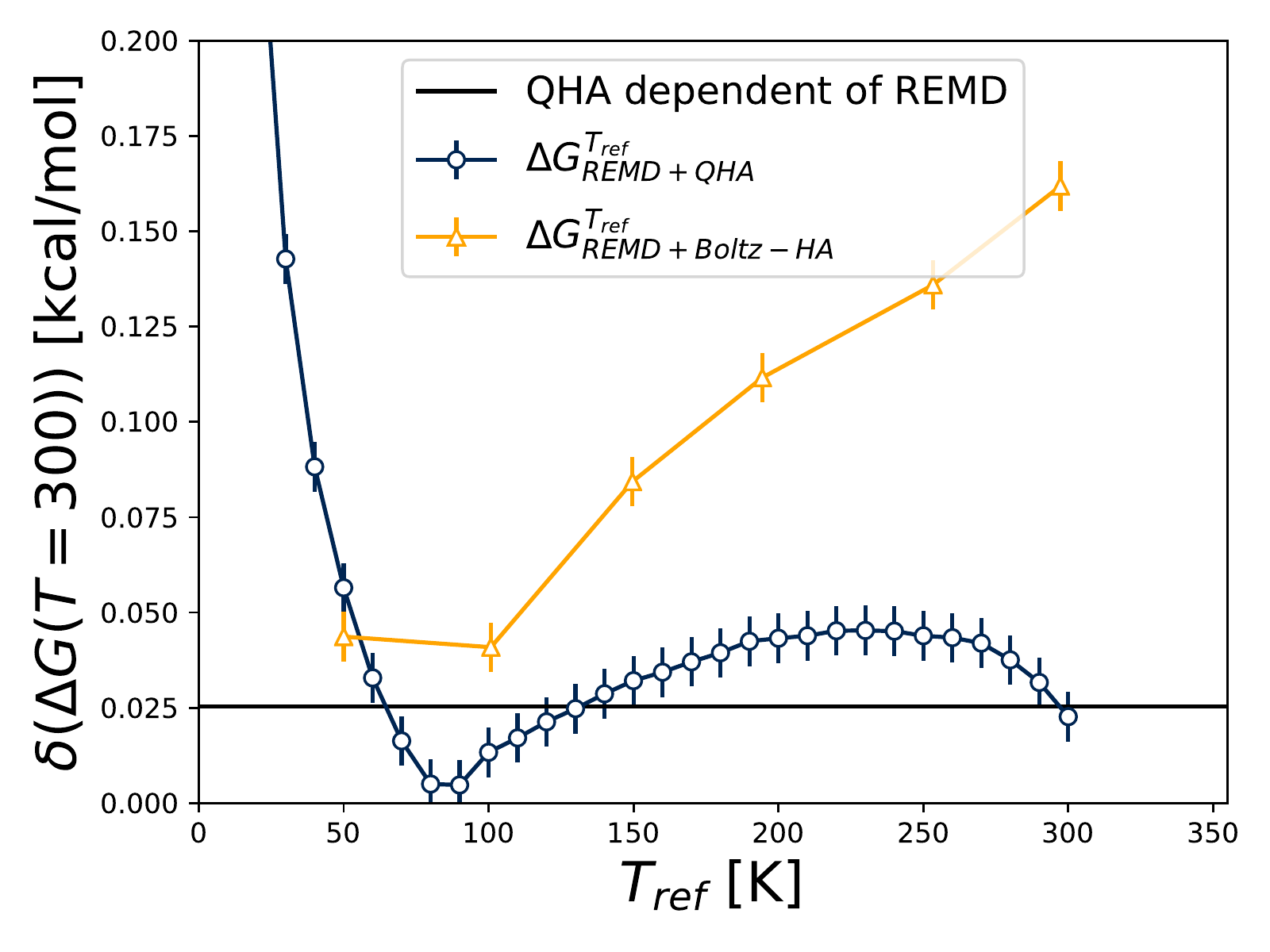}
  \end{center}
  \caption{Left) The polymorph free energy difference of piracetam (form III relative to form I) using MD, QHA, and replacing $\Delta G (T_{ref})$ from PSCP in equation ~\ref{eq:MD_dG} with the QHA values for $T_{ref}$ every 10 K is shown. Right) The error in the $\Delta G(300K)$ is reported for replacing $\Delta G (T_{ref})$ from QHA with PSCP and using Boltzmann weighted HA. Each line of the approximation in the left figure is color coded based off the reference temperature used.
  \label{figure:dG_error_3}}
\end{figure*}

\begin{figure*}
  \begin{center}
    \includegraphics[width=8cm]{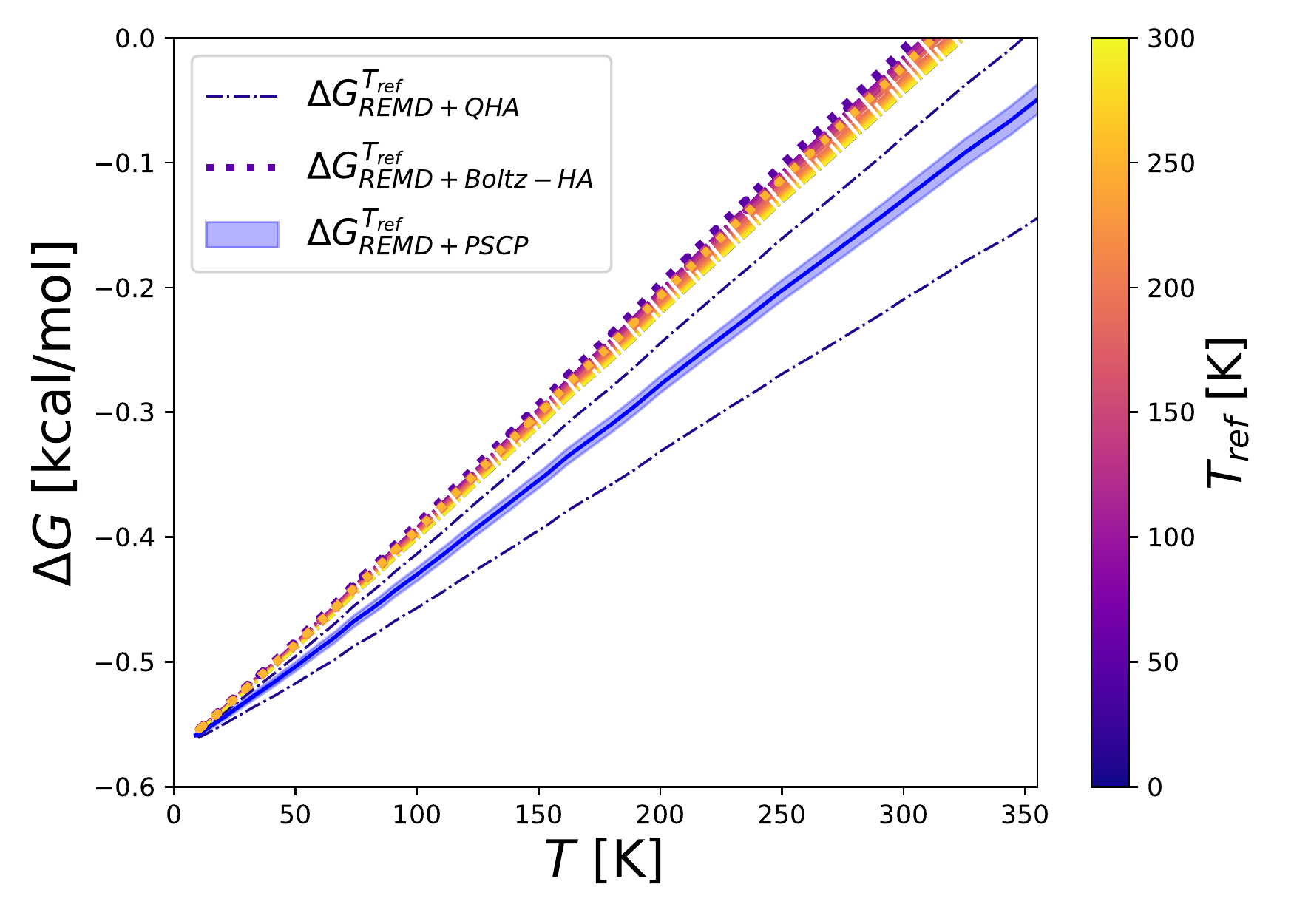}
    \includegraphics[width=8cm]{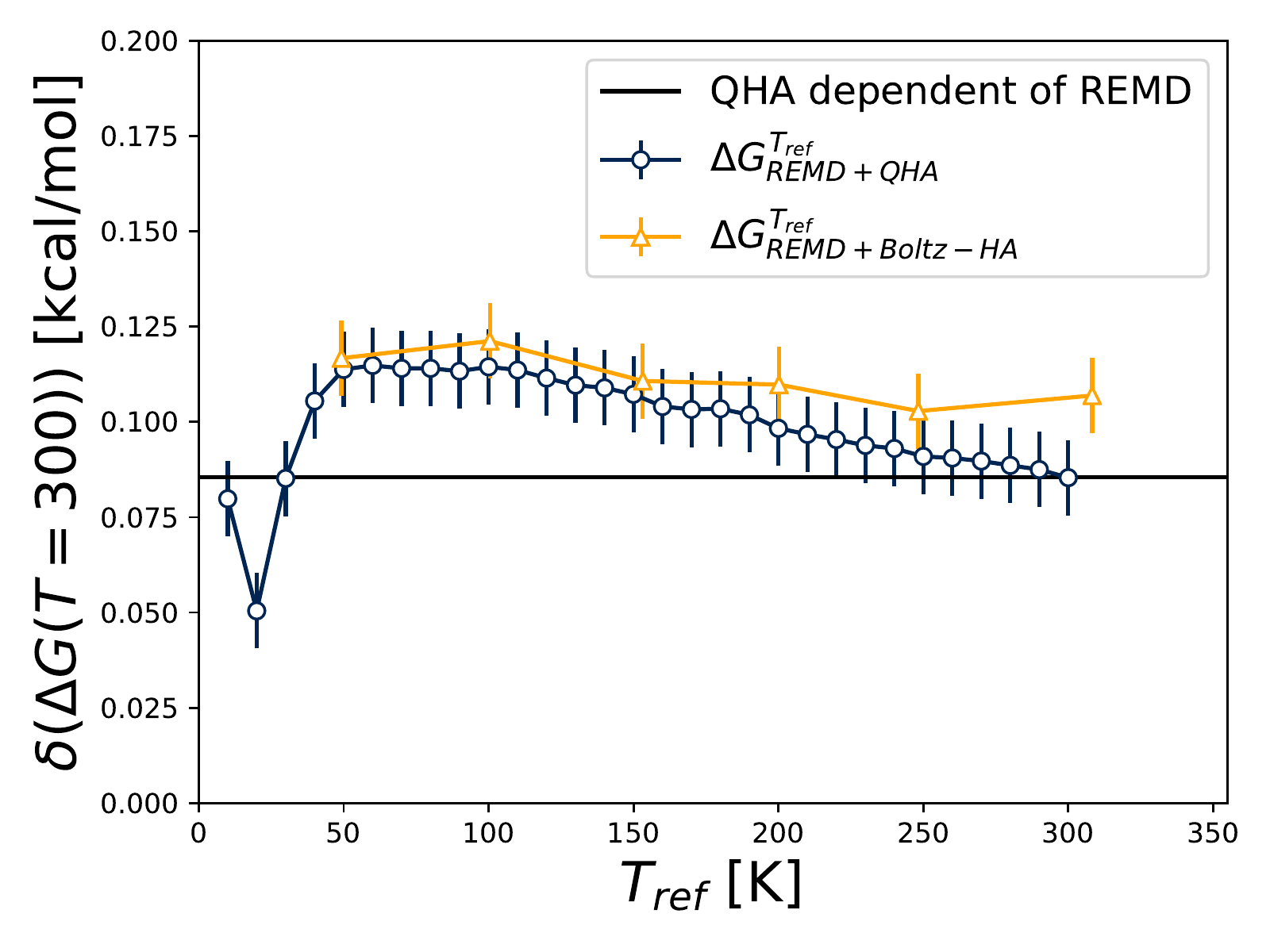}
  \end{center}
  \caption{Left) The polymorph free energy difference of carbamazepine (form III relative to form I) using MD, QHA, and replacing $\Delta G (T_{ref})$ from PSCP in equation ~\ref{eq:MD_dG} with the QHA values for $T_{ref}$ every 10 K is shown. Right) The error in the $\Delta G(300K)$ is reported for replacing $\Delta G (T_{ref})$ from QHA with PSCP and using Boltzmann weighted HA. Each line of the approximation in the left figure is color coded based off the reference temperature used.
  \label{figure:dG_error_4}}
\end{figure*}

\begin{figure*}
  \begin{center}
    \includegraphics[width=8cm]{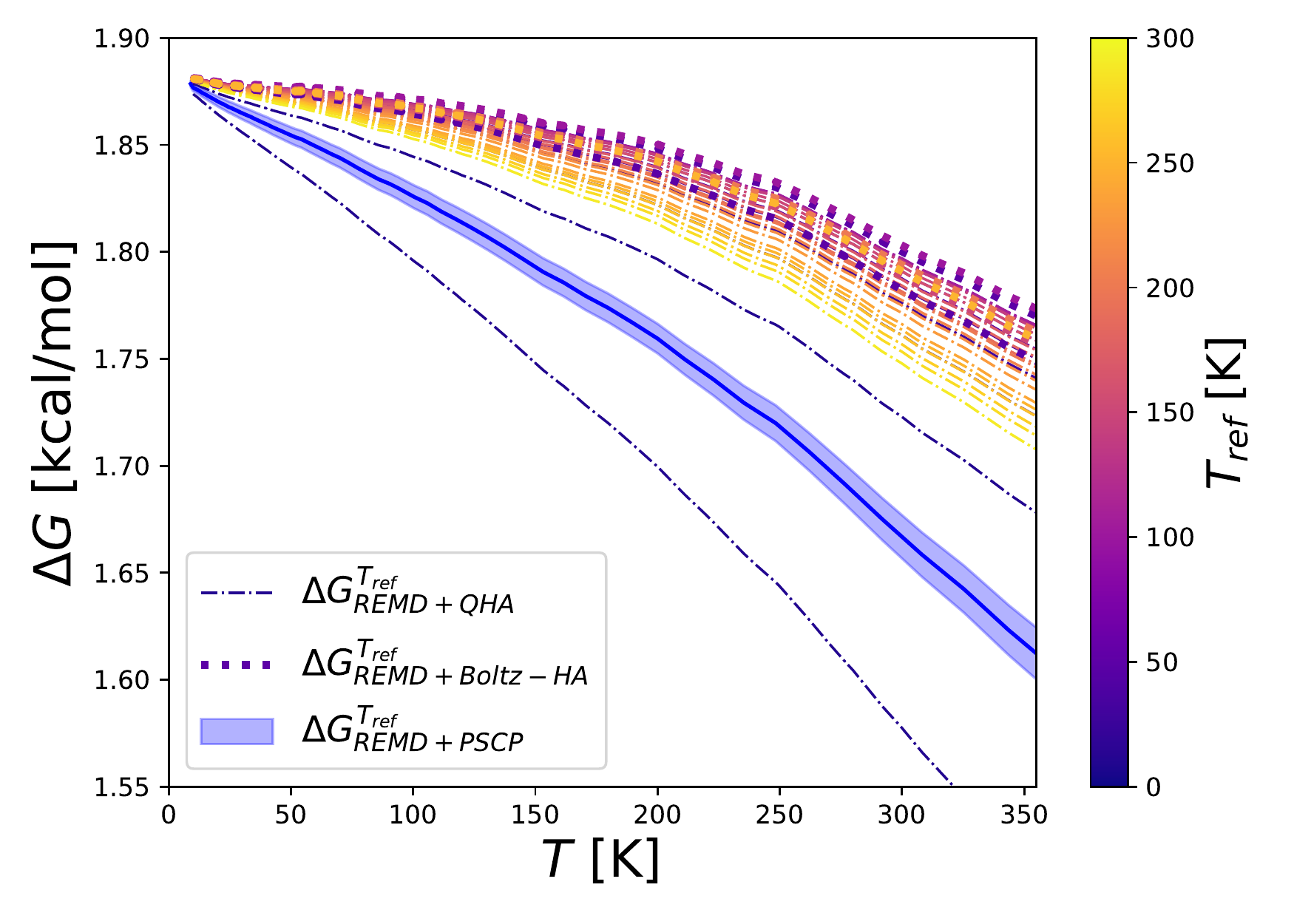}
    \includegraphics[width=8cm]{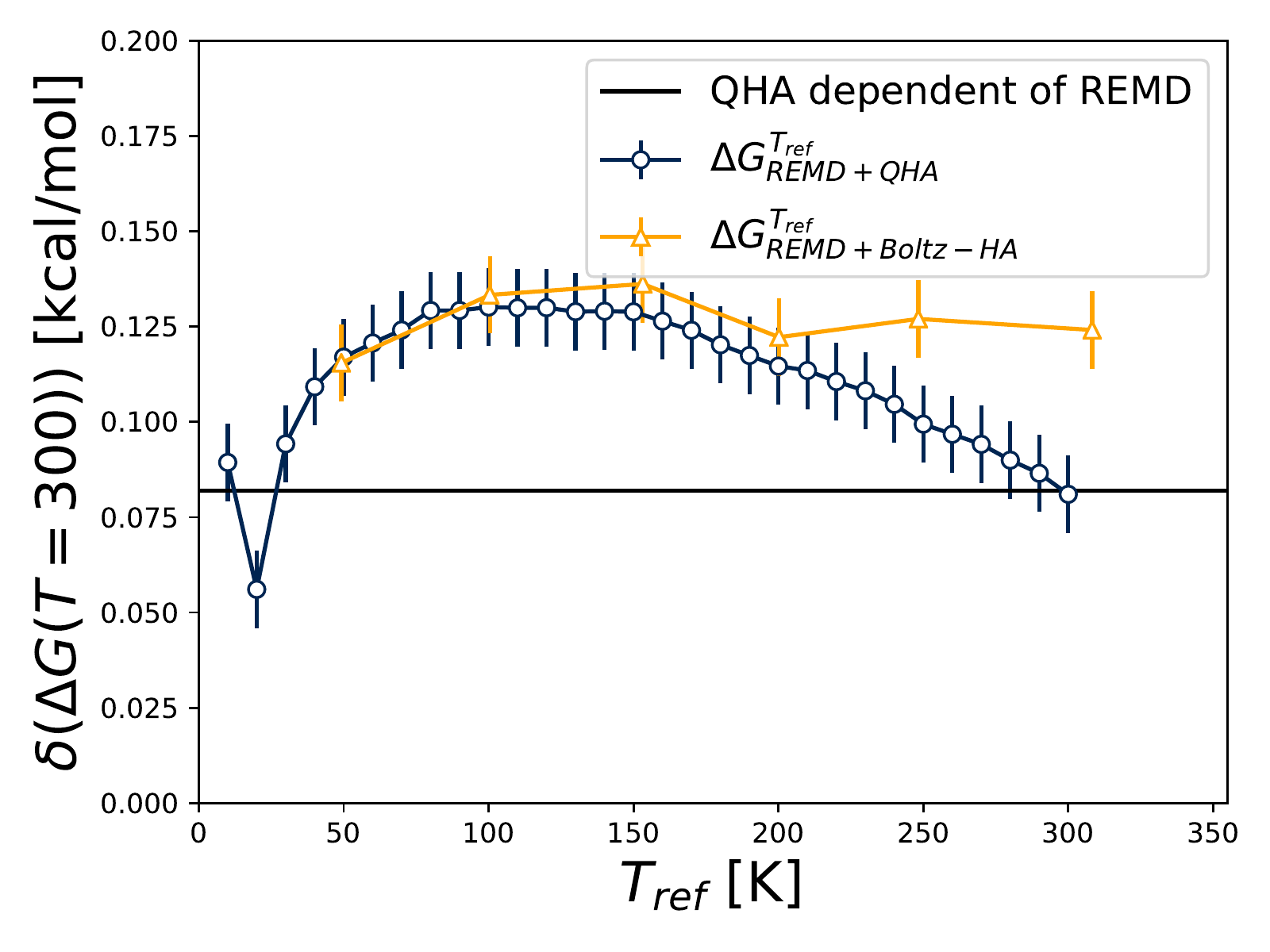}
  \end{center}
  \caption{Left) The polymorph free energy difference of carbamazepine (form IV relative to form I) using MD, QHA, and replacing $\Delta G (T_{ref})$ from PSCP in equation ~\ref{eq:MD_dG} with the QHA values for $T_{ref}$ every 10 K is shown. Right) The error in the $\Delta G(300K)$ is reported for replacing $\Delta G (T_{ref})$ from QHA with PSCP and using Boltzmann weighted HA. Each line of the approximation in the left figure is color coded based off the reference temperature used.
  \label{figure:dG_error_5}}
\end{figure*}

\begin{figure*}
  \begin{center}
    \includegraphics[width=8cm]{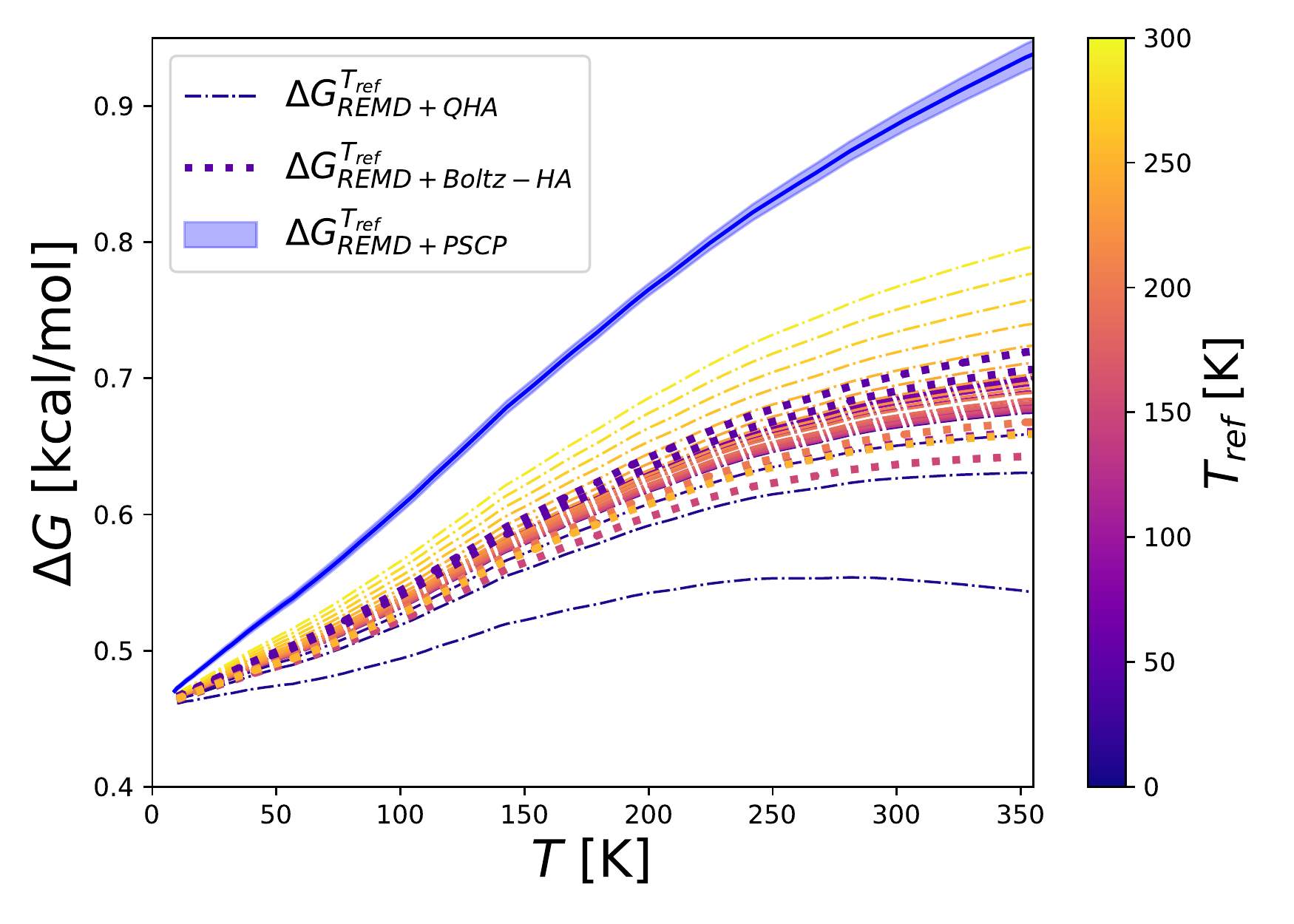}
    \includegraphics[width=8cm]{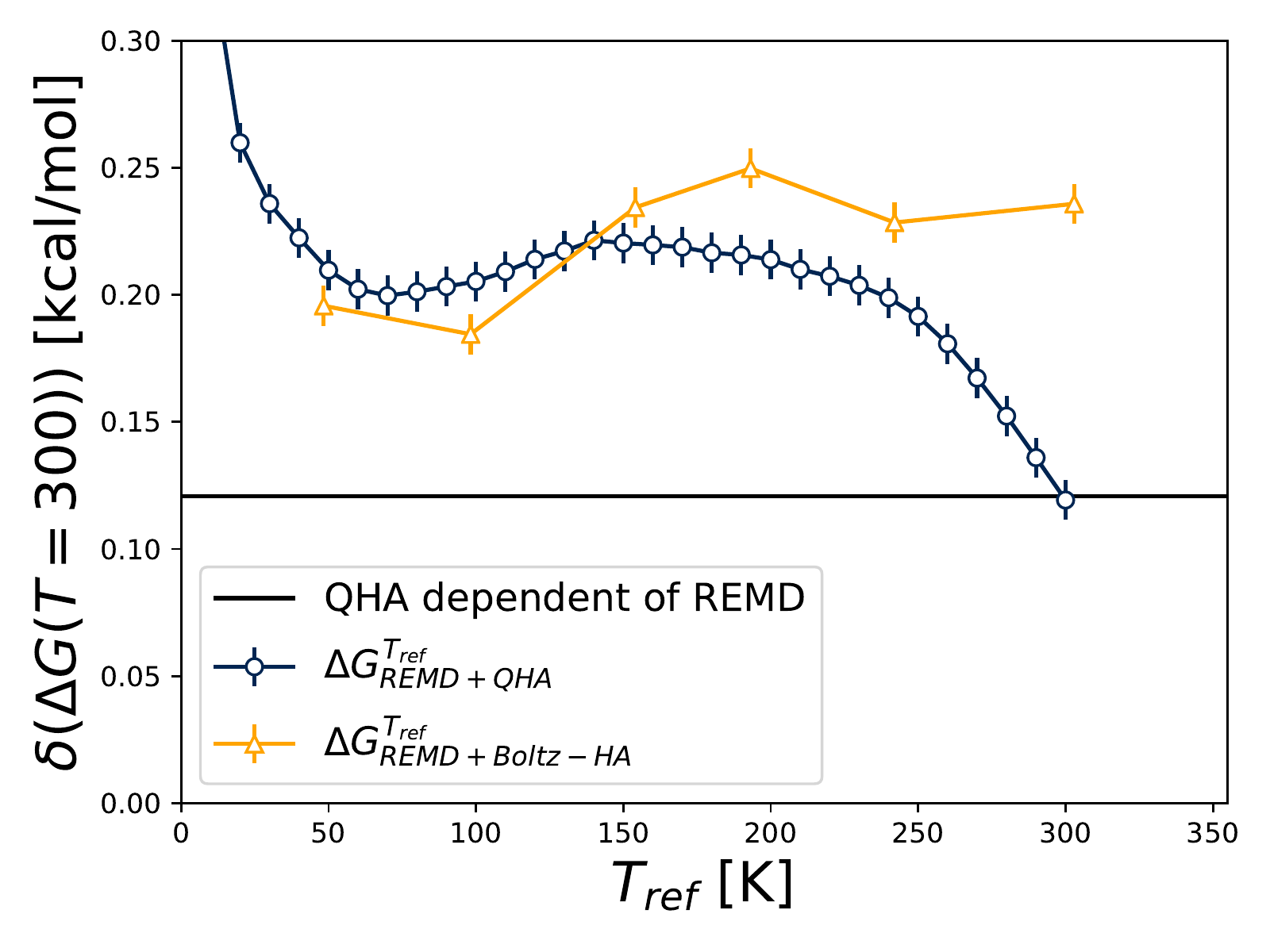}
  \end{center}
  \caption{Left) The polymorph free energy difference of paracetamol (form II relative to form I) using MD, QHA, and replacing $\Delta G (T_{ref})$ from PSCP in equation ~\ref{eq:MD_dG} with the QHA values for $T_{ref}$ every 10 K is shown. Right) The error in the $\Delta G(300K)$ is reported for replacing $\Delta G (T_{ref})$ from QHA with PSCP and using Boltzmann weighted HA. Each line of the approximation in the left figure is color coded based off the reference temperature used.
  \label{figure:dG_error_7}}
\end{figure*}

\begin{figure*}
  \begin{center}
    \includegraphics[width=8cm]{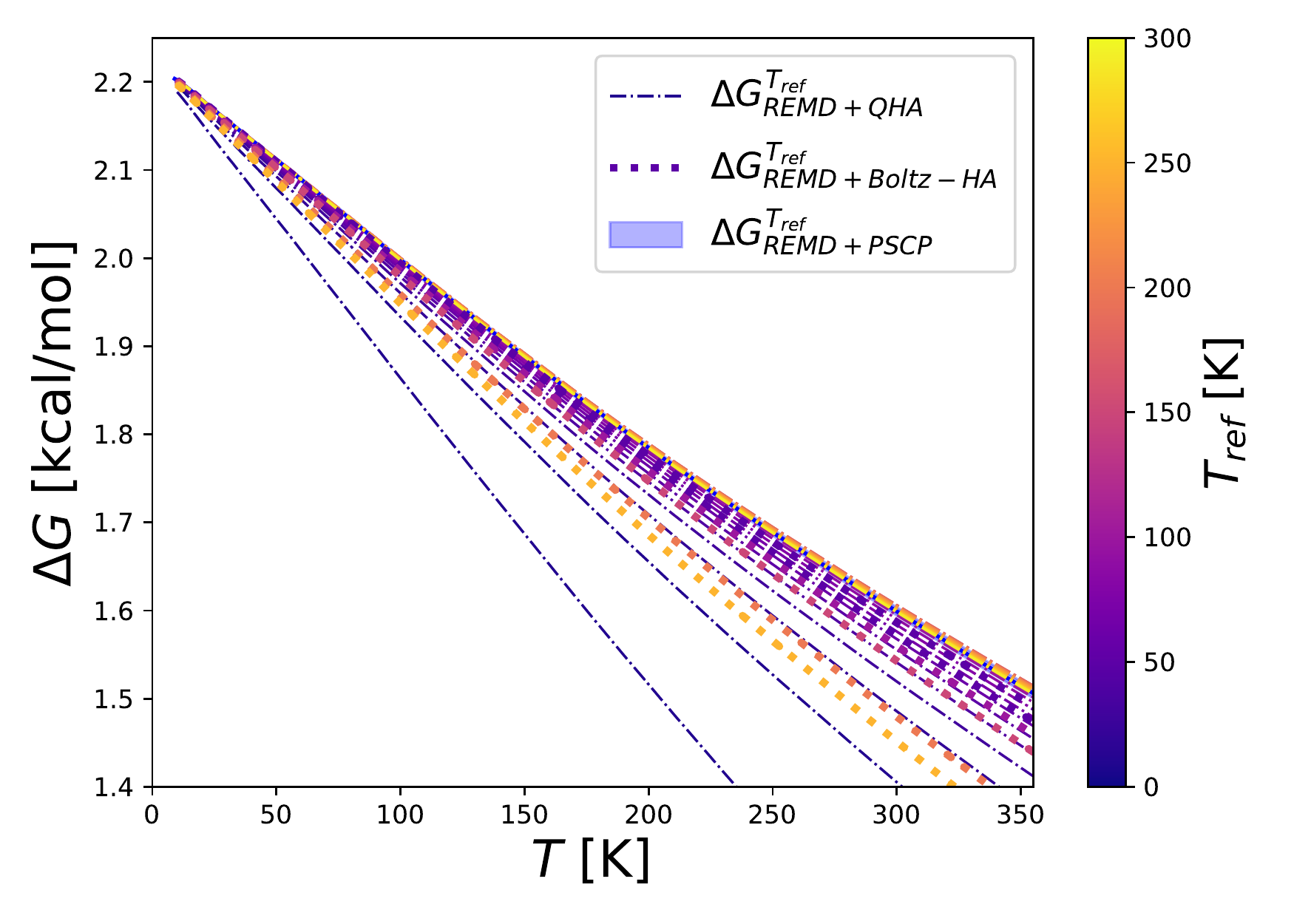}
    \includegraphics[width=8cm]{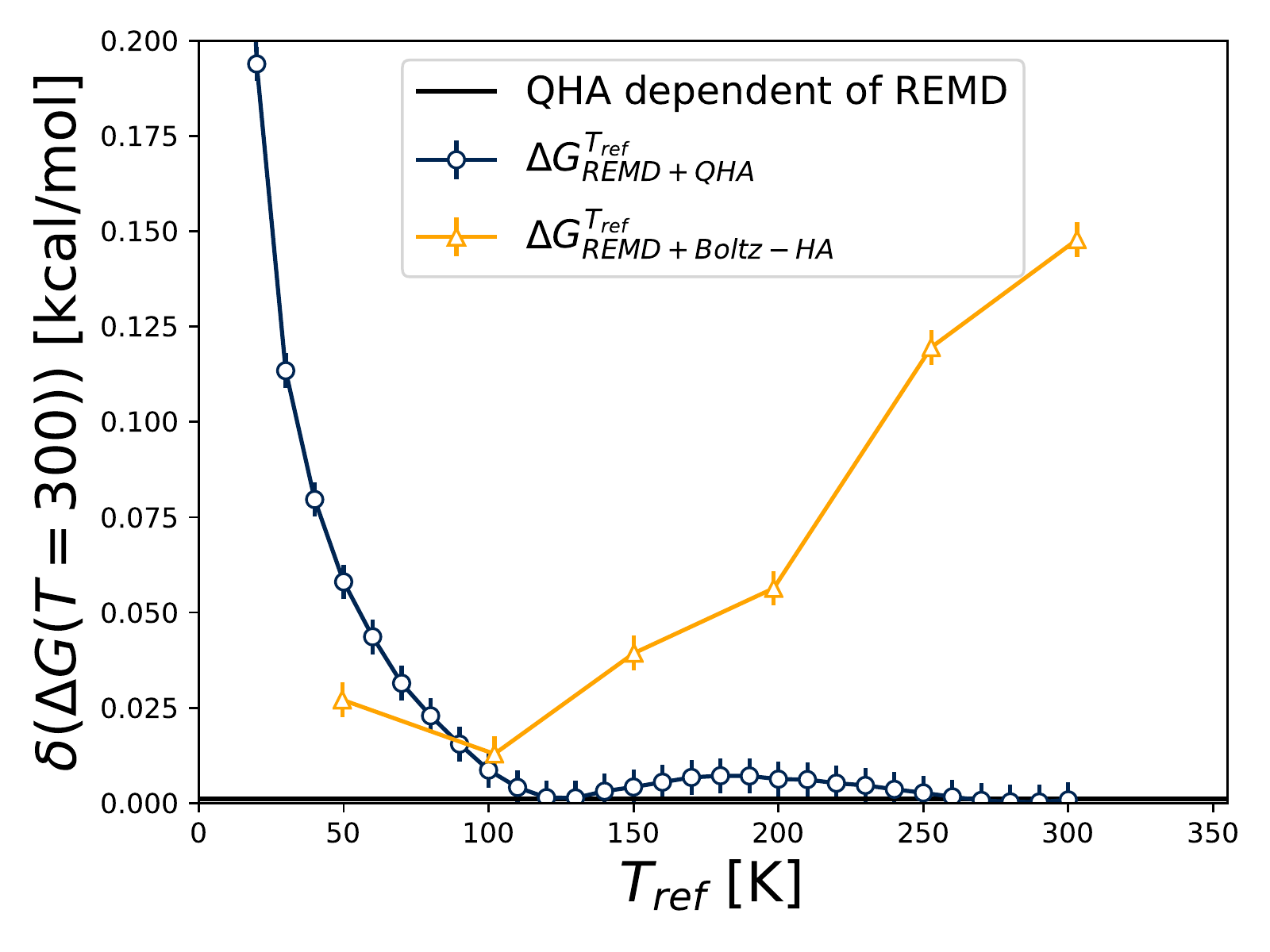}
  \end{center}
  \caption{Left) The polymorph free energy difference of adenine (form $\beta$ relative to form $\alpha$) using MD, QHA, and replacing $\Delta G (T_{ref})$ from PSCP in equation ~\ref{eq:MD_dG} with the QHA values for $T_{ref}$ every 10 K is shown. Right) The error in the $\Delta G(300K)$ is reported for replacing $\Delta G (T_{ref})$ from QHA with PSCP and using Boltzmann weighted HA. Each line of the approximation in the left figure is color coded based off the reference temperature used.
  \label{figure:dG_error_8}}
\end{figure*}

\begin{figure*}
  \begin{center}
    \includegraphics[width=8cm]{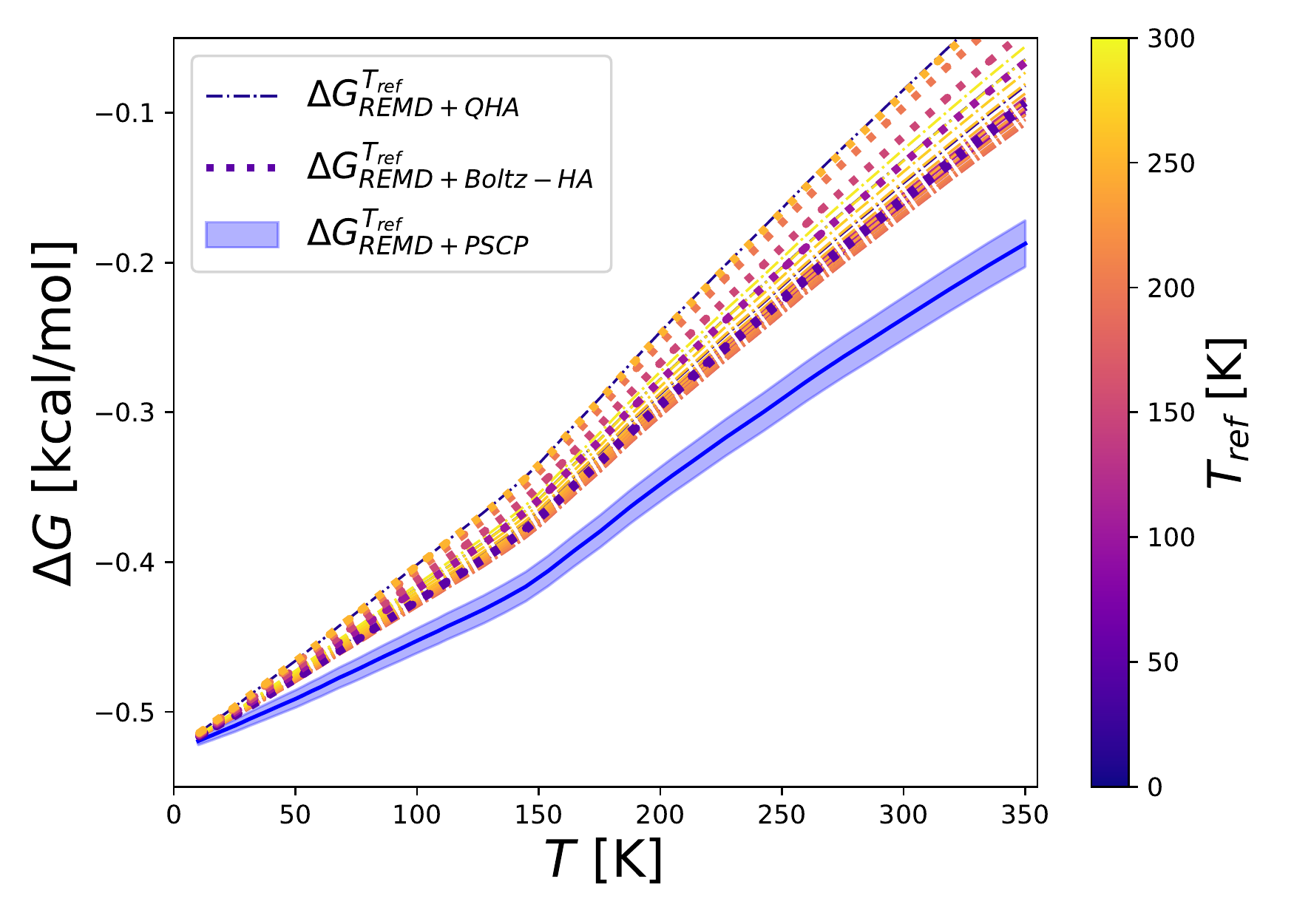}
    \includegraphics[width=8cm]{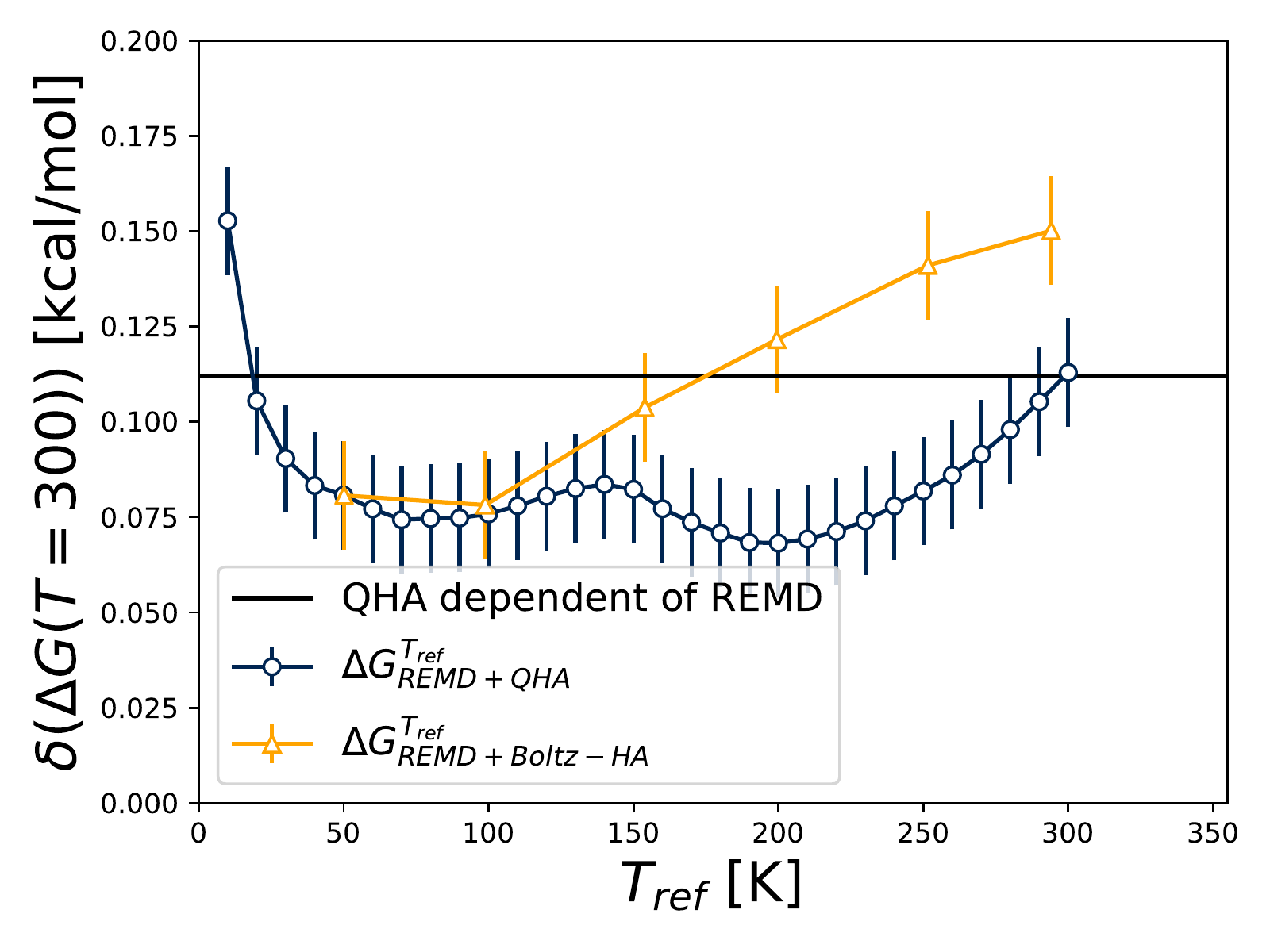}
  \end{center}
  \caption{Left) The polymorph free energy difference of pyrazinamide (form $\beta$ relative to form $\alpha$) using MD, QHA, and replacing $\Delta G (T_{ref})$ from PSCP in equation ~\ref{eq:MD_dG} with the QHA values for $T_{ref}$ every 10 K is shown. Right) The error in the $\Delta G(300K)$ is reported for replacing $\Delta G (T_{ref})$ from QHA with PSCP and using Boltzmann weighted HA. Each line of the approximation in the left figure is color coded based off the reference temperature used.
  \label{figure:dG_error_10}}
\end{figure*}

\begin{figure*}
  \begin{center}
    \includegraphics[width=8cm]{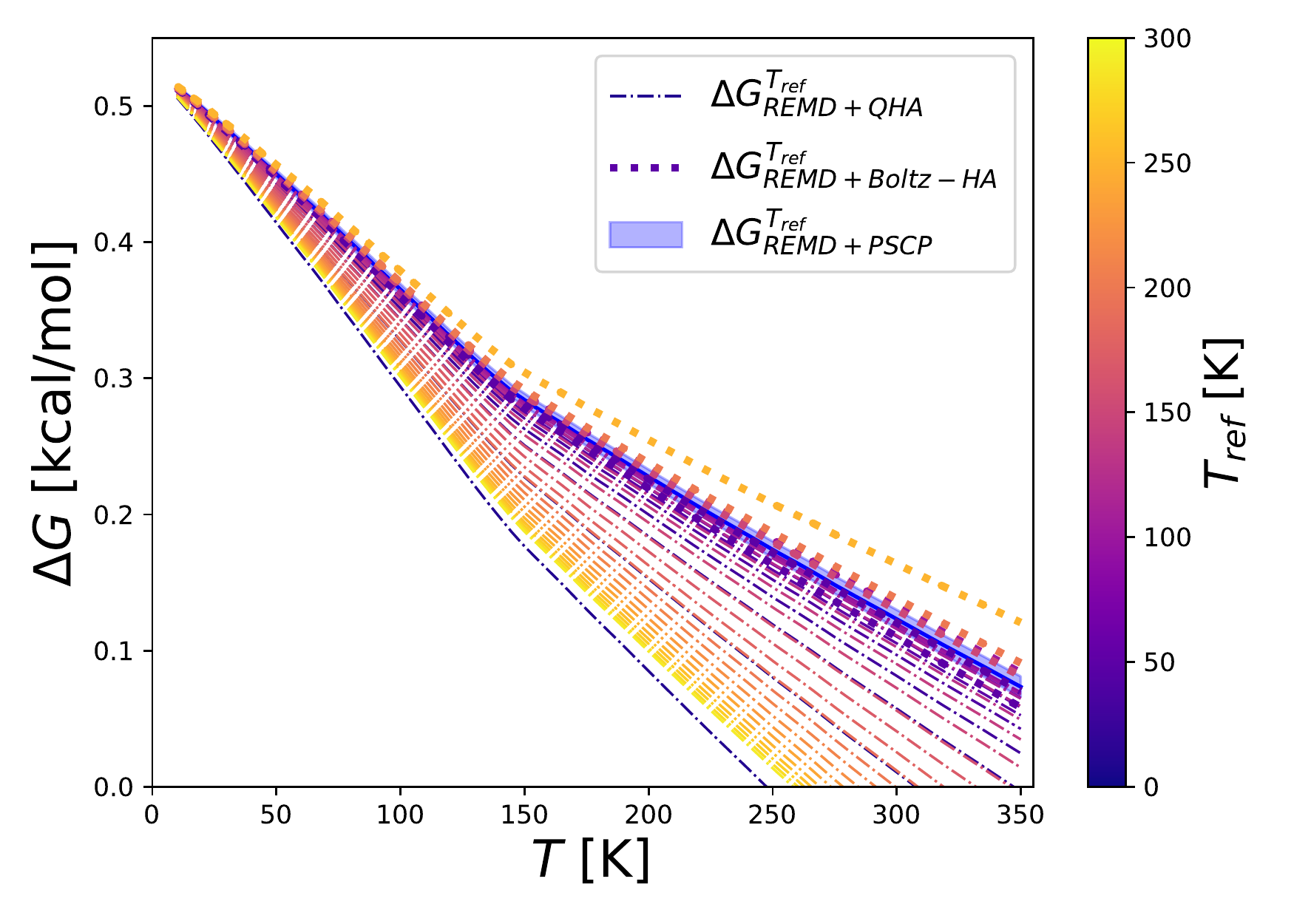}
    \includegraphics[width=8cm]{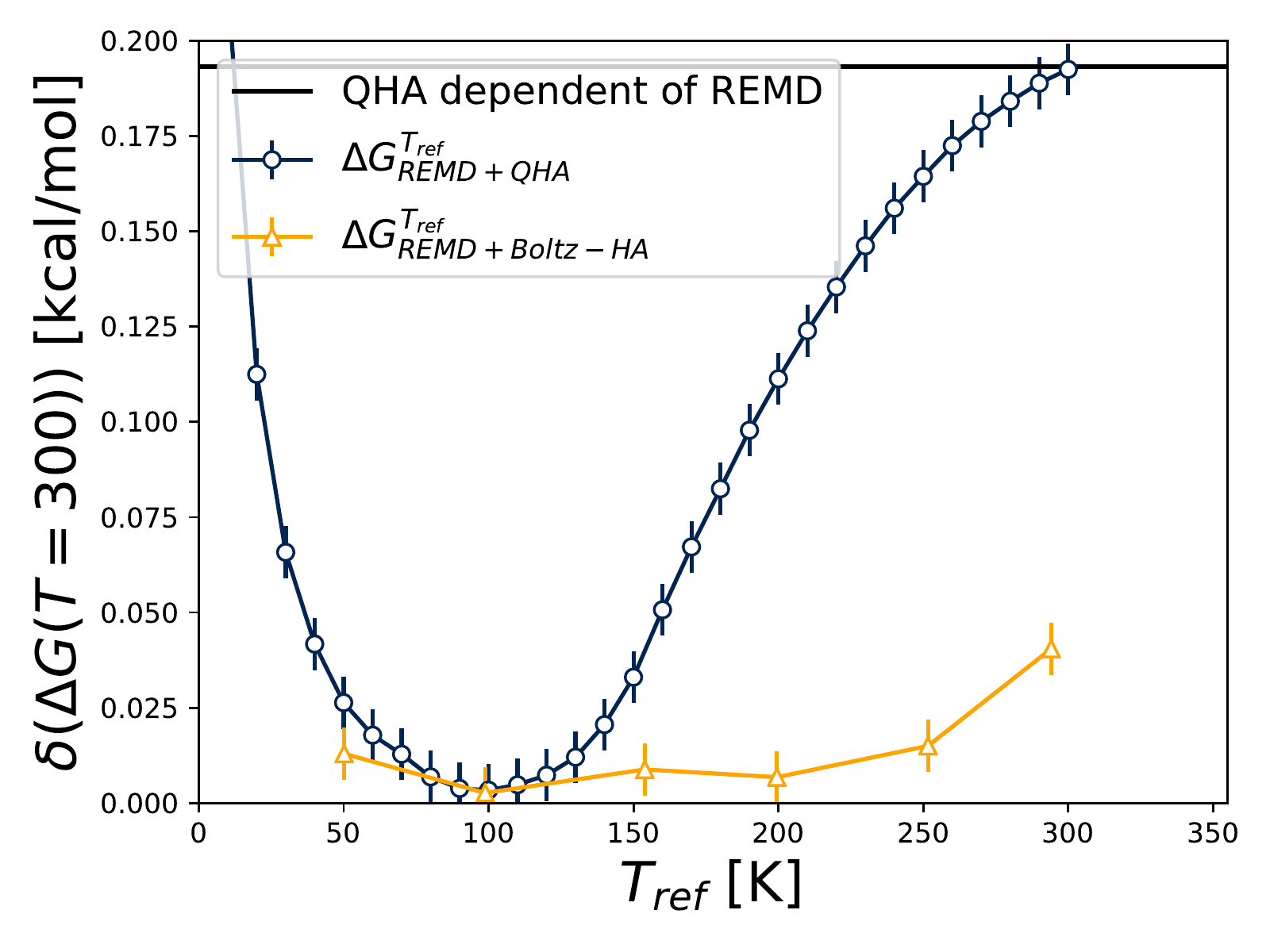}
  \end{center}
  \caption{Left) The polymorph free energy difference of pyrazinamide (form $\gamma$ relative to form $\alpha$) using MD, QHA, and replacing $\Delta G (T_{ref})$ from PSCP in equation ~\ref{eq:MD_dG} with the QHA values for $T_{ref}$ every 10 K is shown. Right) The error in the $\Delta G(300K)$ is reported for replacing $\Delta G (T_{ref})$ from QHA with PSCP and using Boltzmann weighted HA. Each line of the approximation in the left figure is color coded based off the reference temperature used.
  \label{figure:dG_error_11}}
\end{figure*}

\begin{figure*}
  \begin{center}
    \includegraphics[width=8cm]{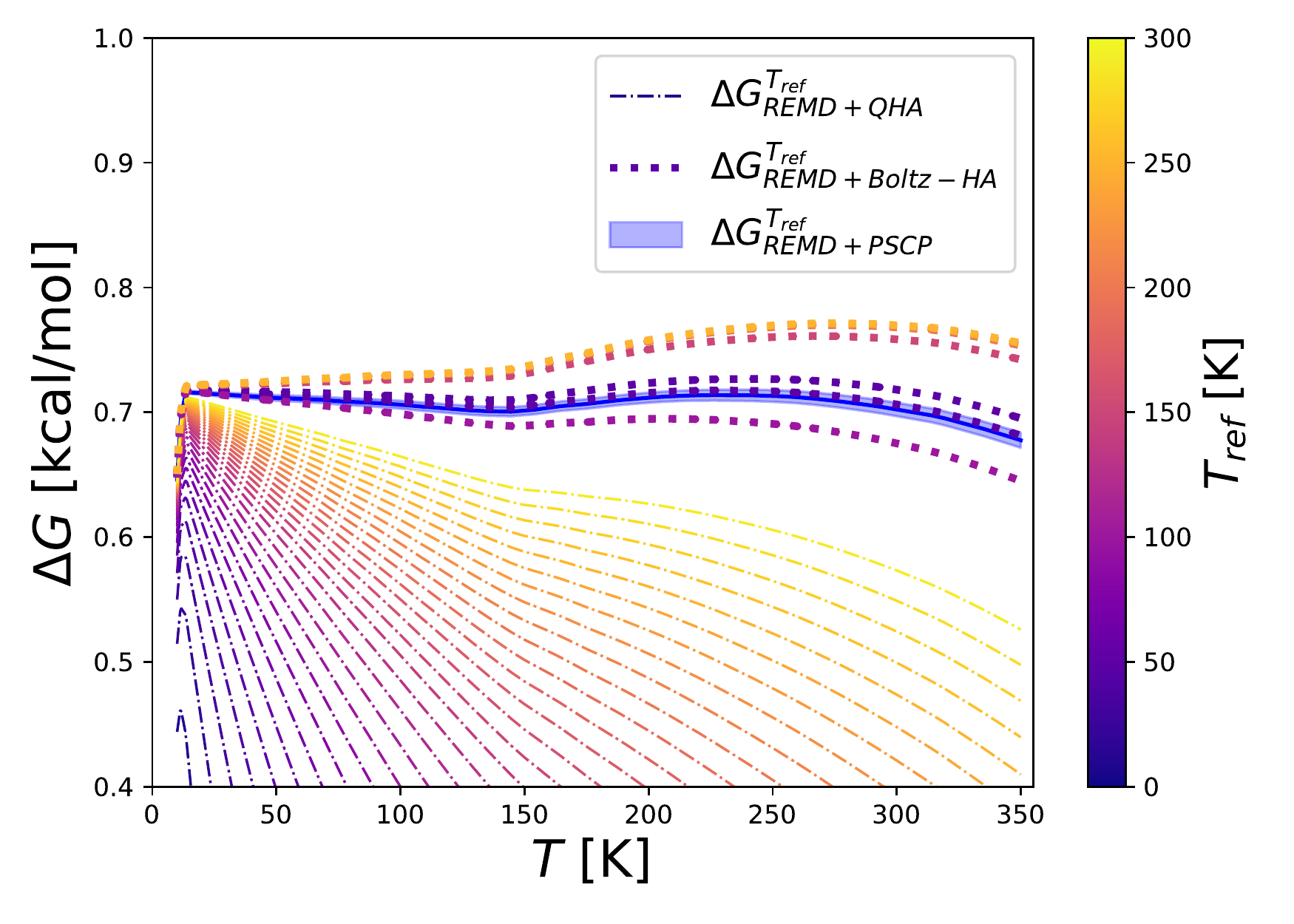}
    \includegraphics[width=8cm]{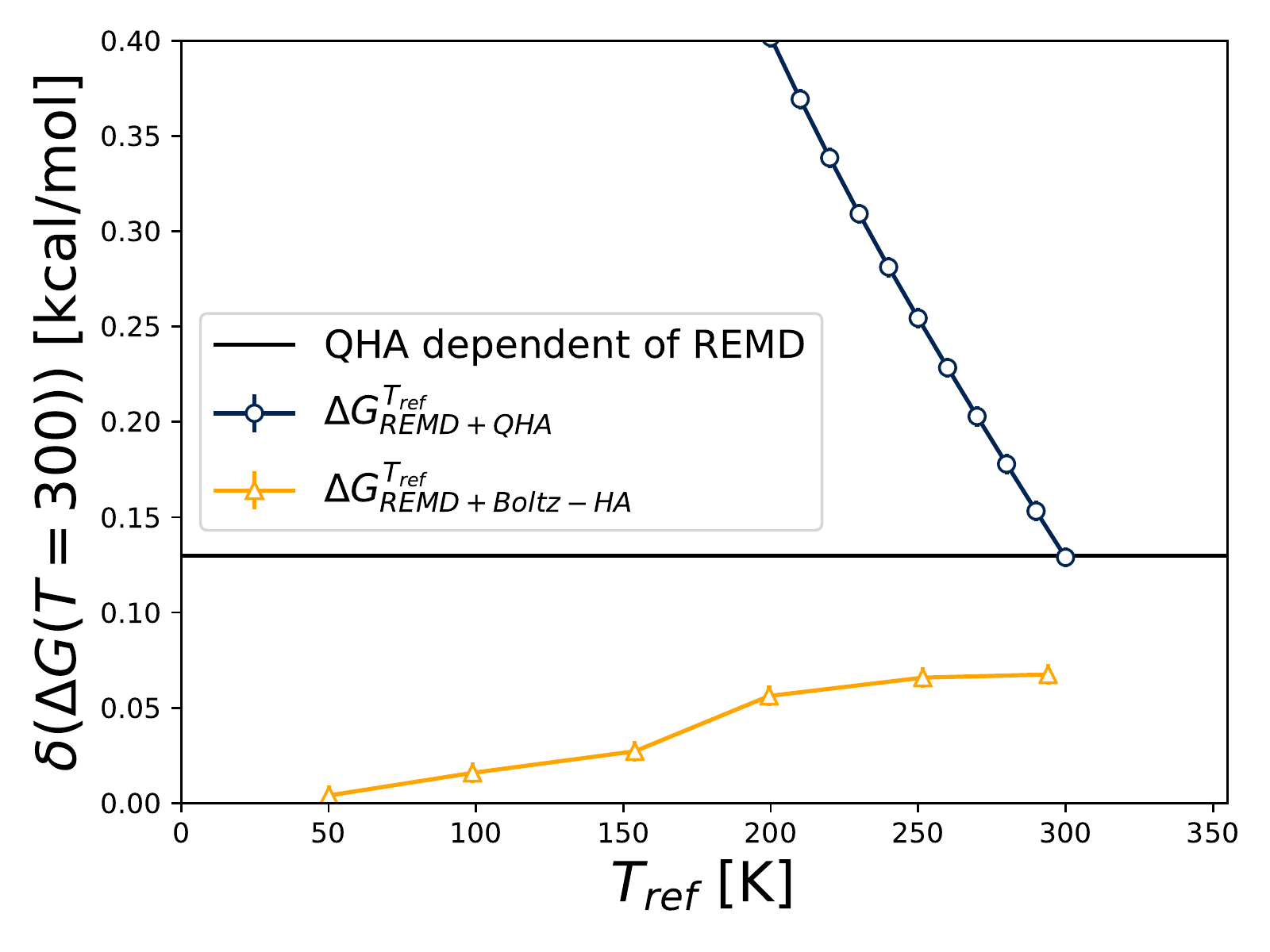}
  \end{center}
  \caption{Left) The polymorph free energy difference of pyrazinamide (form $\delta$ relative to form $\alpha$) using MD, QHA, and replacing $\Delta G (T_{ref})$ from PSCP in equation ~\ref{eq:MD_dG} with the QHA values for $T_{ref}$ every 10 K is shown. Right) The error in the $\Delta G(300K)$ is reported for replacing $\Delta G (T_{ref})$ from QHA with PSCP and using Boltzmann weighted HA. Each line of the approximation in the left figure is color coded based off the reference temperature used.
  \label{figure:dG_error_12}}
\end{figure*}

\begin{figure*}
  \begin{center}
    \includegraphics[width=8cm]{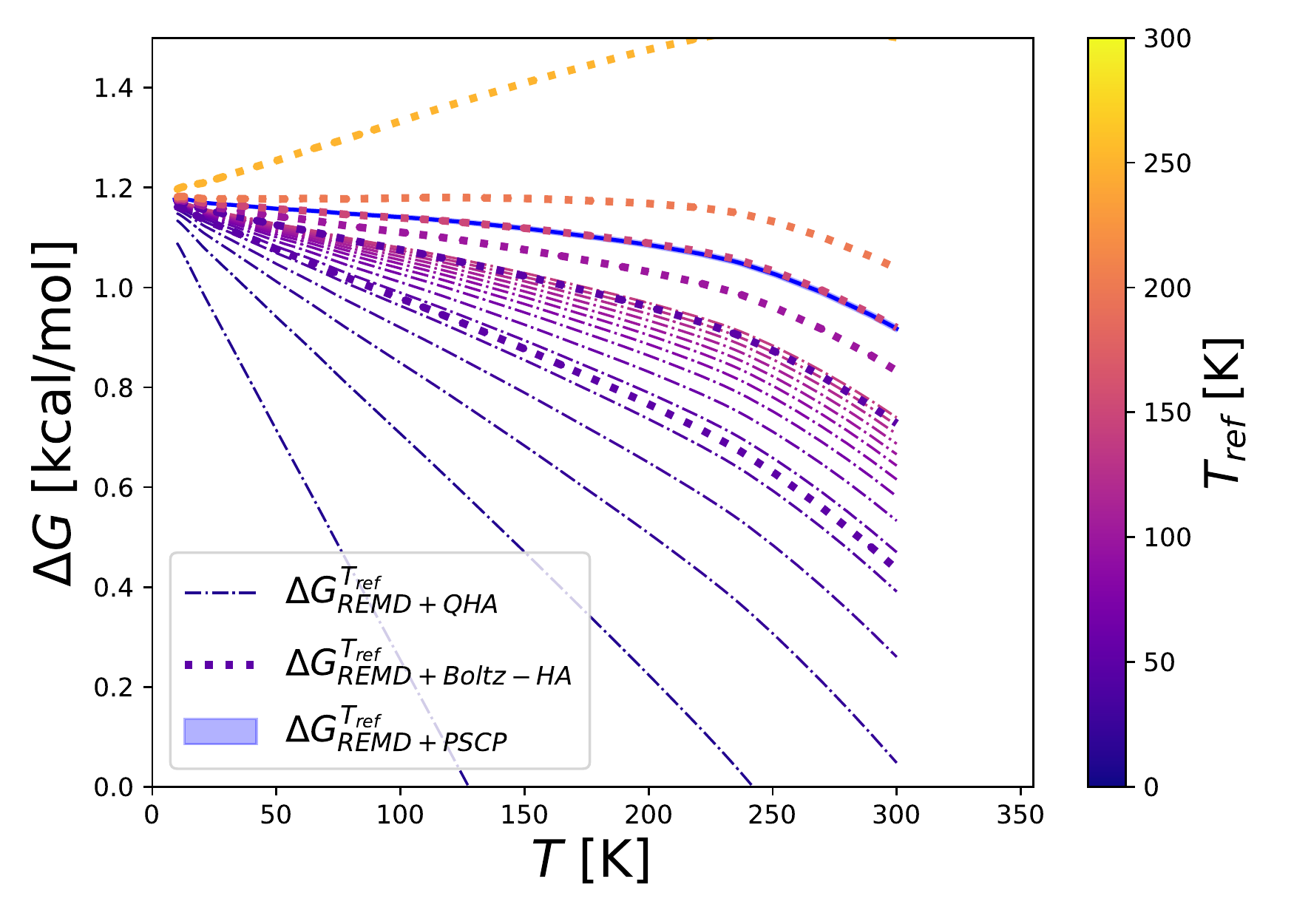}
    \includegraphics[width=8cm]{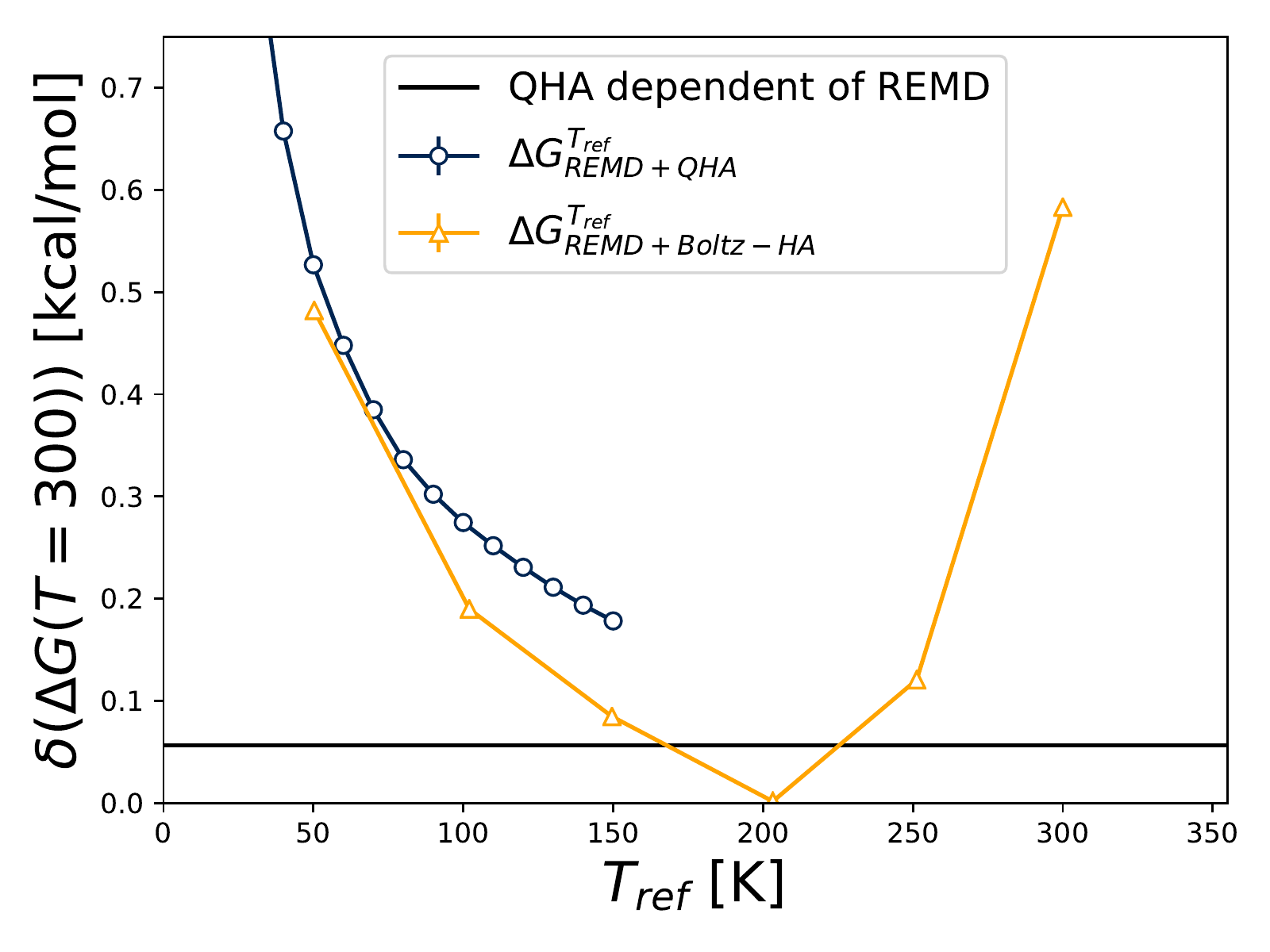}
  \end{center}
  \caption{Left) The polymorph free energy difference of succinonitrile (form III relative to form I) using MD, QHA, and replacing $\Delta G (T_{ref})$ from PSCP in equation ~\ref{eq:MD_dG} with the QHA values for $T_{ref}$ every 10 K is shown. Right) The error in the $\Delta G(300K)$ is reported for replacing $\Delta G (T_{ref})$ from QHA with PSCP and using Boltzmann weighted HA. Each line of the approximation in the left figure is color coded based off the reference temperature used.
  \label{figure:dG_error_9}}
\end{figure*}

\begin{figure*}
  \begin{center}
    \includegraphics[width=8cm]{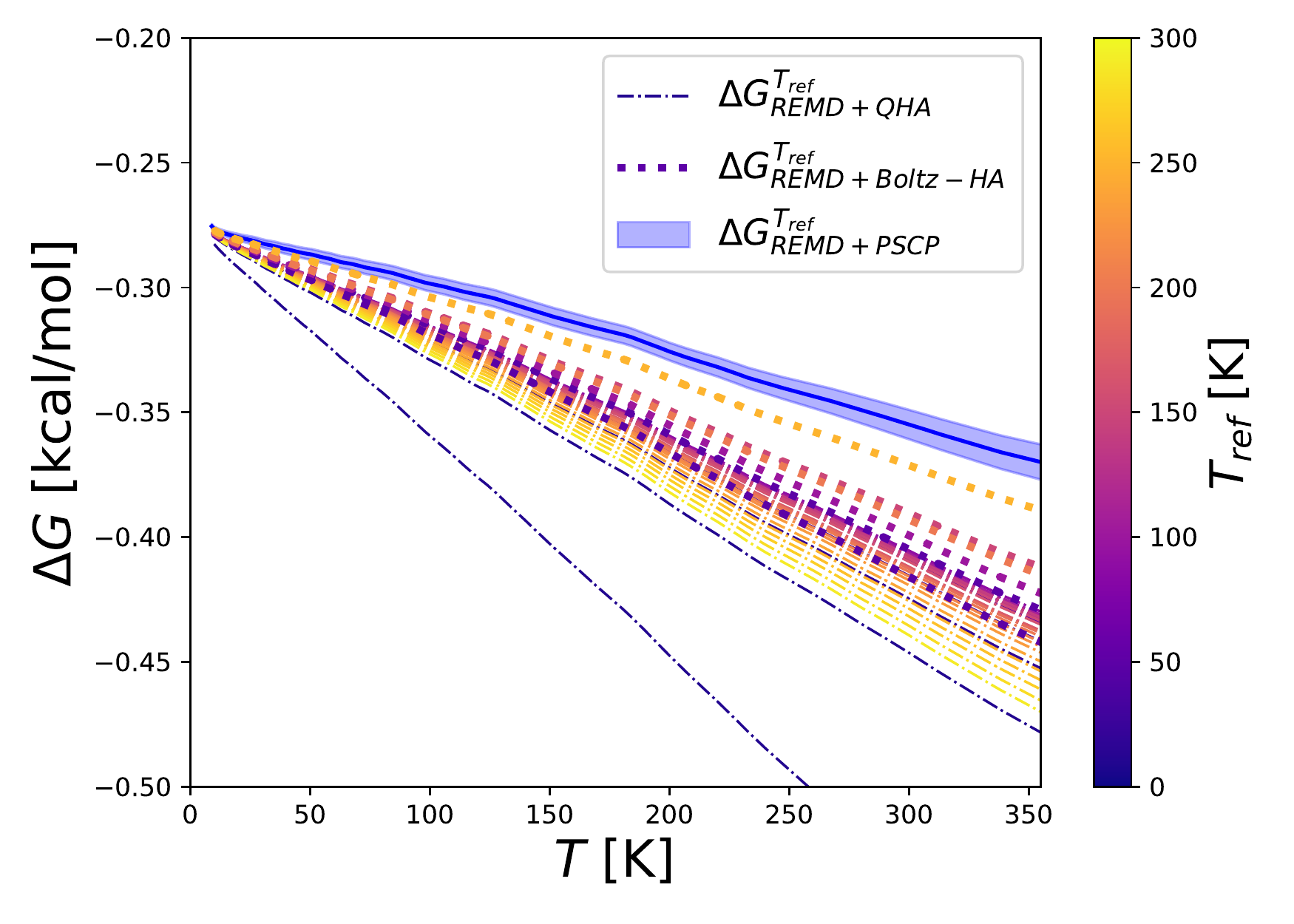}
    \includegraphics[width=8cm]{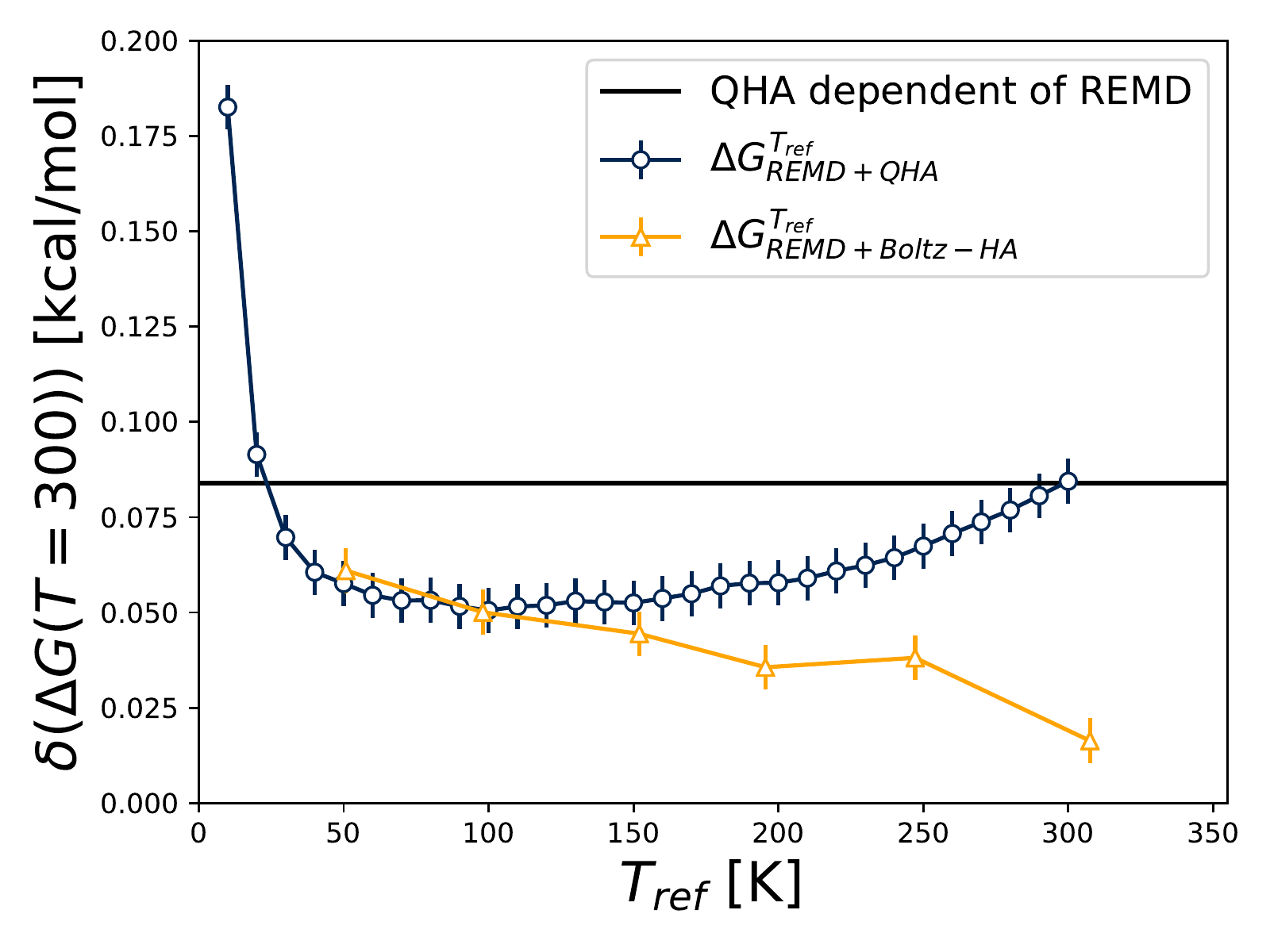}
  \end{center}
  \caption{Left) The polymorph free energy difference of resorcinol (form $\beta$ relative to form $\alpha$) using MD, QHA, and replacing $\Delta G (T_{ref})$ from PSCP in equation ~\ref{eq:MD_dG} with the QHA values for $T_{ref}$ every 10 K is shown. Right) The error in the $\Delta G(300K)$ is reported for replacing $\Delta G (T_{ref})$ from QHA with PSCP and using Boltzmann weighted HA. Each line of the approximation in the left figure is color coded based off the reference temperature used.
  \label{figure:dG_error_6}}
\end{figure*}

\begin{figure*}
  \begin{center}
    \includegraphics[width=8cm]{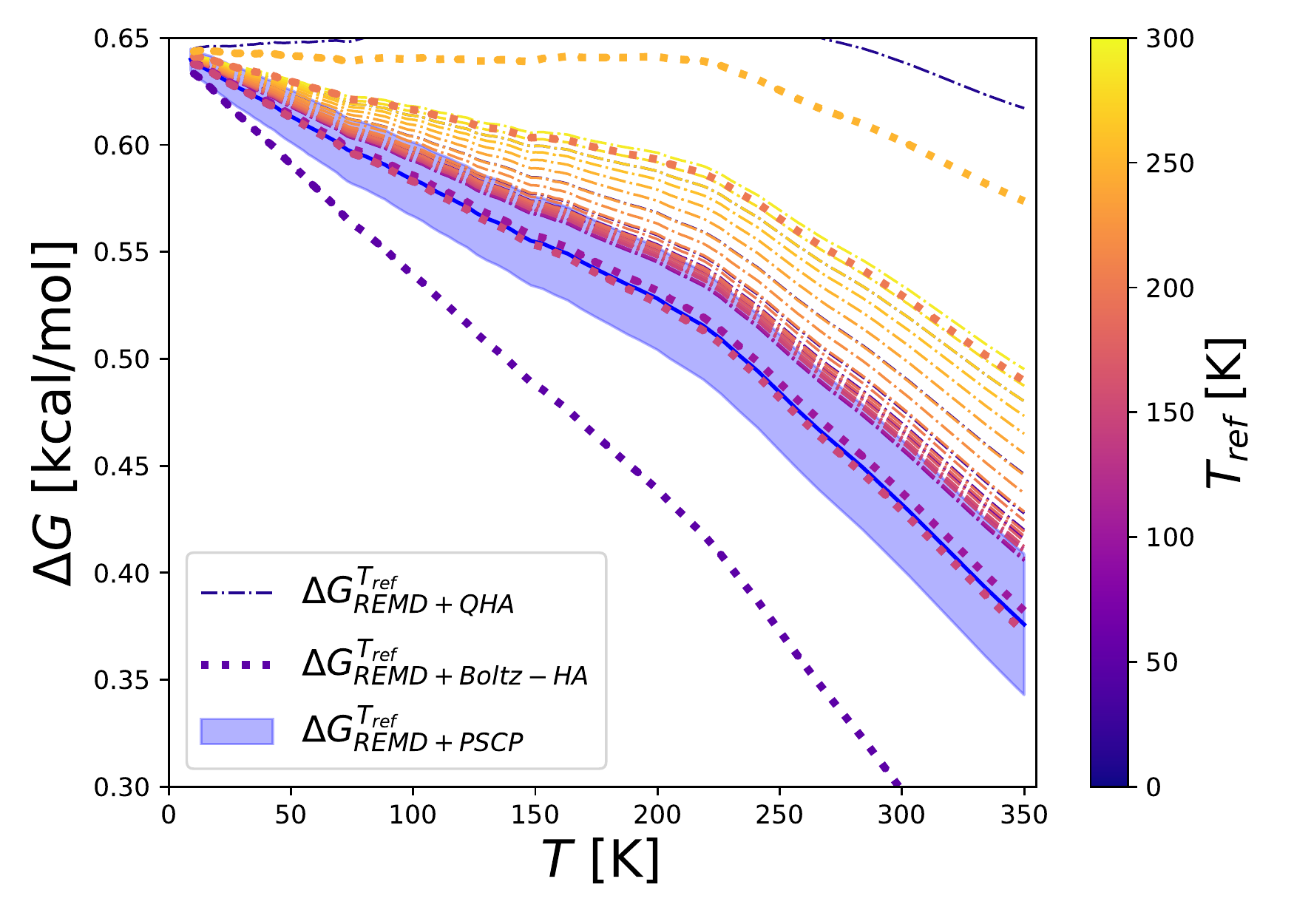}
    \includegraphics[width=8cm]{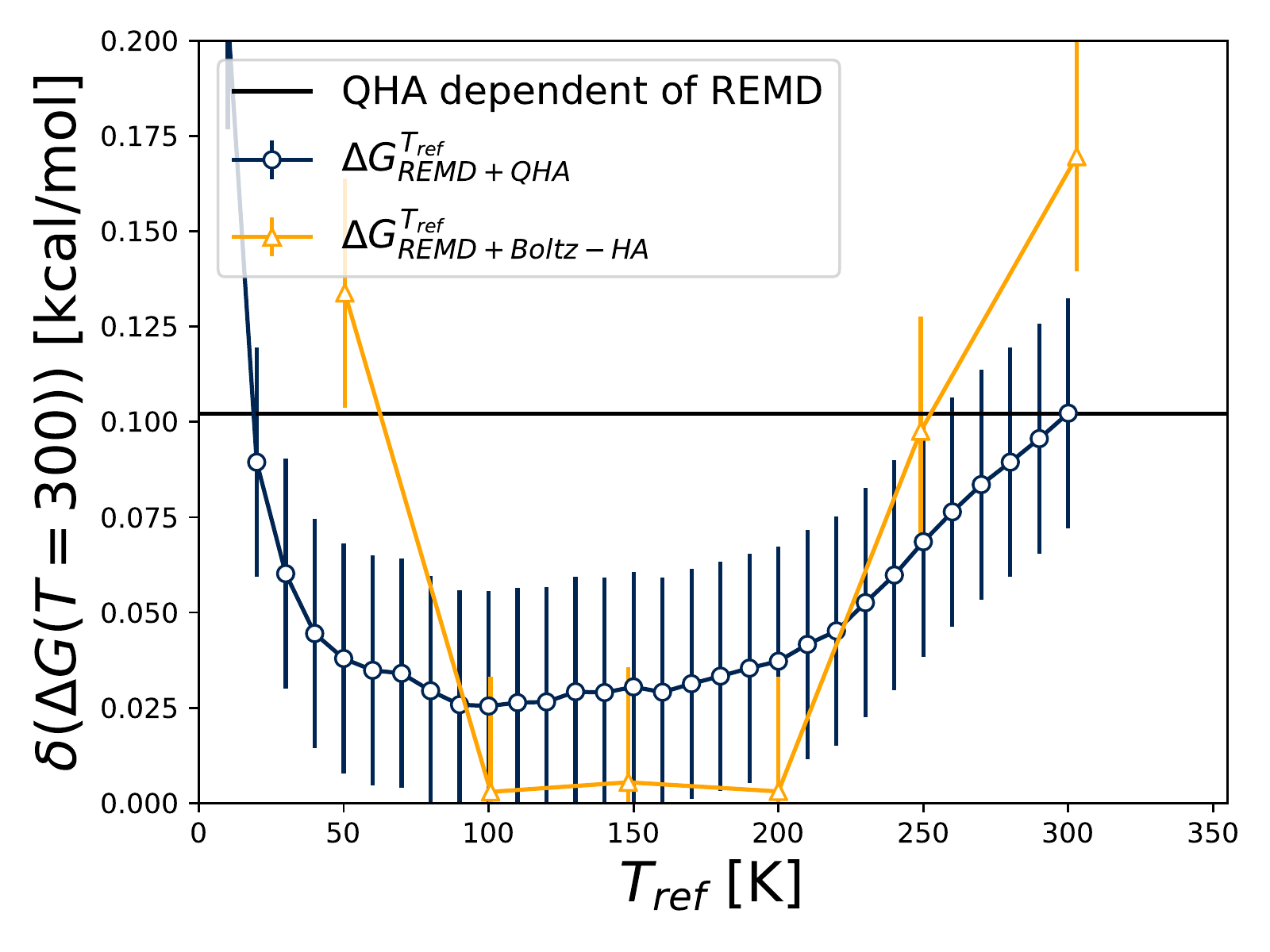}
  \end{center}
  \caption{Left) The polymorph free energy difference of aripiprazole (form X relative to form I) using MD, QHA, and replacing $\Delta G (T_{ref})$ from PSCP in equation ~\ref{eq:MD_dG} with the QHA values for $T_{ref}$ every 10 K is shown. Right) The error in the $\Delta G(300K)$ is reported for replacing $\Delta G (T_{ref})$ from QHA with PSCP and using Boltzmann weighted HA. Each line of the approximation in the left figure is color coded based off the reference temperature used.
  \label{figure:dG_error_14}}
\end{figure*}

\begin{figure*}
  \begin{center}
    \includegraphics[width=8cm]{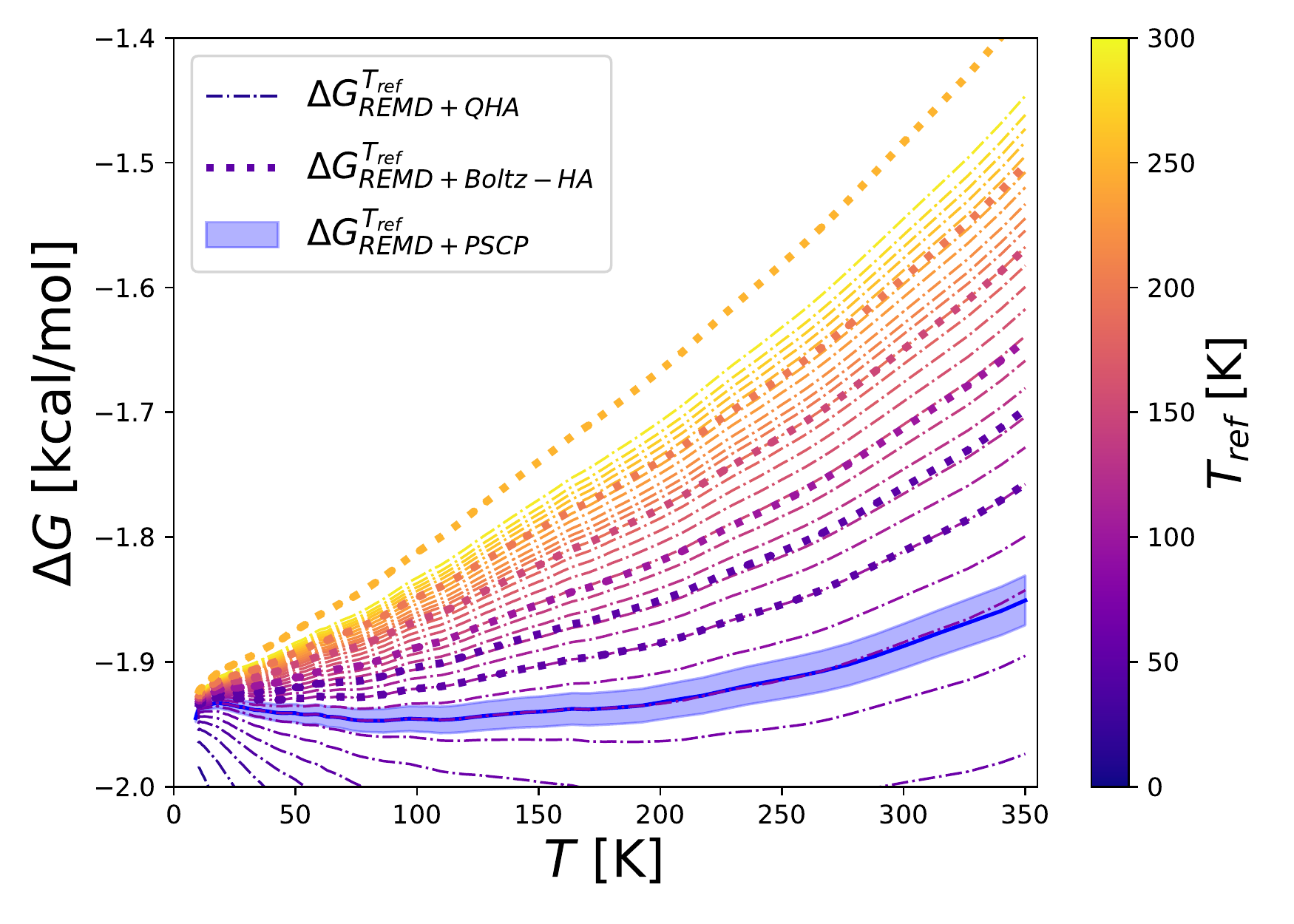}
    \includegraphics[width=8cm]{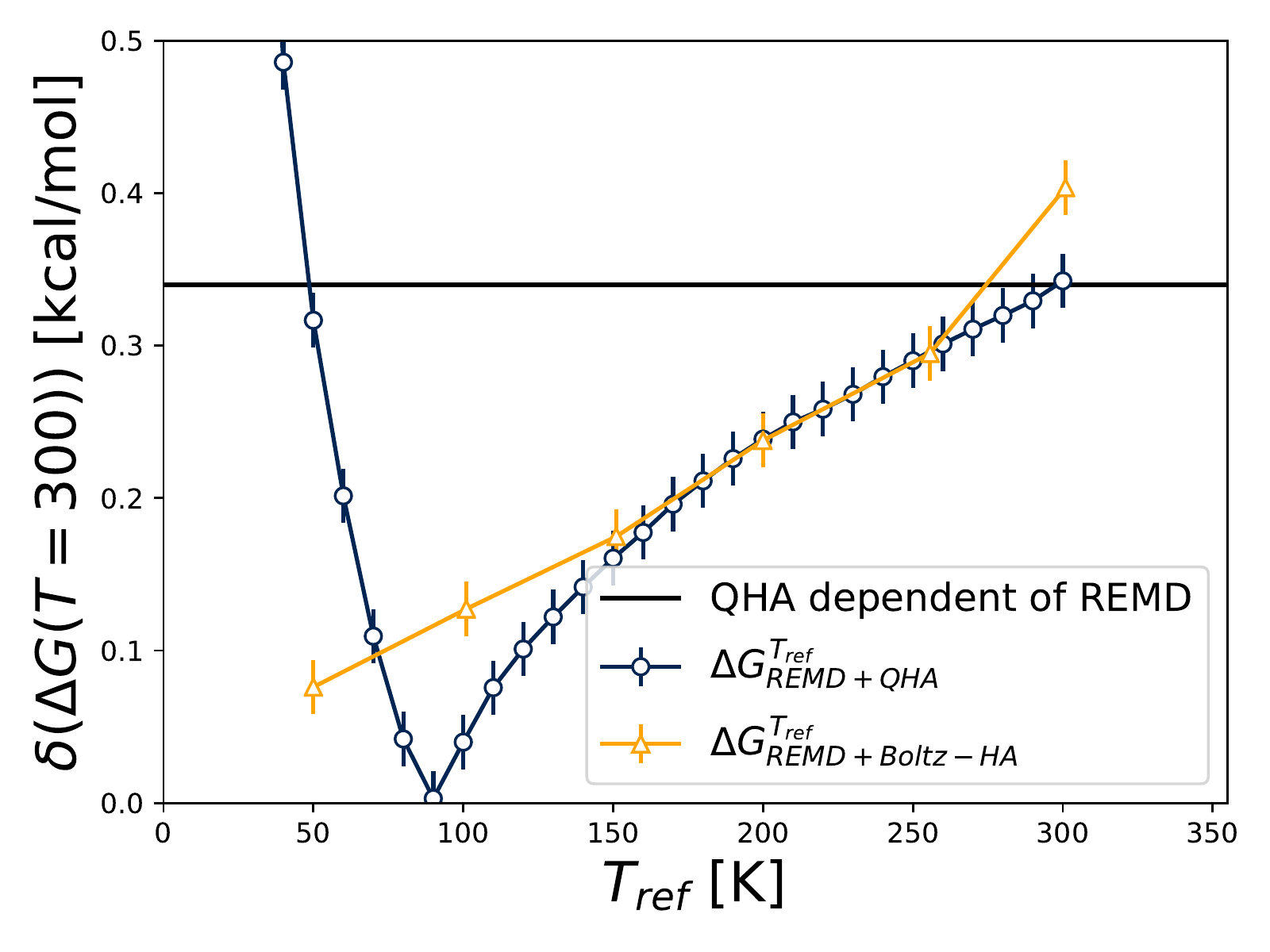}
  \end{center}
  \caption{Left) The polymorph free energy difference of tolbutamide (form II relative to form I) using MD, QHA, and replacing $\Delta G (T_{ref})$ from PSCP in equation ~\ref{eq:MD_dG} with the QHA values for $T_{ref}$ every 10 K is shown. Right) The error in the $\Delta G(300K)$ is reported for replacing $\Delta G (T_{ref})$ from QHA with PSCP and using Boltzmann weighted HA. Each line of the approximation in the left figure is color coded based off the reference temperature used.
  \label{figure:dG_error_15}}
\end{figure*}

\begin{figure*}
  \begin{center}
    \includegraphics[width=8cm]{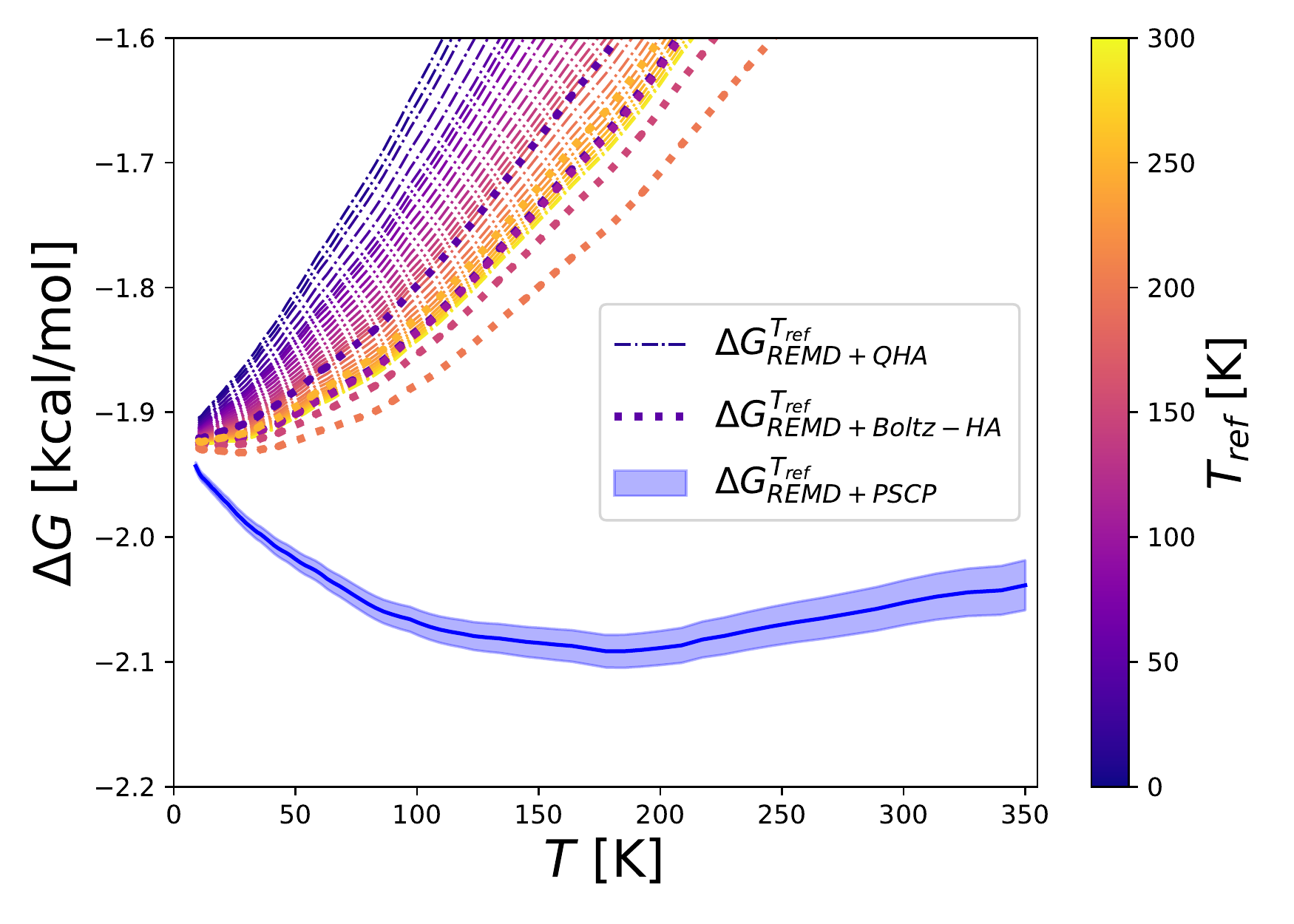}
    \includegraphics[width=8cm]{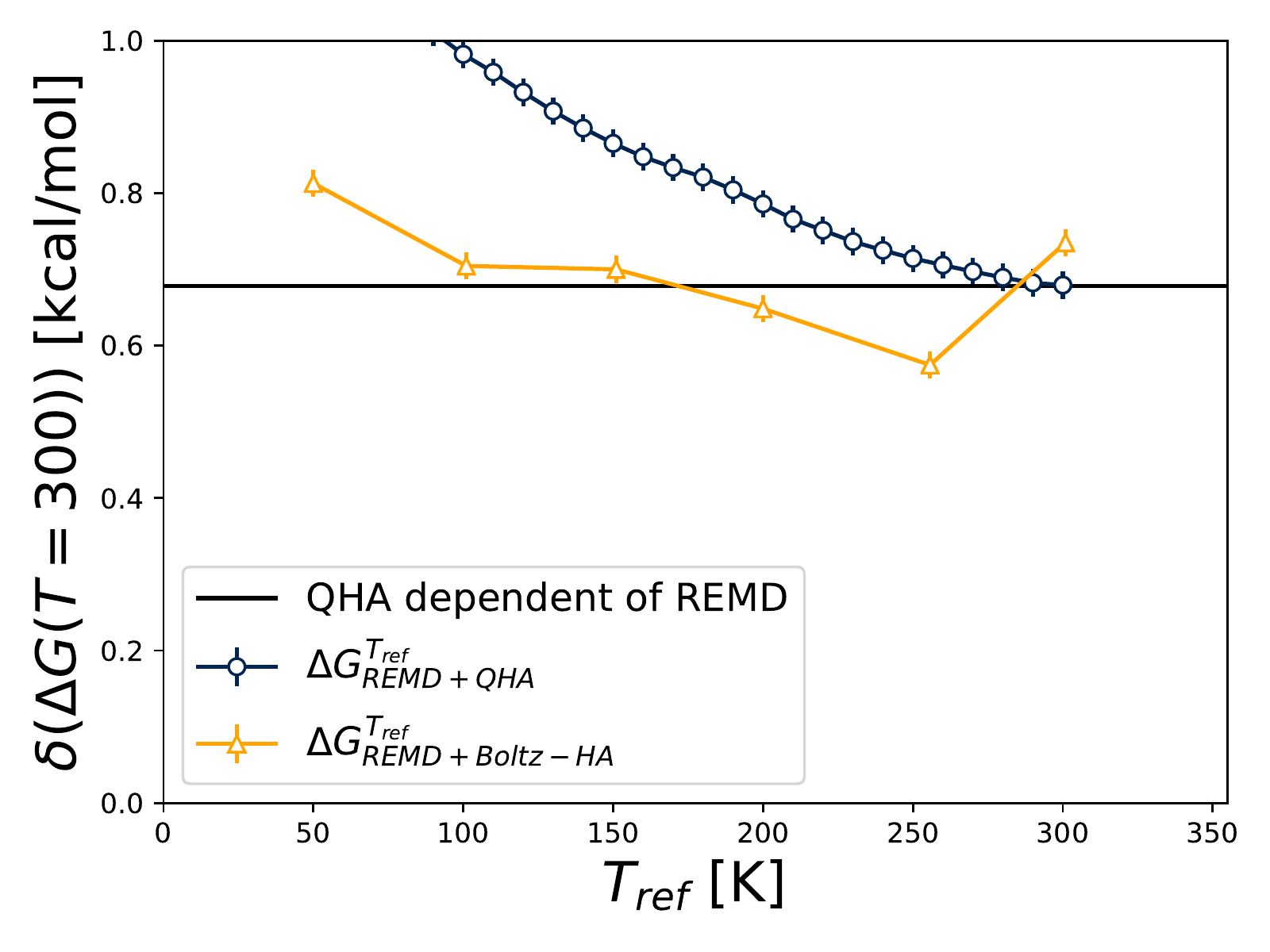}
  \end{center}
  \caption{Left) The polymorph free energy difference of tolbutamide (form III relative to form I) using MD, QHA, and replacing $\Delta G (T_{ref})$ from PSCP in equation ~\ref{eq:MD_dG} with the QHA values for $T_{ref}$ every 10 K is shown. Right) The error in the $\Delta G(300K)$ is reported for replacing $\Delta G (T_{ref})$ from QHA with PSCP and using Boltzmann weighted HA. Each line of the approximation in the left figure is color coded based off the reference temperature used.
  \label{figure:dG_error_16}}
\end{figure*}

\end{singlespace}

\newpage
\newpage

\end{document}